\documentclass[fleqn,usenatbib]{mnras}

\usepackage{newtxtext,newtxmath}

\usepackage[T1]{fontenc}

\DeclareRobustCommand{\VAN}[3]{#2}
\let\VANthebibliography\thebibliography
\def\thebibliography{\DeclareRobustCommand{\VAN}[3]{##3}\VANthebibliography}

\usepackage{graphicx}	
\graphicspath{{Figures/}}
\usepackage{amsmath}	
\usepackage{color}
\usepackage[dvipsnames]{xcolor}
\usepackage{comment}
\usepackage{threeparttable}
\usepackage{subfig}

\newcommand\dd{\mathrm{d}}
\newcommand\DD{\mathrm{D}}
\newcommand\ee{\mathrm{e}}

\newcommand\f{\frac}
\newcommand\p{\partial}
\newcommand\cst{\mathrm{constant}}

\def\XXint#1#2#3{{\setbox0=\hbox{$#1{#2#3}{\int}$}
     \vcenter{\hbox{$#2#3$}}\kern-.5\wd0}}

\def\XXiint#1#2#3{{\setbox0=\hbox{$#1{#2#3}{\iint}$}
     \vcenter{\hbox{$#2#3$}}\kern-.5\wd0}}

\title[Disc breaking in the local model]{Disc breaking and parametric instability in warped accretion discs}

\author[L. E. Held and G. I. Ogilvie]{
Loren E. Held$^{1}$\thanks{E-mail: leh50@cam.ac.uk (LEH)} and Gordon I. Ogilvie$^{1}$
\\
$^{1}$Department of Applied Mathematics and Theoretical Physics, University of Cambridge, Centre for Mathematical Sciences, Wilberforce Road,\\ 
Cambridge CB3 0WA, United Kingdom\\}

\date{Accepted XXX. Received YYY; in original form ZZZ}

\pubyear{2024}

\begin{document}
\label{firstpage}
\pagerange{\pageref{firstpage}--\pageref{lastpage}}
\maketitle

\begin{abstract}
We present the first local simulations of disc breaking/tearing in a warped accretion disc. Warps can arise due to a misalignment between the disc and the rotation axis of the central object, or a misalignment with the orbital plane of a binary (or planetary) companion. Warped discs can break into rings, as found in observations of circumbinary protoplanetary discs and global simulations of tilted discs around spinning black holes. In this work we isolate the mechanism of disc breaking in high-resolution, quasi-2D, local (shearing box), hydrodynamic simulations of a Keplerian disc. We consider the evolution of a free (unforced) warp in the wavelike ($\alpha < H/r$) regime. At large warp amplitudes ($\psi_{\text{max}}  \gtrsim 1$) the disc breaks into four rings on timescales of around 20 orbits which are separated by gaps of around $\sim 10H_0$. The warp exhibits a rich tapestry of small-scale dynamics, including horizontal sloshing motions, vertical oscillations or bouncing, warp steepening, and shocks. The shocks act as a source of enhanced dissipation which facilitates gap opening and thus disc breaking. At smaller warp amplitudes $\psi_{\text{max}} \lesssim 1$, for which we also develop a quasi-linear theory, the disc does not break, but instead exhibits hydrodynamic parametric instability. We also investigate the effect of viscosity: at small warp amplitudes the parametric instability is damped and the warp propagates as a pure bending wave, while at large warp amplitudes the emerging gaps are partially filled by viscous diffusion.
\end{abstract}

\begin{keywords}
accretion, accretion discs -- hydrodynamics -- instabilities, turbulence
\end{keywords}

\section{Introduction}
\label{INTRO}

\subsection{Background}
Simple models of accretion discs often treat the disc as being flat (planar). However, many discs in nature are found to be warped. This could be due to a misalignment in the system, such a circumstellar disc around a tilted spinning black hole, a tilted circumbinary disc, or a circumstellar disc with a tilted external binary (or planetary) companion. Warped discs exhibit various dynamics and processes not found in their planar counterparts.\footnote{Certain processes—specifically precession, parametric instability, and vertical bouncing—are expected to occur in other types of distorted disc as well, e.g. eccentric and tidally distorted discs.} Examples include apsidal and nodal \textit{precession} \citep{bardeen1975lense,papaloizou1995dynamicsoftilted,fragner2010evolution,deng2022non}, as well as \textit{horizontal sloshing motions} \citep{papaloizou1983time, dullemond2022equations} which in turn can become unstable to a purely hydrodynamic instability known as the \textit{parametric instability} \citep{papaloizou1995dynamicsoftilted, gammie2000linear,paardekooper2019local,held2024instability}. Warped discs are also not in vertical hydrostatic equilibrium and so undergo vertical oscillations or \textit{bouncing} in the direction perpendicular to the local disc plane \citep{lubow1981vertically, held2024instability}.

A particularly interesting process in strongly warped discs which has received a great deal of attention over the last decade is that of \textit{disc breaking} (also referred to as \textit{disc tearing}\footnote{Some authors distinguish between disc \textit{breaking} and \textit{tearing} \citep{drewes2021dynamics}, where the former describes breaking in a free warp (without differential precession), while tearing is more an effect due to twisting of a differentially precessing disc subject to an external torque. We will not distinguish between the two terms.}), in which the disc is unable to support itself and breaks into distinct rings. Such discs have been found in nature, both through direct observations, such as the tilted rings seen in images of the protoplanetary disc GW Ori \citep{mathieu1995submillimeter, fang2017millimeter, kraus2020triple}, or suggested indirectly through shadows cast by a tilted inner disc on the outer disc, such as in the protoplanetary discs HD142527 and HD139614 \citep{marino2015shadows, muro2020shadowing}.

The theory of warped discs dates back several decades, and often boils down to solving one-dimensional evolution equations for the local disc tilt vector (see \cite{nixon2016warp} for a review).  A key parameter governing the evolution of warped discs is the ratio of the dimensionless viscous stress (parametrized by the stress-to-thermal pressure ratio $\alpha$) to the disc aspect ratio $H/r$, where $H$ is the scale-height or thickness of the disc. Warped discs are generally classified according to whether they are in the \textit{diffusive} ($\alpha > H/r$) or \textit{wavelike} regimes ($\alpha < H/r$). Warped disc theory in the diffusive regime is relatively well developed, with theories covering both the linear \citep{papaloizou1983evolution,pringle1992simple, lodato2006evolution, nixon2013tearing}, and non-linear regimes \citep{pringle1992simple, ogilvie1999non}. However, theories for the wavelike regime (which is likely relevant to low viscosity protoplanetary discs, and in which the warp propagates as a bending wave)  are either linear \citep{papaloizou1995dynamics, demianski1997dynamics, lubow2000tilting} or weakly non-linear \citep{ogilvie2006non}.

Although the first simulations to observe disc breaking or tearing were carried out nearly three decades ago \citep{larwood1996tidally, larwood1997hydrodynamical}, interest in the topic did not take off until the 2010s.
Disc breaking has now been observed in many global simulations, in particular smoothed particle hydrodynamics (SPH) simulations \citep{lodato2010diffusive,nixon2012tearing,nixon2013tearing,facchini2013wave,lodato2013wave,dougan2015tearing,facchini2018signatures,nealon2020rocking,raj2021disk, drewes2021dynamics, nealon2022bardeen,young2023conditions,martin2024mergers, rowther2025leading,smallwood2025shedding,young2025high}. Grid-based simulations of this phenomenon are less common on account of the complicated warp geometry and the difficulty of resolving thin discs in Eulerian coordinate systems (a few examples include \cite{fragner2010evolution,zhu2019inclined,liska2021disc,kaaz2023nozzle, rabago2024warps}).

Typical configurations in which a warp is induced in an initially tilted disc by an external torque include simulations of a tilted circumbinary disc \citep{larwood1997hydrodynamical,facchini2013wave,nixon2013tearing, facchini2018signatures, nealon2020rocking}, a circumstellar disc with a tilted external perturber \citep{larwood1996tidally, fragner2010evolution, dougan2015tearing, facchini2018signatures, zhu2019inclined}, or a circumstellar disc around a tilted spinning black hole \citep{nelson2000hydrodynamic,nealon2015bardeen, nealon2015apsidal, liska2021disc,raj2021disk,drewes2021dynamics,kaaz2023nozzle,musoke2023disc}. The simplest approach, however, is to initialize the warp ab initio with no external torque and then allow it to evolve \citep{nelson1999hydrodynamic,lodato2010diffusive}, which is known as a \textit{free warp}. This is the approach we will follow in this paper. 

\begin{figure}
    \centering
    \includegraphics[width=1\linewidth]{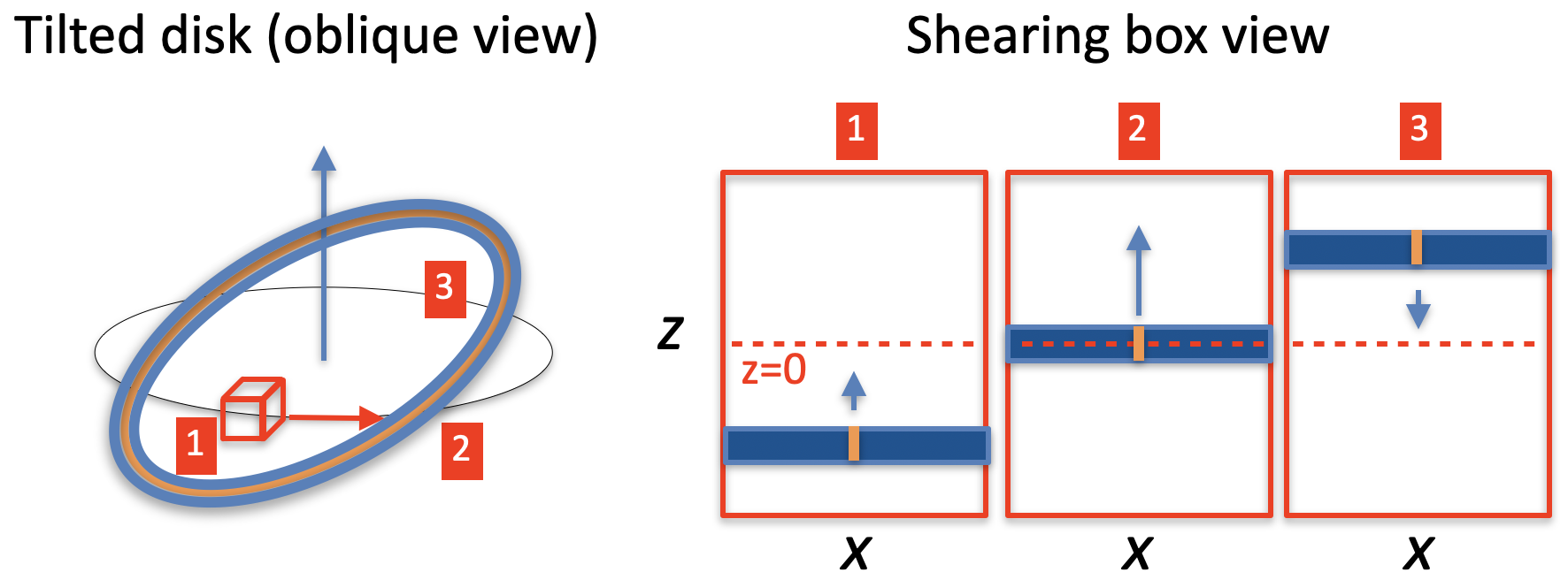}
    \caption{Left: schematic of a tilted disc annulus viewed from a global (inertial) reference frame and a shearing box (red) moving along an untilted circular orbit (the box should be tall enough to intersect the tilted disc, but we have reduced its size for visual clarity). The orange line denotes a single radius inside that annulus. Right: view of the tilted disc from the rotating reference frame of the shearing box. In general, a warped disc will appear in the shearing box as a set of fluid columns oscillating at the local orbital frequency.}
    \label{FIGURE_TiltedDiskShearingBoxView}
\end{figure}

\subsection{Modeling disc breaking using local models}
We aim to study warp propagation (and disc breaking) using the standard shearing box formalism \citep{ogilvie2022hydrodynamics}. See Figure \ref{FIGURE_TiltedDiskShearingBoxView}, which shows a schematic of a tilted disc on the left, and the corresponding local view along an untilted circular orbit on the right. Relative to an observer moving around this untilted circular orbit, a tilted disc will appear to oscillate uniformly about $z = 0$ at the orbital frequency of the radial location about which the box is centred (assuming it is not precessing).\footnote{Note that if the disc is \textit{rigidly} precessing, it will appear to oscillate uniformly about $z = 0$ at some frequency other than the orbital frequency, and if it is \textit{differentially} precessing, then each individual fluid column in the disc will appear to oscillate about $z = 0$ with a different vertical frequency.} The extension to modeling a radially periodic warp rather than a tilted disc is straightforward: instead of a flat disc in which individual fluid columns all start at the same vertical position, we consider a sinusoidal warp in which each fluid column is displaced by a different amount from the shearing box orbital plane.

There are two limitations to this approach. First, in order to model a full wavelength of the warp (say $\lambda \sim 90H_0$) we require boxes that are very large ($L_x = 90H_0$) in the radial direction. Unless the disc is very thin (e.g. $H/r \sim 10^{-3}$), these boxes would span a significant fraction of the radial size of the disc and the radial background structure of the disc would become important. We neglect such background radial density (and pressure) gradients in our model. Second, in order to study large warp amplitudes, we also require a large extent in the vertical direction $z$. At some height above the disc, the shearing box approximation for the vertical component of gravity ($g_z = -\Omega_0^2 z$) will differ from the gravitational potential at that location due to a point source. These limitations notwithstanding, the local model still contains all the dynamics relevant to disc breaking, and has the advantage that it can reach significantly higher resolutions than are currently accessible to global simulations. This allows us to fully resolve both the tenuous regions where disc breaks, as well as small-scale processes such as bouncing and parametric instability.

\subsection{Motivation and aims}
In this work we aim to study the dynamics of disc breaking by means of vertically stratified, quasi-2D (in the $xz$-plane), high-resolution local simulations. For simplicity we will consider the evolution of a free (unforced) and untwisted initial warp with a long wavelength. This set-up is similar to that in the global SPH simulations of \cite{lodato2010diffusive}. We will present results at both small and large warp amplitudes, and at small and large viscosities (all our simulations are in the wavelike regime, as explained in Section \ref{RESULTS_ParameterStudy_ViscosityStudy}).

The structure of the paper is as follows: we present our methods in Section
\ref{METHODS}, including the governing equations, numerical code, initial conditions, and diagnostics. In Section \ref{THEORY} we present a linear theory of small amplitude bending waves, which can be compared to our small warp amplitude simulations. In Section \ref{RESULTS_FIDUCIALSIMULATION} we discuss our fiducial simulation of a free warp with a large initial warp amplitude. To our knowledge this is the first shearing box simulation of disc breaking. In Section \ref{RESULTS_PARAMETERSTUDY} we present a parameter study, first of initial maximum warp amplitudes ($\psi_{\text{max}} \sim 0.1, 0.6, 1.3, 2$) and then a viscosity study (inviscid, $\alpha = 0.003, 0.03$). In Section \ref{RESULTS_DISCUSSION} we summarize our key results, and compare them to existing theories of disc breaking. We conclude in Section \ref{CONCLUSIONS}.

\section{Methods}
\label{METHODS}

\subsection{Governing equations}
\label{METHODS_GoverningEquations}
We work in the standard shearing-box formalism. In this approximation the domain consists of a Cartesian box centered around some cylindrical radius $r_0$ in the disc, and which is orbiting with the angular frequency $\Omega_0$ of the gas at that radius. The governing equations are given by

\begin{align}
&\partial_t \rho + \nabla \cdot (\rho \mathbf{u}) = 0, \label{EQUATION_SB1}\\
&\partial_t \mathbf{u} + \mathbf{u}\cdot\nabla \mathbf{u} = -\frac{1}{\rho} \nabla P - 2\Omega_0 \mathbf{e}_z \times \mathbf{u} + \mathbf{g_\text{eff}}+\frac{1}{\rho}\nabla \cdot \mathbf{T}, \label{EQUATION_SB2}
\end{align}
with the symbols taking their usual meanings (see \cite{heldlatter2018}). We model the gas as being isothermal with an equation of state given by $P = c_{\text{s}0}^2 \rho$, where $c_{\text{s}0}$ is the constant sound speed.

We include vertical stratification by including the $z$-component of gravity. In a standard vertically stratified shearing box, the gravitational acceleration is given by $\mathbf{g_\text{eff}}=2q\Omega_0^2x\,\mathbf{e}_x-\Omega_0^2 z\,\mathbf{e}_z$ (third term on the right-hand side of Equation \ref{EQUATION_SB2}), where $q$ is the dimensionless shear parameter $q \equiv -\left.d\ln{\Omega}/d\ln{r}\right\vert_{r=r_0}$. The background orbital shear flow is given by $u_{y0} = -q \Omega_0 x$. For Keplerian discs $q=3/2$, a value we adopt throughout this paper. 

\subsection{Numerical set-up}
\label{METHODS_NumericalSetup}

\subsubsection{Code}
\label{METHODS_Code}
We use the finite-volume code \textsc{PLUTO} (version 4.4, patch 2) for all our simulations \citep{mignone2007}. The solution is advanced in time using the 2nd-order-in-time Runge--Kutta algorithm. Most of our runs were carried out using the HLLC Riemann solver and 3rd-order-in-space parabolic reconstruction, although we have also tested other solvers and interpolation methods (see Appendix \ref{APPENDIX__ParameterStudy_RiemannSolverInterpolationStudy}). The boundary conditions are implemented using ghost cells which border the active domain. When we include explicit viscosity $\nu$, we also employ the Super-Time-Stepping (STS) scheme \citep{alexiades1996super}, which reduces the computational time. We employ the built-in shearing box module in PLUTO \citep{mignone2012}, which evolves the primitive-variable equations (Equations \ref{EQUATION_SB1}-\ref{EQUATION_SB2}) in conservative form.

\subsubsection{Units}
\label{METHODS_Units}
The time unit is given by the inverse angular frequency $\Omega_0$, speeds are measured in units of the isothermal sound speed at initialization $c_{\text{s}0}$, and the mass unit is given by the initial mid-plane density $\rho_0$. All other units are formed from combinations of these base units, e.g. lengths are measured in units of the initial scale-height $H_0 = c_{\text{s}0}/\Omega_0$.

\subsubsection{Box size, density floor, and resolution}
\label{METHODS_BoxSizeDensityFloorResolution}
We typically employ a quasi-2D domain in $(x,z)$ with radial size $L_x = 90H_0$, and a nominal size of $L_y = 0.08H_0$ in the azimuthal direction. The size of the domain in the vertical direction depends on the initial (maximum) warp amplitude in the disc (see below), but ranges from $L_z = 68H_0$ for our largest warp amplitude runs to $L_z = 48H_0$ for smaller warp amplitude runs. The large vertical box size means we need to employ a density floor, which we set to $\rho_{\text{floor}} \sim 1.3\times 10^{-14}\rho_0$. Our typical resolution is $64$ cells-per-scale-height $H_0$, although we have also carried out select runs at a lower resolutions of $16/H_0$ and $32/H_0$, and also at a higher resolution of $128/H_0$. 

\subsubsection{Boundary conditions}
\label{METHODS_BoundaryConditions}
In the radial direction we employ standard shear-periodic boundary conditions. In the azimuthal direction we employ periodic boundary conditions. In the vertical direction we typically employ \textit{reflective} boundary conditions ($\rho \rightarrow \rho$, $u_z \rightarrow -u_z$, and $u_t \rightarrow u_t$, where $t$ denotes the transverse (i.e. the $x$- or $y$-) components of the velocity. We have also checked the effect of \textit{outflow} boundary conditions (see Appendix \ref{APPENDIX__ParameterStudy_BoundaryConditionStudy}), whereby the vertical gradients of all velocity components are zero, i.e. $\partial_z u_i = 0\,\,(i\in{x,y,z})$ (numerically, we set variables in the ghost zones equal to those in the active cells bordering the ghost zones), and the density in the ghost zones is kept in vertical isothermal hydrostatic equilibrium.

\subsubsection{Initial condition}
\label{METHODS_InitialCondition}
Our initial condition consists of a vertically stratified disc that has been given an artificial sinusoidal warp 
with radial wavelength $\lambda = 2 \pi/k = L_x = 90H_0$. The initial density profile is given by     

\begin{equation}
\rho = \rho_0\exp\left({-\frac{(z+A\sin{k x})^2}{2H_0^2}}\right).
\label{EQUATION_FreeWarpDensityProfile}
\end{equation}
This initial condition corresponds to applying an $x$-dependent vertical displacement $-A\sin kx$ to a standard hydrostatic disc. An example for $A=28H_0$ is shown in the top-left-hand panel of Figure \ref{FIGURE_FiducialSimulationFlowField}. The maximum initial warp amplitude is related to this parameter through $\psi_{\text{max}} = Ak$, and thus the amplitude of the warp can be controlled through the parameter $A$. This set-up is not in equilibrium by design as the warp is free to evolve. Our set-up resembles that found in the global SPH simulations of \cite{lodato2010diffusive} which started with a disc that was given a large warp of radial wavelength about $90H_0$ at initialization and which was then allowed to evolve freely.

In global simulations tilted discs are often initialized by means of a \textit{tilt angle}.\footnote{Note that even in a global simulation of a free warp around a point mass, the absolute tilt angle is not significant and only differential tilts are meaningful. This is in contrast to discs surrounding, say, a central binary or spinning black hole, with a preferred plane in which the disc is rigidly precessing  by some angle with respect to that plane.} In the local approach we lose significance of absolute tilt angle in degrees or radians. The important parameter is the ratio of the tilt angle to the angular thickness of the disc, which, in the local model, is effectively a comparison of the vertical displacement of the disc to its scale-height, i.e. $A/H_0$.

\subsection{Diagnostics}
\label{METHODS_Diagnostics}

\subsubsection{Energetics}
\label{METHODS_Energetics}
The total energy density is given by

\begin{equation}
    E_{\text{total}} = \frac{1}{2}\rho u^2 + \rho \Phi + c_{\text{s}0}^2\rho\ln{\rho}. 
    \label{EQUATION_TotalEnergy}
\end{equation}
Here $E_{\text{kin}}$ is the total kinetic energy density, $\rho \Phi$ the gravitational potential energy density, and $\Phi =-\frac{3}{2}\Omega_{0}^2x^2+\frac{1}{2}\Omega_{0}^2z^2$ is the standard effective gravitational potential in the shearing-box approximation for a Keplerian disc. The term $c_{s0}^2 \rho \ln{\rho}$ embodies the variable part of the Helmholtz free
energy, which replaces the internal energy in an isothermal gas (see footnote 5 in \cite{held2024instability} for a brief derivation).

\subsubsection{Averaged quantities}
\label{METHODS_Averages}

The volume-average of a quantity $X$ is denoted $\langle X \rangle$ and is defined as 
\begin{equation}
\langle X \rangle(t) \equiv \frac{1}{V} \int_V X(x, y, z, t) \, \dd V,
\label{EQUATION_VolumeAverageDefinition}
\end{equation}
where $V=L_x L_y L_z$ is the volume of the box. 

The orbit-average of a quantity $X$ is denoted as $\langle X \rangle_{\tau}$ and is defined as

\begin{equation}
\langle X \rangle_{\tau} \equiv \frac{1}{\tau} \int X(x,y,z,t)\,\dd t,
\label{EQUATION_OrbitAverageDefinition}
\end{equation}
where $\tau = 2\pi/\Omega_0$ is an orbital period. By orbit-averaging various diagnostics (see below), we can eliminate any contaminating signals which are not at the orbital frequency.\footnote{For various diagnostics defined below, the notation $\langle X \rangle_{\tau}$ will also imply an additional mass-weighted average $(1/\Sigma)\int \rho X\,\,\dd z$ or a vertical-average $ \langle X \rangle_z \equiv (1/L_z)\int X\,\,\dd z$, and possibly multiplication by $e^{i\Omega_0 t}$.}

\subsubsection{Surface density, centre of mass, and disc thickness}
\label{METHODS_SurfaceDensityCOMDiskThickness}
The profile of the surface density at time $t$ (in units of $\rho_0 H_0$) is defined as:\footnote{Here and elsewhere, the $y$-average of the integrand is implied, i.e. $\langle \rho \rangle_{y}$, $\langle \rho Z \rangle_{y}$, etc, where $\langle X\rangle_y \equiv (1/L_y)\int X\,\, \dd y$. For visual clarity, and because our domain is essentially two-dimensional, we drop the angle brackets $\langle \cdot \rangle_y$.}

\begin{equation}
    \Sigma(x,t) \equiv \int_{-L_z/2}^{L_z/2}  \rho\,\,\dd z.
    \label{EQUATION_SurfaceDensity}
\end{equation}
The orbit-averaged radial profile of surface density is $\langle \Sigma \rangle_{\tau}(x) \in \mathbb{R}$.

The instantaneous radial profile of the vertical centre of mass coordinate (in units of $H_0$) is given by

\begin{equation}
    \bar{z}(x,t) \equiv \frac{1}{\Sigma(x,t)}\int_{-L_z/2}^{L_z/2}  \rho z\,\,\dd z.
    \label{EQUATION_COM}
\end{equation}

Warped discs are expected to bounce twice per orbit, so we expect the thickness of the disc to change in time. An intrinsic measure of disc thickness is the dynamical scale-height $H(x,t)$ (in units of $H_0$) defined as:
\begin{equation}
    H^2(x,t) = \frac{1}{\Sigma(t,x)}\int_{-L_z/2}^{L_z/2}  \rho (z-\bar{z})^2 \dd z.
     \label{EQUATION_DiskThickness}
\end{equation}
Note that this definition differs slightly from that used in \cite{held2024instability}. There $H^2(t) \equiv (1/\Sigma)\int \langle \rho \rangle_{x} z^2 \dd z$, where $\langle \rho \rangle_{x}$ denotes a radial average, which made sense for a purely bouncing disc for which we did not expect $\bar z\ne0$. For our warped disc set-up here, it is more prudent to use $z-\tilde{z}$ (in order to pick up the location of the local disc mid-plane), and to calculate a radial profile (because the thickness varies with radius as well as time).

\subsubsection{Reynolds stress}
\label{METHODS_ReynoldsStress}
In purely hydrodynamic discs, the radial transport of angular momentum is related to the $xy$-component of the Reynolds stress tensor $R_{xy} \equiv \rho u_x \delta u_y$, where $\delta u_y \equiv u_y + q\Omega_0 x$ is the perturbation of the $y$-component of the total velocity $u_y$ about the background Keplerian flow $u_{y0} = -(3/2) \Omega_0 x$. Note that the Reynolds stress is related to the classical dimensionless angular momentum transport parameter $\alpha$ by $\alpha \equiv \langle R_{xy} \rangle / \langle P \rangle$.

The orbital- and vertically-averaged radial profile of the $xy$-component of Reynolds stress is defined as

\begin{equation}
    \langle R_{xy} \rangle_{\tau}(x) \equiv \left\langle \frac{1}{L_z}\int \rho u_x \delta u_y\,\, \dd  z\right\rangle_{\tau},
    \label{EQUATION_ReyonldsStress}
\end{equation}
where $\langle R_{xy} \rangle_{\tau} \in \mathbb{R}$ and has units of $\rho_0 c_{\text{s}0}^2$. 

\subsubsection{Complex tilt}
\label{METHODS_ComplexTilt}

Following \cite{ogilvie2022hydrodynamics}, we can define the local equivalent of a complex tilt, $Z = (z+iu_z/\Omega_0)e^{i\Omega_0t}$.
This quantity can be thought of as the complex amplitude of a harmonic vertical oscillation and uses information from the vertical position and velocity. The mass weighted and orbit-averaged oscillation amplitude is then

\begin{equation}
   \langle  Z \rangle_{\tau} = \left\langle \frac{1}{\Sigma} \int \rho Z\,\,\dd z \right\rangle_{\tau},
   \label{EQUATION_ComplexTilt}
\end{equation}
where $\langle Z \rangle_{\tau} \in \mathbb{C}$ and has units of $H_0$. Note that in the literature,  the tilt is often defined in terms of a tilt vector $\mathbf{l} = [l_x, l_y, l_z] = [\cos{\gamma}\sin{\beta}, \sin{\gamma}\sin{\beta}, \cos{\beta}]$, where $\beta(r,t)$ and $\gamma(r,t)$ are the tilt and twist angles, respectively \citep{nixon2016warp}. The horizontal (misaligned) components of the tilt vector can then be combined into a complex tilt vector $l_x+il_y$, and there is a correspondence between this quantity and its local equivalent via $-Z/r_0 \leftrightarrow l_x + il_y = \sin{\beta} \, e^{i\gamma}$, where $r_0$ is the radius at which the local model is constructed.

\subsubsection{Complex warp amplitude}
\label{METHODS_ComplexWarpAmplitude}
The orbit-averaged complex warp amplitude can be defined in terms of the complex tilt $Z$ as

\begin{equation}
    \langle \psi \rangle_{\tau}(x) = \left\langle -\frac{\partial}{\partial x}\left(\frac{1}{\Sigma} \,\int \rho Z \,\,\dd z \right) \right\rangle_{\tau},
    \label{EQUATION_ComplexWarpAmplitude}
\end{equation}
where $\langle \psi \rangle_{\tau} \in \mathbb{C}$ and is dimensionless. The correspondence between this quantity and the warp amplitude in cylindrical coordinates is $(\psi/r_0)e^{i\gamma} \leftrightarrow (l_x + il_y)' = (\beta'\cos{\beta} + i\gamma'\sin{\beta})e^{i\gamma}$, where here (and only here) the prime denotes a radial derivative.

\subsubsection{Complex torque}
\label{METHODS_ComplexTorque}
The orbit-averaged complex `torque' is defined as

\begin{equation}
 \langle G \rangle_{\tau}(x) =  \left\langle \int  -\rho u_x Z\,\, \dd z \right\rangle_{\tau},
 \label{EQUATION_ComplexTorque}
 \end{equation}
 where $\langle G \rangle_{\tau} \in \mathbb{C}$ and has units of $\rho_0 H_0^{3} \Omega_0$. We associate this quantity with the radial flux of the misaligned ($x$ and $y$) components of angular momentum. Note that $\rho u_x Z$ itself has units of $\rho_0 \Omega_0 H_0^2$, while an angular momentum flux (or torque) has units of $\rho_0 H_0^5 \Omega_0^2$. However, the quantity $\int -\rho u_x Z \dd z \cdot 2\pi r_0 \cdot r_0 \Omega_0$ does have the same dimensions as an angular momentum flux or torque (units of $\rho_0 H_0^5 \Omega_0^2$), so, for simplicity, we simply refer to Equation \ref{EQUATION_ComplexTorque} as a complex torque.

\subsubsection{Amplitude of horizontal sloshing motions}
The warp gives rise to a vertical shear $u_x(z)$ and $\delta u_y(z)$, also referred to as horizontal `sloshing' motions \citep{dullemond2022equations}. These sloshing motions play an important role in mediating the torque. Since the sloshing motions consist of a horizontal velocity proportional to the height above the warped midplane, the orbit-averaged amplitudes of these motions can be defined as

\begin{equation}
    \langle U_x \rangle_{\tau}(x) \equiv \left\langle \left(\frac{1}{\Sigma H^2} \int \rho u_x (z-\bar{z})^2\,\,\dd z\right)e^{i\Omega_0 t}  \right\rangle_{\tau},
    \label{EQUATION_ComplexShearAmplitudeXcomponent}
\end{equation}
\begin{equation}
    \langle U_y \rangle_{\tau}(x) \equiv \left\langle \left(\frac{1}{\Sigma H^2} \int \rho \, \delta u_y (z-\bar{z})^2\,\,\dd z \right)e^{i\Omega_0 t} \right\rangle_{\tau},
     \label{EQUATION_ComplexShearAmplitudeYcomponent}
\end{equation}
where $U_x,U_y \in \mathbb{C}$ and have units of $\Omega_0$.

\section{Theory}
\label{THEORY}
In this section we present an analytical theory for the evolution of small-amplitude bending waves without dissipation. Key predictions of the model include the modulation of fluid quantities at the bending-wave frequency (which appears as a beat in the kinetic energy), the expected fractional change in surface density at different locations in the disc, and also the sign and magnitude of the Reynolds stress which is driven by the horizontal sloshing motions. We find that these predictions are in excellent agreement with the corresponding small-warp-amplitude ($A = 2H_0, \psi_{\text{max}} \sim 0.14)$ simulation which we discuss in Section~\ref{RESULTS_ParameterStudy_InitialWarpAmplitudeStudy}. 

For simplicity, we limit the discussion here to the inviscid case, and quote key results. A more detailed derivation is given in Appendix~\ref{APPENDIX_InviscidBendingWaveTheory}.

\subsection{Governing equations}
\label{THEORY_GoverningEquations}
Consider axisymmetric motions of an inviscid, isothermal gas in a Keplerian shearing box. The governing equations, equivalent to equations~(\ref{EQUATION_SB1}) and~(\ref{EQUATION_SB2}), are
\begin{align}
  &\DD u_x-2\Omega_0\,\delta u_y=-\p_xh,\label{theory1}\\
  &\DD\,\delta u_y+\f{1}{2}\Omega_0 u_x=0,\label{theory2}\\
  &\DD u_z=-\Omega_0^2z-\p_zh,\label{theory3}\\
  &\DD h=-c_{\text{s}0}^2(\p_xu_x+\p_zu_z),\label{theory4}
\end{align}
where $\delta u_y=u_y+q\Omega_0 x$, the enthalpy per unit mass of an isothermal gas is given by $h=c_{\text{s}0}^2\ln\rho$ \citep{held2024instability}, and $\DD=\p_t+u_x\p_x+u_z\p_z$ is the Lagrangian time-derivative for axisymmetric motions. The basic state is
\begin{equation}
  h_0=\cst-\f{1}{2}\Omega_0^2z^2,\qquad
  u_{x0}=\delta u_{y0}=u_{z0}=0,
  \label{EQUATION_TheoryNoDissipationBackgroundState}
\end{equation}
corresponding to the usual hydrostatic equilibrium with a Gaussian distribution of density and pressure with scale-height $H_0=c_{\text{s}0}/\Omega_0$.

\subsection{Linearized equations}
\label{THEORY_LinearizedEquations}
Decompose each fluid variable $X\in\{u_x,u_y,u_z,h\}$ into a background and perturbed part, i.e. $X =X_{0}+X'$. The linearized equations for small perturbations are
\begin{align}
  &\p_tu'_x-2\Omega_0 u'_y=-\p_xh',\label{linearized1}\\
  &\p_tu'_y+\f{1}{2}\Omega_0 u'_x=0,\label{linearized2}\\
  &\p_tu'_z=-\p_zh',\label{linearized3}\\
  &\p_th'-\Omega_0^2zu'_z=-c_{\text{s}0}^2(\p_xu'_x+\p_zu'_z),\label{linearized4}
\end{align}
where $h'=p'/\rho=c_{\text{s}0}^2\rho'/\rho$ is the Eulerian enthalpy perturbation.

\subsection{Sinusoidal bending wave ansatz}
\label{THEORY_BendingWaveAnsatz}
We wish to study the evolution of a free warp, which can be done in the context of a periodic box by considering a sinusoidal bending wave. The initial condition in our simulations (see Equation \ref{EQUATION_FreeWarpDensityProfile}) can be interpreted as a corrugation in the form of a vertical displacement $\xi_z=-A\sin kx$, corresponding to an initial enthalpy perturbation $h'=-\xi_z\p_zh_0=-A\Omega_0^2z\sin kx$. In the time-dependent bending wave, the vertical displacement $\xi_z$ is related to the velocity perturbation by $u_z' = \partial_t \xi_z$. The maximum initial warp amplitude is $\psi_{\text{max}} =Ak$, which is the maximum value of $|\p\xi_z/\p x|$.

We search for standing-wave solutions of the linearized equations (Section~\ref{THEORY_LinearizedEquations}), having the form

\begin{align}
  &u_x'=A_xz\cos kx\sin\omega t,\label{ansatz1}\\
  &u_y'=A_yz\cos kx\cos\omega t,\label{ansatz2}\\\
  &u_z'=A_z\sin kx\sin\omega t,\label{ansatz3}\\\
  &h'=A_hz\sin kx\cos\omega t,\label{ansatz4}\
\end{align}
where $A_x$, etc., are constant amplitudes. The motivation behind this ansatz relates to the ideas behind the `sloshing' motion mentioned above. The horizontal pressure gradient in the perturbed/warped disc, divided by the density, is proportional to the vertical coordinate $z$ (see Section 4 of \cite{papaloizou1983time}). Thus, in linear theory, we are looking at the $n=1$ mode of the disc (where $n$ is the vertical mode number of a perturbation), which involves a vertical displacement (or velocity) independent of $z$ and a horizontal one proportional to $z$ \citep{ogilvie2022hydrodynamics}.

\subsection{Dispersion relation}
\label{THEORY_DispersionRelation}
Substituting the above ansatz into the linearized equations (\ref{linearized1})--(\ref{linearized4}) and eliminating $A_y$ and $A_z$, we obtain the following equations:
\begin{align}
  &(\omega^2-\Omega_0^2)A_x=-k\omega A_h,
    \label{CoupledOscillator1}\\
  &(\omega^2-\Omega_0^2)A_h=-c_{\text{s}0}^2k\omega A_x.\label{CoupledOscillator2}
\end{align}

Equation~(\ref{CoupledOscillator1}) says that the horizontal epicyclic oscillator, with a natural frequency of $\kappa=\Omega_0$ in a Keplerian disc, is forced by the horizontal pressure gradient (product of $k$ and $A_h$). Equation~(\ref{CoupledOscillator2}) says that the vertical oscillator, also with a natural frequency of $\Omega_0$, is forced by the horizontal velocity divergence (product of $k$ and $A_x$). In these ways the two oscillators are coupled together. Eliminating $A_x$ and $A_h$ we obtain the dispersion relation
\begin{equation}
  (\omega^2-\Omega_0^2)^2=c_{\text{s}0}^2k^2\omega^2,
  \label{EQUATION_TheoryNoDissipationDispersionRelation}
\end{equation}
which relates the frequency $\omega$ of the bending wave to its radial wavenumber $k$. For standing waves, we can take $\omega$ and $k$ to be positive without loss of generality. Using the quadratic formula we can write the two positive roots of Equation~(\ref{EQUATION_TheoryNoDissipationDispersionRelation}) as
\begin{equation}
  \omega=\omega_\pm=\left[\Omega_0^2+\f{1}{4}(c_{\text{s}0}k)^2\right]^{1/2}\pm\f{1}{2}c_{\text{s}0}k. \label{EQUATION_Theory_DispesionRelation2}
\end{equation}

\subsection{Mode amplitudes}
\label{THEORY_ModeAmplitudes}
To derive expressions for the amplitudes $A_x$ and $A_h$ we then back-substitute Equation~(\ref{EQUATION_Theory_DispesionRelation2}) for $\omega$ into Equations (\ref{CoupledOscillator1})--(\ref{CoupledOscillator2}) to work out the ratio of $A_x$ to $A_h$ (and similarly for $A_y$ and $A_z$ for the coupled oscillator equations [not shown] relating those two amplitudes). After choosing a normalization that is convenient for our initial condition on $h'$, we obtain
\begin{equation}
  A_h=-\Omega_0^2H_0,\quad
  A_x=\pm\Omega_0,\quad
  A_y=\pm\f{\Omega_0^2}{2\omega_\pm},\quad
  A_z=\f{\Omega_0^2H_0}{\omega_\pm}.
\end{equation}

Let us denote a particular mode as $\mathcal{S} = \{u_x', u_y', u_z', h'\}$. If the standing wave has a frequency $\omega_{+}$ (i.e. greater than the orbital frequency $\Omega_0$) we denote the wave as $\mathcal{S}_{+}$, and if it has a frequency $\omega_{-}$ (i.e. less than the orbital frequency) we denote it as $\mathcal{S}_{-}$. Each of our standing wave solutions has a non-zero $u_y'$ at $t=0$. If the initial condition used in our simulations (Equation \ref{EQUATION_FreeWarpDensityProfile}) perturbs only the density and not the velocity, then we will not get a pure standing wave but rather a superposition of the two standing waves such that $u_y'$ vanishes at $t=0$.\footnote{The sinusoidal density profile which we use as the initial condition in our simulations (Equation \ref{EQUATION_FreeWarpDensityProfile}) can be thought of as a perturbation of the density. The velocity perturbations which we seed the simulations with, on the other hand, are random noise which does not directly affect the resultant bending wave, and whose purpose is simply to seed any possible instabilities. Thus, in the simulations, we initialize an initial vertical displacement (a warp), but not sloshing motions.}
Thus the combination that we need need in order to match the initial condition $u_y'=0$ and with an initial vertical displacement $\xi_z=-A\sin kx$ is $(C_{+}\mathcal{S}_{+} + C_{-}\mathcal{S}_{-})(A/H_0)$.

To work out the coefficients $C_{\pm}$ we first require
$u_y'$ to vanish at $t=0$, which implies $(C_+/\omega_+)-(C_-/\omega_-)=0$ and so determines the ratio $C_+/C_-=\omega_+/\omega_-$. Next we require $h'$ to equal $-A\Omega_0^2z\sin kx$ at $t=0$, which implies $C_++C_-=1$.
Thus the coefficients are given by
\begin{equation}
  C_\pm=\f{\omega_\pm}{\omega_++\omega_-}=\f{1}{2}\left[1\mp\f{c_{\text{s}0}k}{\sqrt{4\Omega_0^2+(c_{\text{s}0}k)^2}}\right].
\end{equation}

\begin{figure}
    \centering
    \includegraphics[width=1\linewidth]{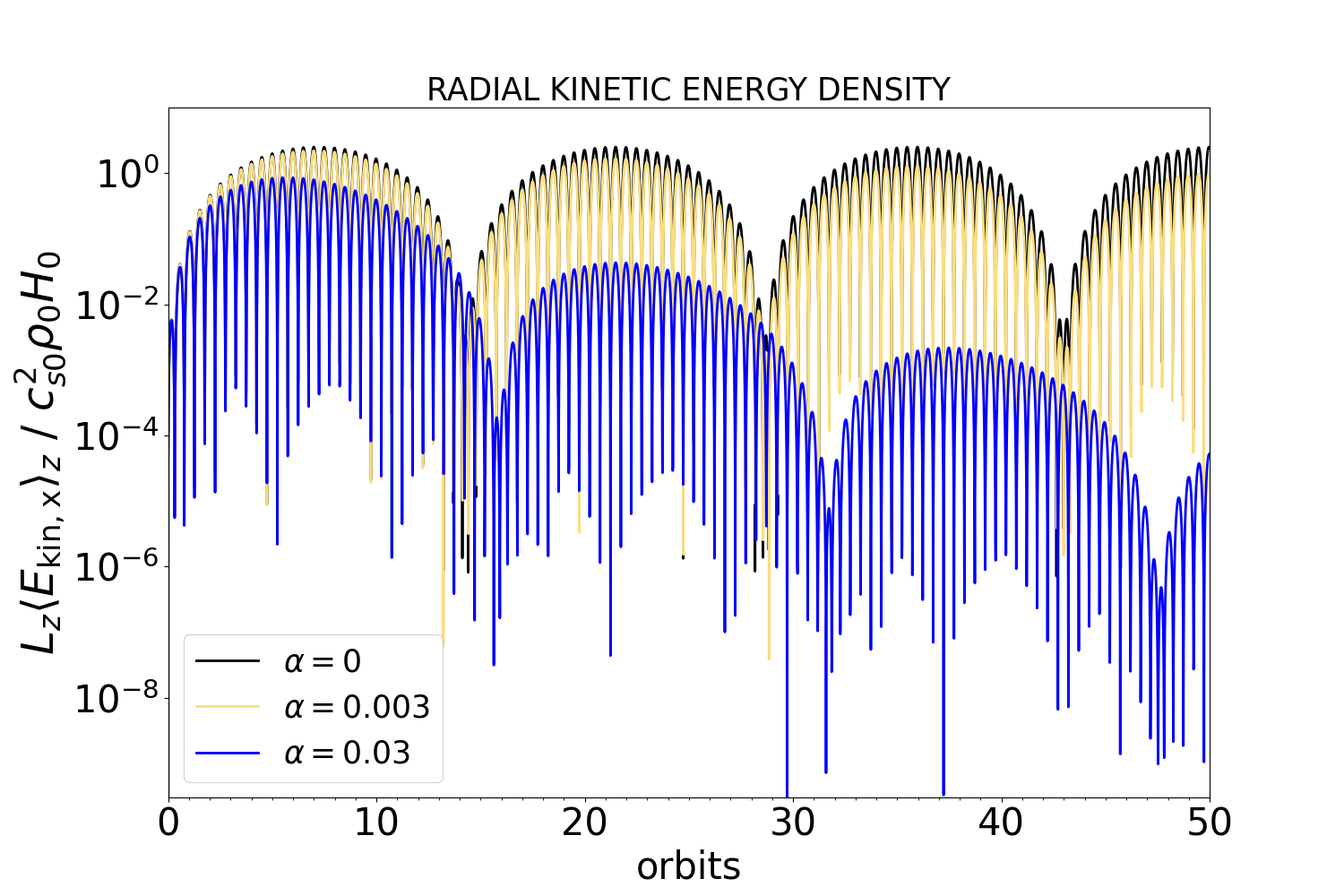}
    \includegraphics[width=1\linewidth]{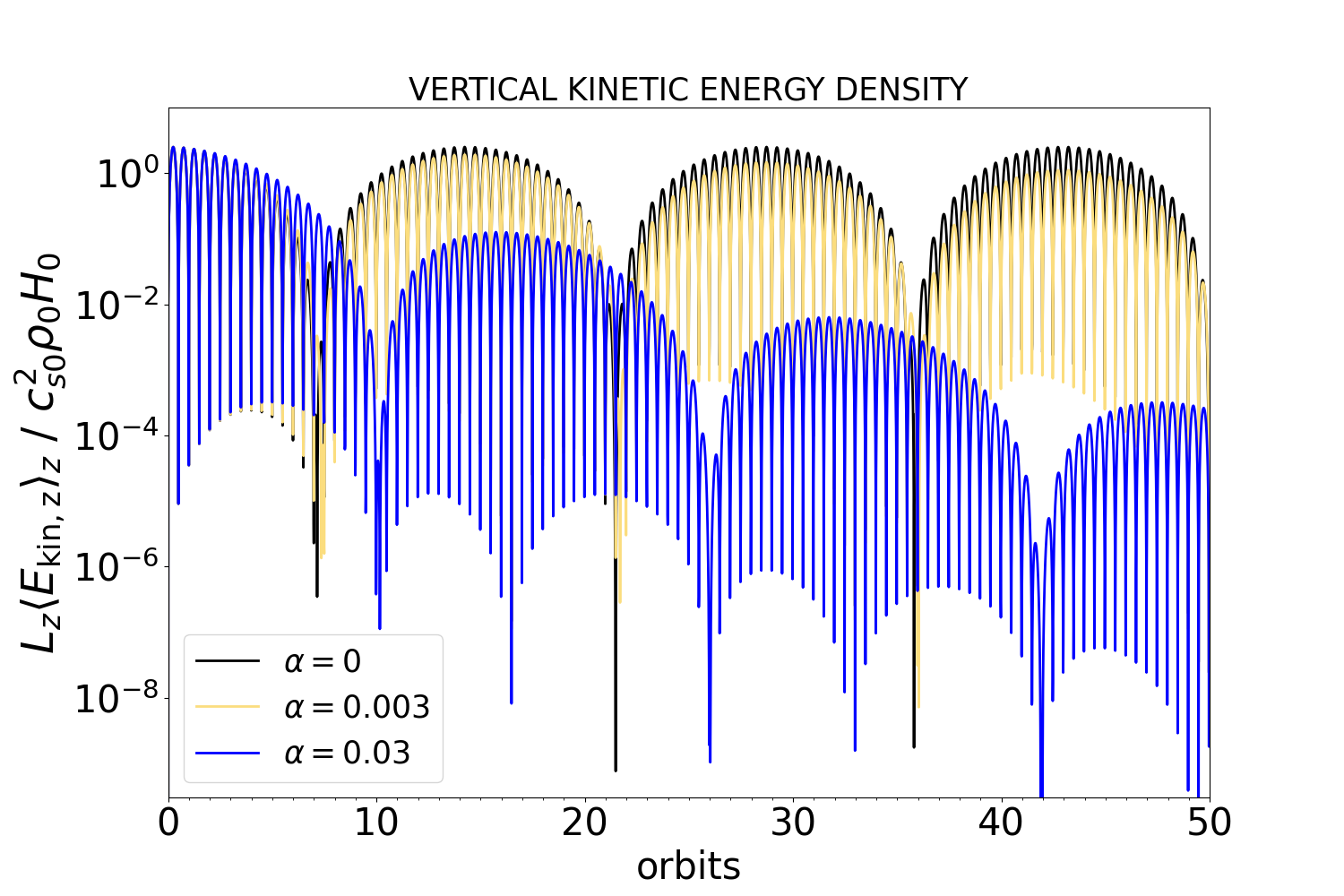}
    \caption{Theoretical prediction of time-evolution of vertically-integrated radial kinetic energy density (top panel) and vertical kinetic energy (bottom panel) for three different different viscosities: inviscid (black curve: wavelike regime), $\alpha = 0.003$ (gold curve), $\alpha = 0.03$ (blue curve).}
    \label{FIGURE_Theory_EkinxEkinzTimeSeries}
\end{figure}

\subsection{Structure of the bending wave (long wavelength limit)}
\label{THEORY_LongWavelengthLimit}
The usual long-wavelength bending-wave theory found in papers on warped discs corresponds to the limit $kH_0\ll1$ (i.e.\ $c_{\text{s}0}k\ll\Omega_0$) and the approximations
\begin{equation}
  \omega_\pm\approx\Omega_0\pm\f{1}{2}c_{\text{s}0}k
  \label{THEORY_dispersionrelationlongwavelengthlimit}
\end{equation}
of the dispersion relation. The leading approximations to the modes are
\begin{equation}
  A_h=-\Omega_0^2 H_0,\quad
  A_x=\pm\Omega_0,\quad
  A_y\approx\pm\f{\Omega_0}{2},\quad
  A_z\approx\Omega_0 H_0.
\end{equation}
The leading approximation to the coefficients of the two standing waves is $C_\pm\approx\f{1}{2}$, meaning that the initial condition excites the modes $\mathcal{S}_\pm$ with equal amplitude. 

\begin{figure*}
\centering
\includegraphics[scale=0.19]{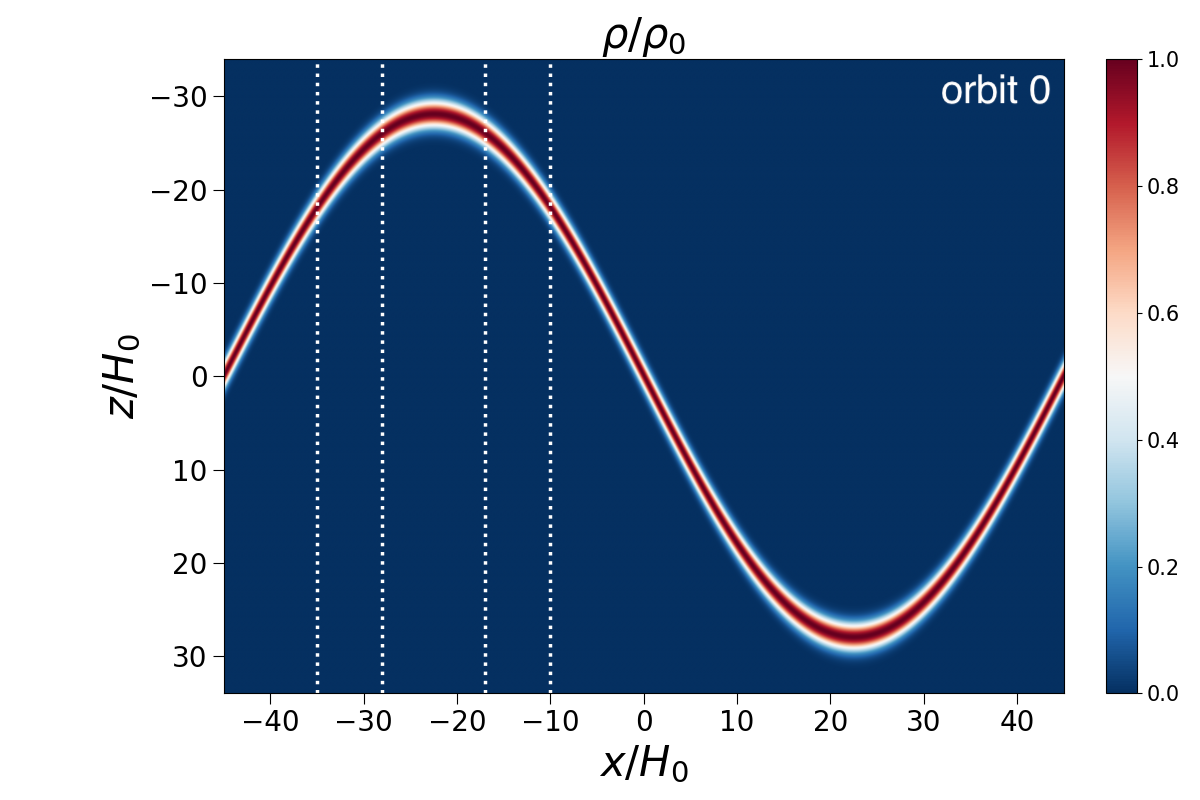}
\includegraphics[scale=0.19]{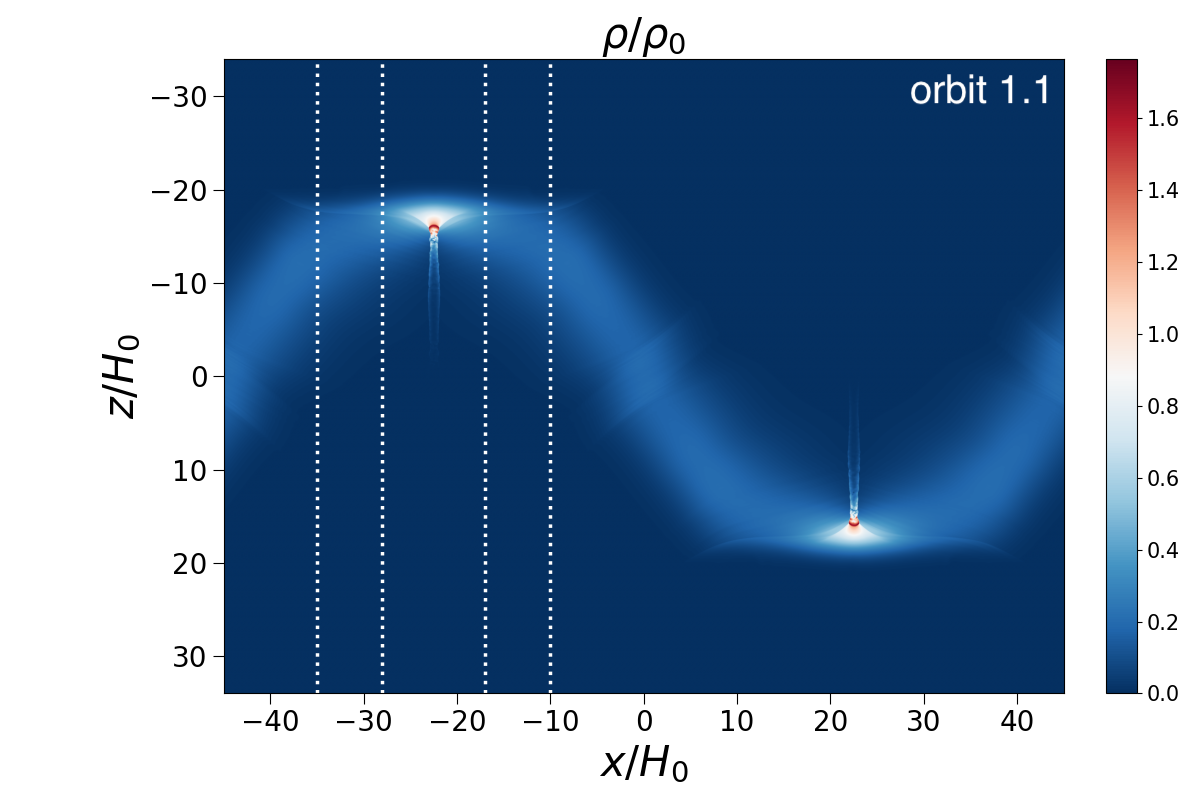}
\includegraphics[scale=0.19]{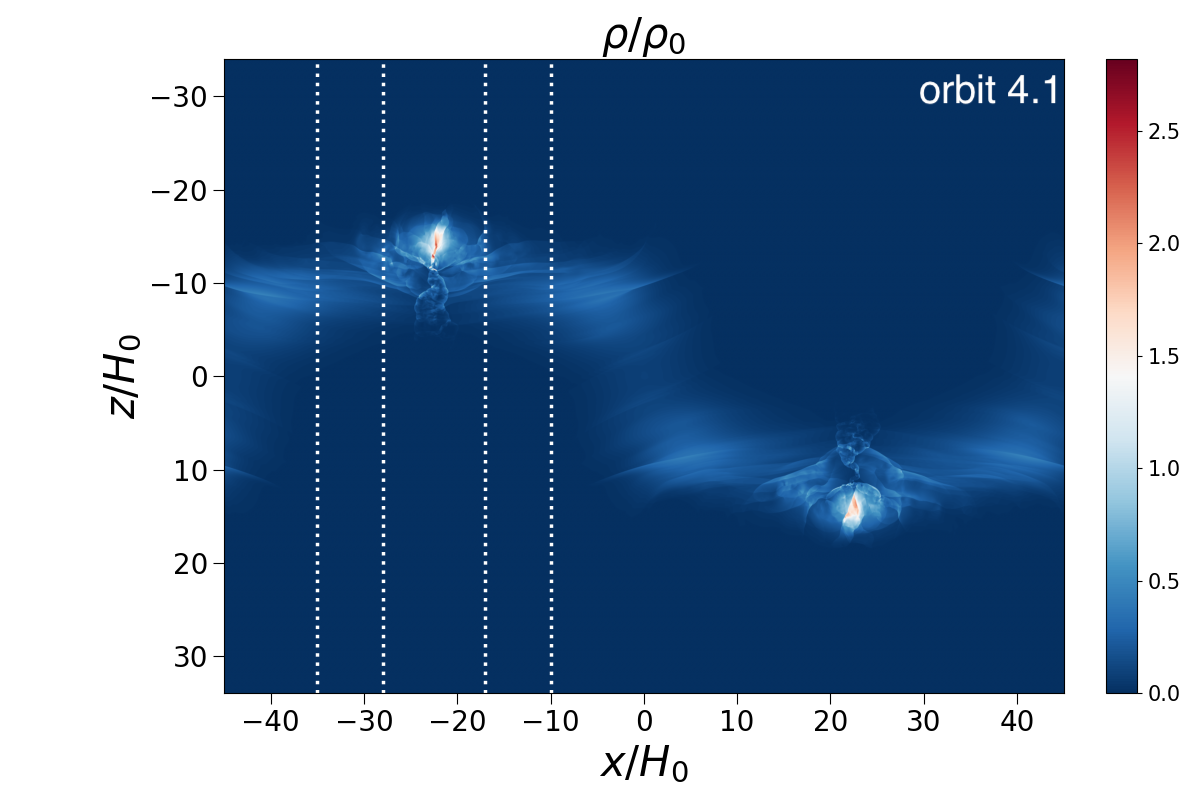}
\includegraphics[scale=0.19]{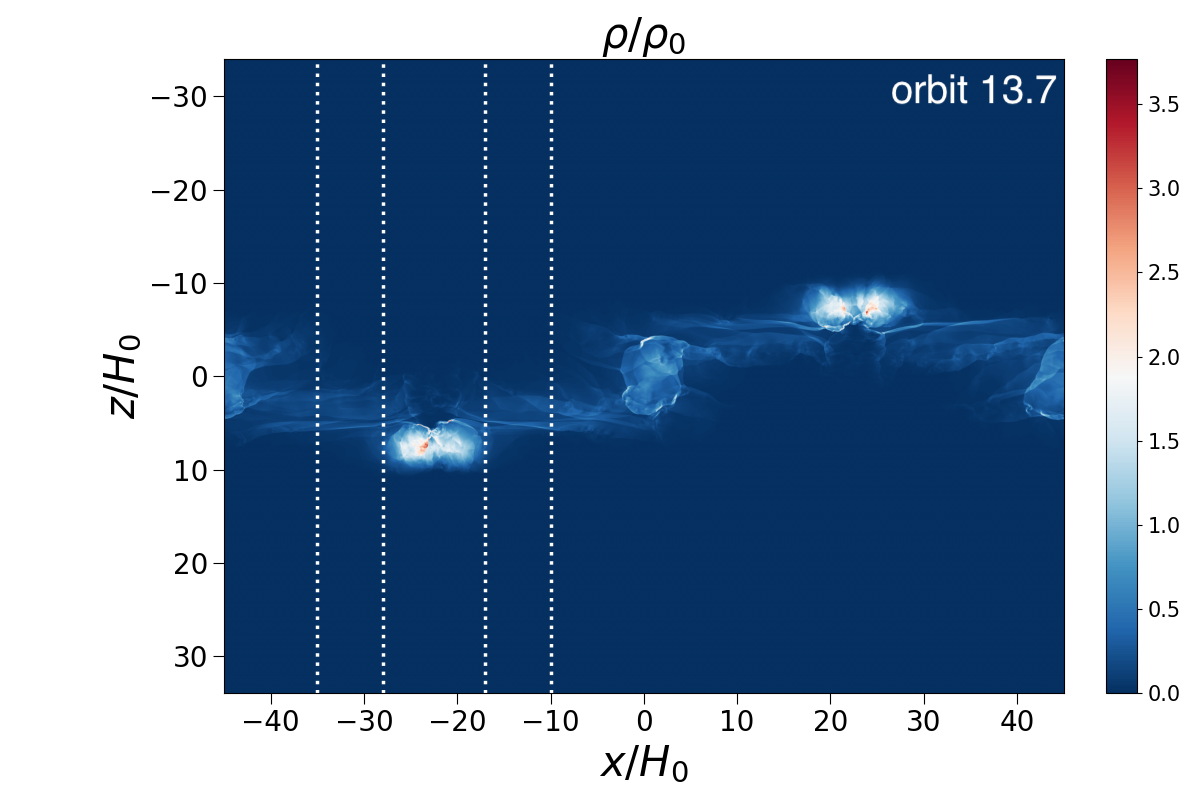}
\includegraphics[scale=0.19]{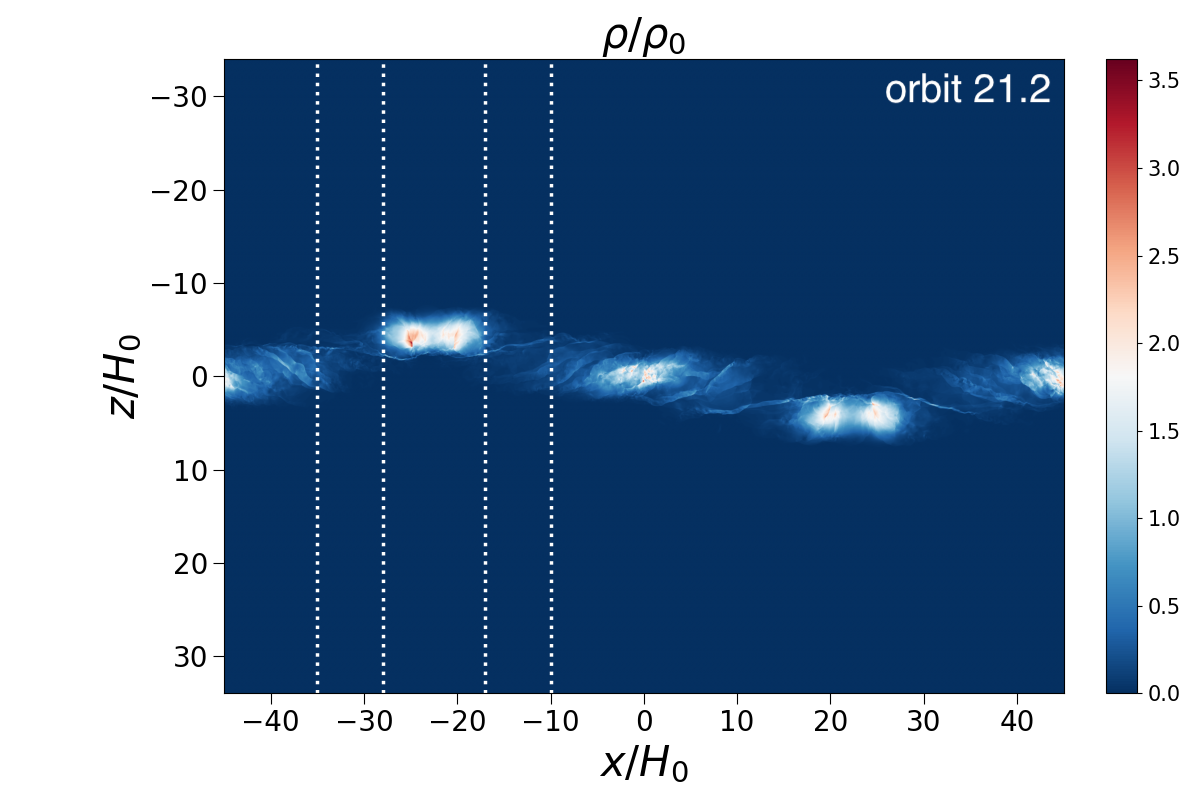}
\includegraphics[scale=0.19]{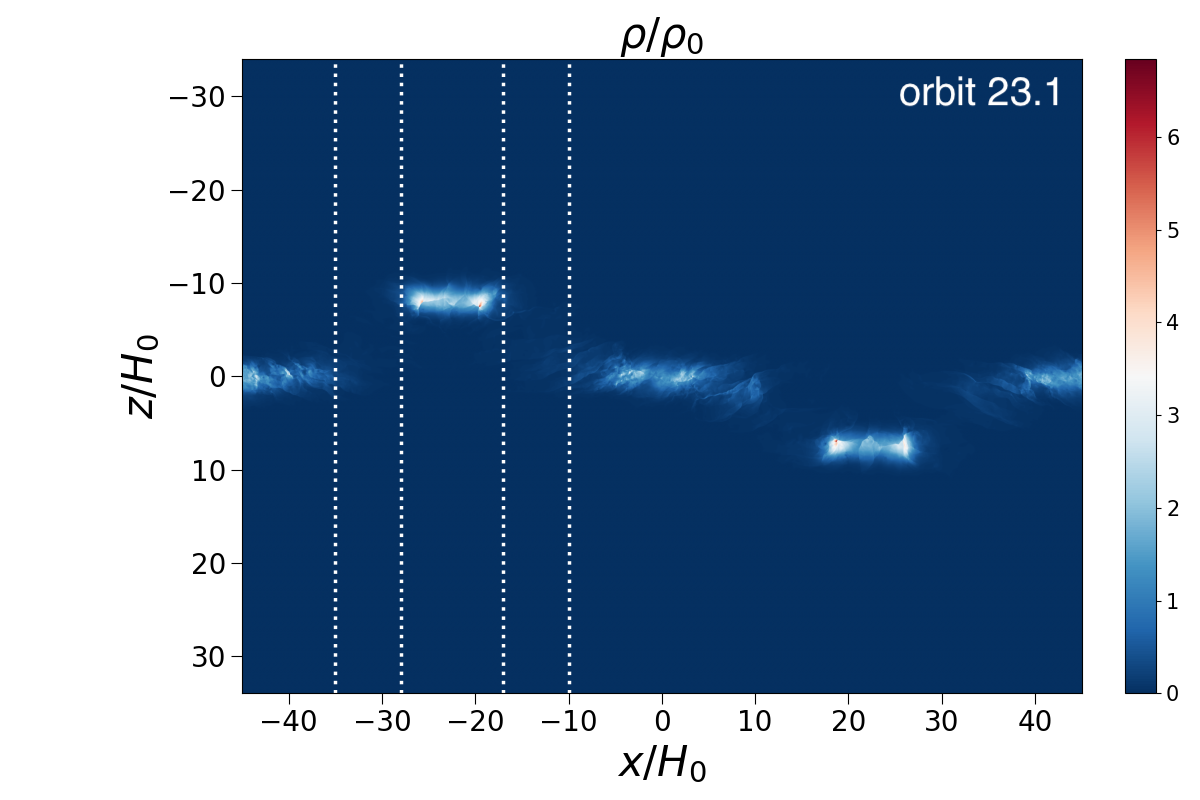}
\caption{Density snapshots in the $xz$-plane in the fiducial simulation with a large warp amplitude ($A=28H_0, \psi \sim 1.95$). Top-left: the initial condition (orbit 0) is a free warp in the shape of a sinusoid. Top-middle (orbit 1.1): growing horizontal sloshing motions converge near the peaks of the warp leading to shocks that dampen the tilt. Top-right (orbit 4.1): after four orbits the warp has been noticeably flattened around the peaks of the sine curve. Bottom-left: by orbit 13.7 the disc appears to have broken into two rings. Bottom-middle (orbit 21.2): the disc begins to break into two more rings. Bottom-right (orbit 23.1): the disc has broken into four flat rings (the two rings at the radial boundaries are actually the same ring). The dashed vertical white lines show the approximate locations at which gaps open in the left-hand part of the disc.}
\label{FIGURE_FiducialSimulationFlowField}
\end{figure*}

Finally, using these approximations and applying trigonometric identities, we can write the combined bending wave $(C_{+}\mathcal{S}_{+} + C_{-}\mathcal{S}_{-})(A/H_0)$ in the long-wavelength limit as
\begin{align}
  &u_x'\approx\f{A\Omega_0z}{H_0}\cos kx\cos\Omega_0 t\sin t_\text{b},\label{vx}\\
  &u_y'\approx-\f{A\Omega_0z}{2H_0}\cos kx\sin\Omega_0 t\sin t_\text{b},\label{vy}\\
  &u_z'\approx A\Omega_0\sin kx\sin\Omega_0 t\cos t_\text{b},\label{vz}\\
  &h'\approx-A\Omega_0^2z\sin kx\cos\Omega_0 t\cos t_\text{b},\label{hp}
\end{align}
which corresponds to a horizontal displacement ($\partial_t \xi_x \equiv u_x'$) of
\begin{equation}
  \xi_x\approx\f{Az}{H_0}\cos kx\sin\Omega_0t\sin t_\text{b}\label{xix}
\end{equation}
and a vertical displacement ($\partial_t \xi_z \equiv u_z'$) of
\begin{equation}
  \xi_z\approx-A\sin kx\cos\Omega_0t\cos t_\text{b},\label{xiz}
\end{equation}
where
\begin{equation}
  t_\text{b}=\Omega_bt
  \label{EQUATION_PhaseOfBendingWave}
\end{equation}
is the phase of the bending wave with frequency $\Omega_\text{b} = (1/2)c_{s0} k$. The solution (\ref{vx})--(\ref{hp}) can be understood as a combined vertical oscillation (corrugation) and horizontal oscillation (sloshing) at the orbital frequency $\Omega_0$, modulated slowly in space and in time at the bending-wave frequency $\f{1}{2}c_{\text{s}0}k$ (see Figure \ref{FIGURE_Theory_EkinxEkinzTimeSeries}). This interpretation is compatible with the local, axisymmetric view of warps developed in \cite{ogilvie2022hydrodynamics}. The modulation takes the form of a single standing wave of wavenumber $k$ and frequency $\f{1}{2}c_{\text{s}0}k$ (even though in another sense it consists of a superposition of two standing waves).

The solution in this long-wavelength limit can be related to the usual linearized equations for bending waves [e.g.\ equations~8 and~9 of \cite{lubow2000tilting}]. Specifically, if we define the complex tilt $Z$ and torque $G$ as in Section~\ref{METHODS_Diagnostics}, then the solution (\ref{vx})--(\ref{hp}) can be expressed as 
\begin{align}
  &Z=-A\sin kx\cos t_\text{b},\\
  &G=-\frac{1}{2}A\Sigma H_0\Omega_0\cos kx\sin t_\text{b},
\end{align}
and satisfies the linear bending-wave equations
\begin{align}
  &\Sigma\frac{\partial Z}{\partial t}=\frac{\partial G}{\partial x},\\
  &\frac{\partial G}{\partial t}=\frac{1}{4}\Sigma H_0^2\Omega_0^2\frac{\partial Z}{\partial x},
\end{align}
which can readily be understood to be the local analogues of equations~8 and~9 of \cite{lubow2000tilting} in the absence of viscosity and precession.

A key prediction (see Appendix ~\ref{APPENDIX_InviscidBendingWaveTheory}) is that, in the long wavelength limit, the orbit-averaged and vertically-integrated Reynolds stress is given by
\begin{equation}
  \int\rho\langle u_x'u_y'\rangle_{\tau}\,\dd z\approx-\f{\psi_{\text{max}} ^2\Sigma\Omega_0^2H_0}{16k}\cos^2kx\sin2t_\text{b}.
  \label{EQUATION_ReynoldsStressxy}
\end{equation}
to second-order in the warp amplitude $\psi_{\text{max}} $, where $\Sigma = \sqrt{2\pi}\rho_0 H_0$.

The fractional change in surface density due to the bending wave (also to second-order in $\psi_{\text{max}} $) is given by
\begin{equation}
  \f{\langle A^2 \Sigma_2 \rangle_{\tau}}{\Sigma}\approx -\f{\psi_{\text{max}}^2}{4}\cos(2kx)(1-\cos2t_\text{b}).
  \label{EQUATION_FractionalChangeSurfaceDensity}
\end{equation}

Thus the surface density is enhanced at $k x= (n+
1/2)\pi$ (which is where the initial vertical displacement is largest in magnitude) and depleted at $k x= n\pi$ (which is where the initial warp is strongest). Note that these locations of enhancement are the same as those at which shocks occur in large-amplitude simulations, leading to density spikes on top of peaks (e.g. see top right-hand panel of Figure \ref{FIGURE_Fiducial_OrbitAveragedRadialProfilesOrb0Orb1}). 

There are several important caveats to consider. First, these changes in the surface density and Reynolds stress are oscillatory and \textit{reversible} in the inviscid case treated here. This is in contrast to the case where dissipation is included, which results in a permanent change in the surface density after the bending wave is damped out. The frequency of the bending wave $\Omega_\text{b} = (1/2)c_{s0} k$ (see Equation \ref{EQUATION_PhaseOfBendingWave}) is also changed to $\Omega_\text{b} = (1/2)\sqrt{c_{s0}^2 k^2 - \gamma^2}$ when dissipation is included, where $\gamma = \alpha \Omega_0$. (The derivation including dissipation will be included in a future publication.) Second, a surprising and remarkable result is that the $xy$-component of the Reynolds stress alone \textit{does not} drive the expected change in surface density, as explained in Appendix \ref{APPENDIX_InviscidBendingWaveTheory}. Finally, note that the theory on which our linear bending wave analysis is based will break down if $\psi_{\text{max}}$ is not sufficiently small.

\begin{figure}
    \centering
    \includegraphics[width=1.0\linewidth]{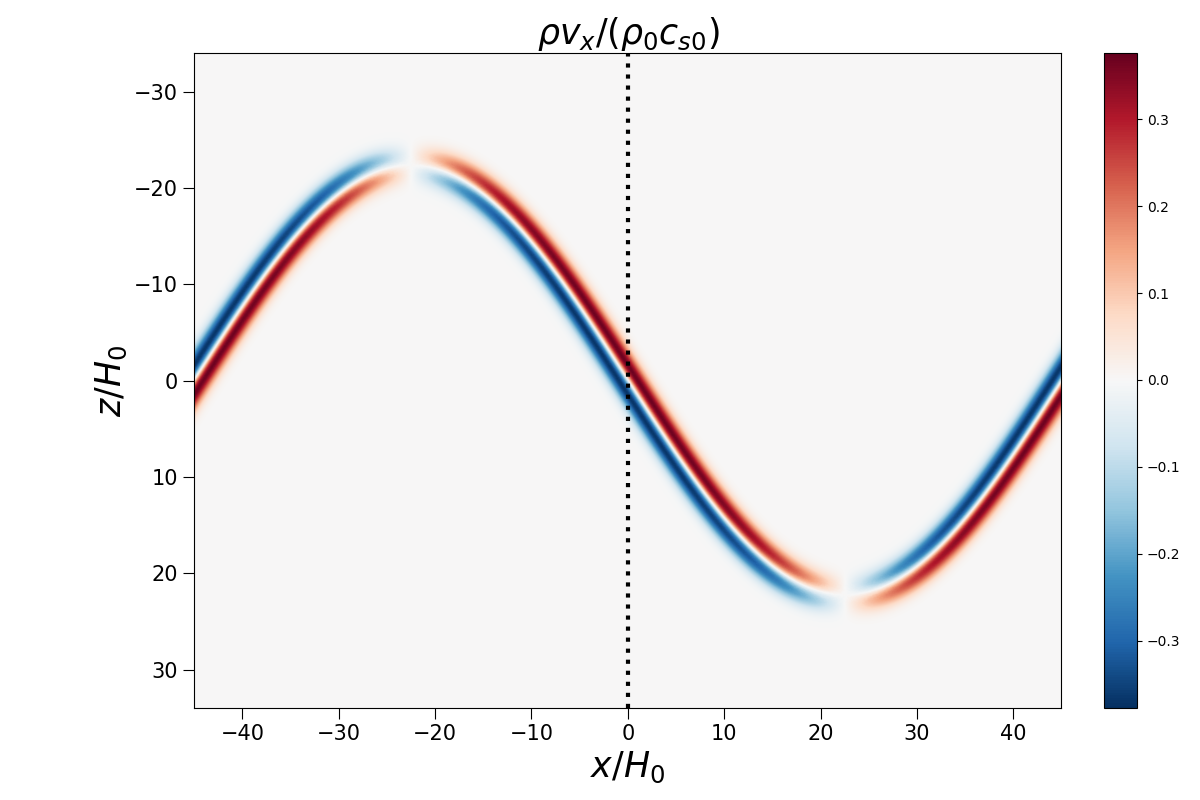}
    \includegraphics[width=1\linewidth]{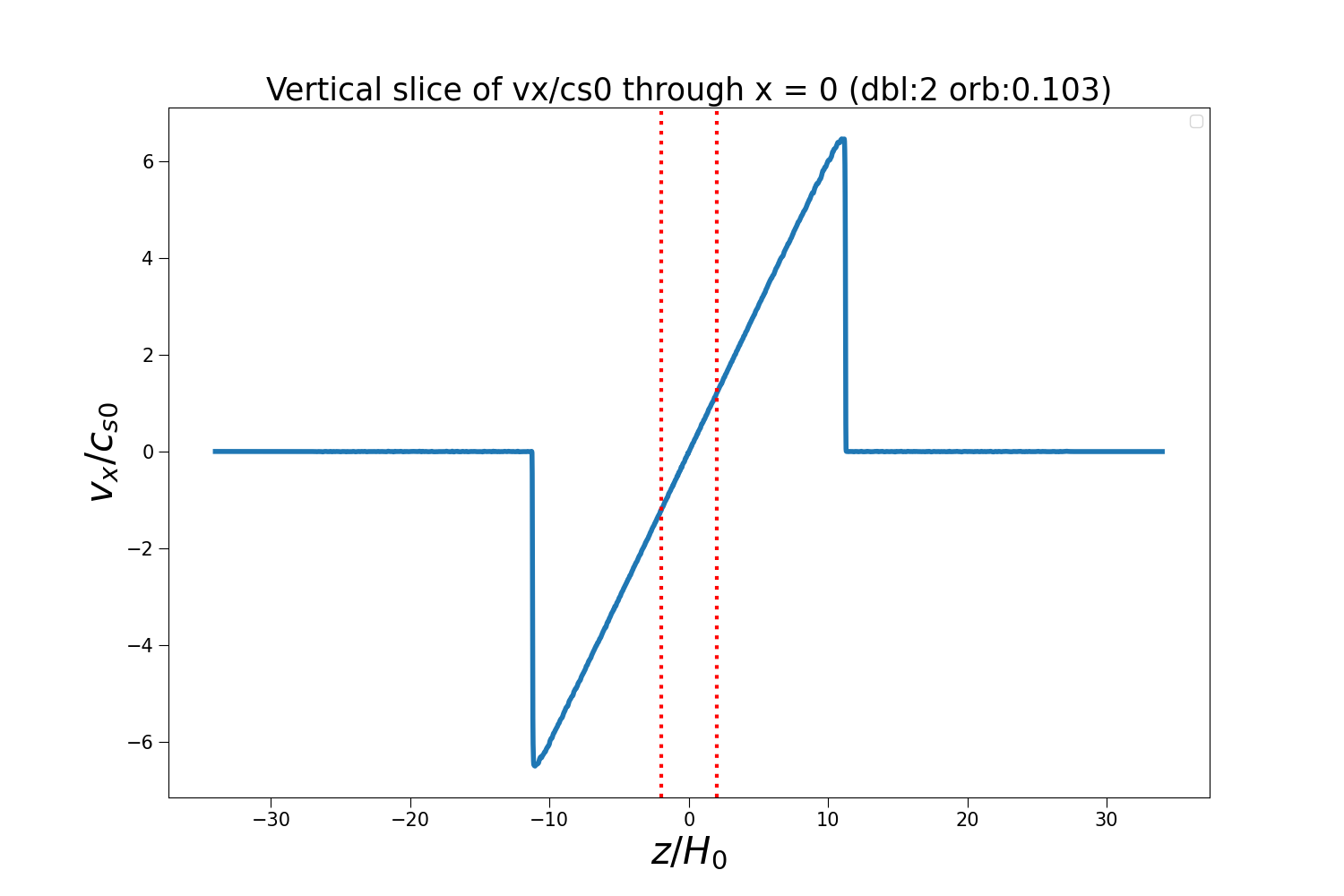}
    \caption{Top: $\rho u_x$ (in units of $\rho_0 c_{\text{s}0}^2$) in $xz$-plane shortly after initialization (orbit 0.1), showing the development of the horizontal sloshing motions. The vertical black dotted line through $x/H_0 = 0$ marks the radial location of the vertical slice of $u_x$ (bottom panel). The dotted red vertical lines at $z=-2H_0$ and $z=2H_0$ in the bottom panel show the approximate location of the bulk of the disc.}
    \label{FIGURE_FiducialSimRhoVxAndVerticalSliceOfVx}
\end{figure}

\section{Fiducial simulation of disc breaking}
\label{RESULTS_FIDUCIALSIMULATION}
In this section we describe the results of our fiducial simulation with a large initial warp. The warp is initialized by means of a sinusoidal density profile with a radial wavenumber of $k = 2\pi/L_x$ and amplitude $A=28H_0$ (see Equation \ref{EQUATION_FreeWarpDensityProfile}). Thus the maximum warp amplitude at initialization is $\psi_{\text{max}}  = kA \approx 1.95$. The simulation is inviscid (and in the wavelike regime), and was run for 50 orbits. The simulation is listed as FreeWarp17 in Table \ref{TABLE_LargeWarpAmplitudeSimulations} in Appendix \ref{APPENDIX_TablesOfSimulations}.

\subsection{Flow field evolution}
\label{RESULTS_FiducialSimuulation_SummaryOfEvolution}
We start by describing the evolution of disc from a continuous warp to a series of broken rings. See Figure \ref{FIGURE_FiducialSimulationFlowField}, which shows the density field in the $xz$-plane at different times. 

\subsubsection{Growth of sloshing motions and shocks}
At initialization the disc is continuous and the warp has a sinusoidal shape, as shown in the upper left-hand panel of Figure \ref{FIGURE_FiducialSimulationFlowField}. (Note that from now on we will mostly restrict our discussion to the left-hand side of the domain, $x/H_0 \in [-45,0]$, since the warp is anti-symmetrical about $x=0$.)

Although the disc has not been initialized with any mean velocity in the $x$-direction, a vertical shear (horizontal sloshing motion) begins to grow immediately after initialization due to the unbalanced radial pressure gradient. This shear can be clearly seen in the 2D flow field of $\rho u_x$ in the $xz$-plane (see top panel of Figure \ref{FIGURE_FiducialSimRhoVxAndVerticalSliceOfVx} which shows $\rho u_x$ at orbit 0.1). If we take the left-hand peak of the sinusoid ($x/H_0 \sim -22.5$) as a reference point, the flow diverges immediately above the local disc mid-plane and converges below this point. The amplitude of the shear velocity (as measured by $u_x$) increases linearly with vertical coordinate $z$, as shown in the bottom panel of Figure \ref{FIGURE_FiducialSimRhoVxAndVerticalSliceOfVx}, and becomes supersonic about $\pm 2H_0$ above and below the local disc mid-plane. Note that velocity profile is truncated around $|z| \sim 10H_0$ because this is where the density (and pressure) floor sets in (see Section \ref{METHODS_BoxSizeDensityFloorResolution}).

While the disc remains continuous during this early stage of evolution, an important non-linear feature is the formation of converging shocks twice per orbit immediately below the peaks of the sinusoidal warp. These can be seen in the left-half of the domain in the middle panel of Figure \ref{FIGURE_FiducialSimulationFlowField} as thin vertical features just below the peak of the warp. The shocks result from the converging horizontal sloshing motions and are a source of dissipation. This has important consequences for the surface density evolution of the disc, and allows for the sloshing motions to transport mass and tilt (i.e. the misaligned component of the angular momentum) in an irreversible manner \textit{from} the regions where the disc will eventually break (marked by the dashed vertical white lines in Figure \ref{FIGURE_FiducialSimulationFlowField} ) \textit{to} the peaks of the warp.

\subsubsection{Warp steepening}
\begin{figure}
    \centering
    \includegraphics[width=1\linewidth]{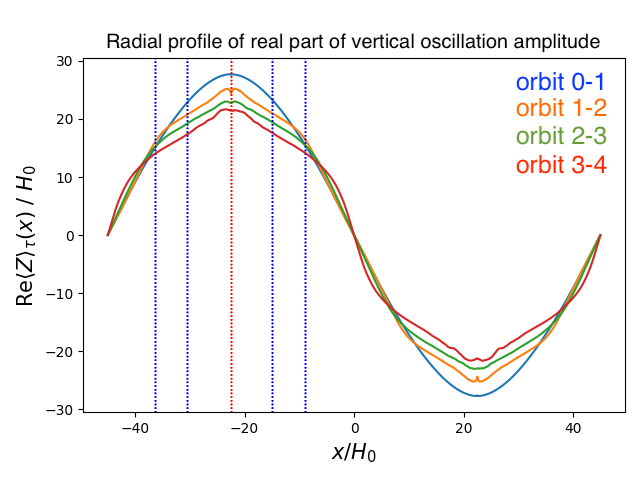}
    \caption{Radial profile of the real part of the orbit-averaged vertical oscillation amplitude ( Equation \ref{EQUATION_ComplexTilt}). Different colors correspond to orbit-averages taken over different intervals. Over the first few orbits, there is clear evidence of warp steepening where the warp amplitude was initially greatest. The dashed blue lines mark the edges of the gaps that will form at later times (in the left-hand part of the domain), while the dashed red line marks the location of the peak of the sine curve.}
    \label{FIGURE_FiducialSimulationCenterOfMassProfile}
\end{figure}

The horizontal sloshing motions grow in amplitude between orbits 0 and 4. During this time we also observe a noticeable steepening of the warp in the vicinity of locations of maximum warp amplitude ($x/H_0 \sim  -45,0$ and $+45$). This steepening can be seen in the density field around $x=0$ (top-right panel of Figure \ref{FIGURE_FiducialSimulationFlowField}), and even more clearly in Figure \ref{FIGURE_FiducialSimulationCenterOfMassProfile}, which shows the orbit-averaged radial profile of the real part of the complex tilt $\langle Z \rangle_{\tau}(x)$ over the first four orbits.

\subsubsection{Vertical oscillations (bouncing)}
\begin{figure}
    \centering
    \includegraphics[width=1.1\linewidth]{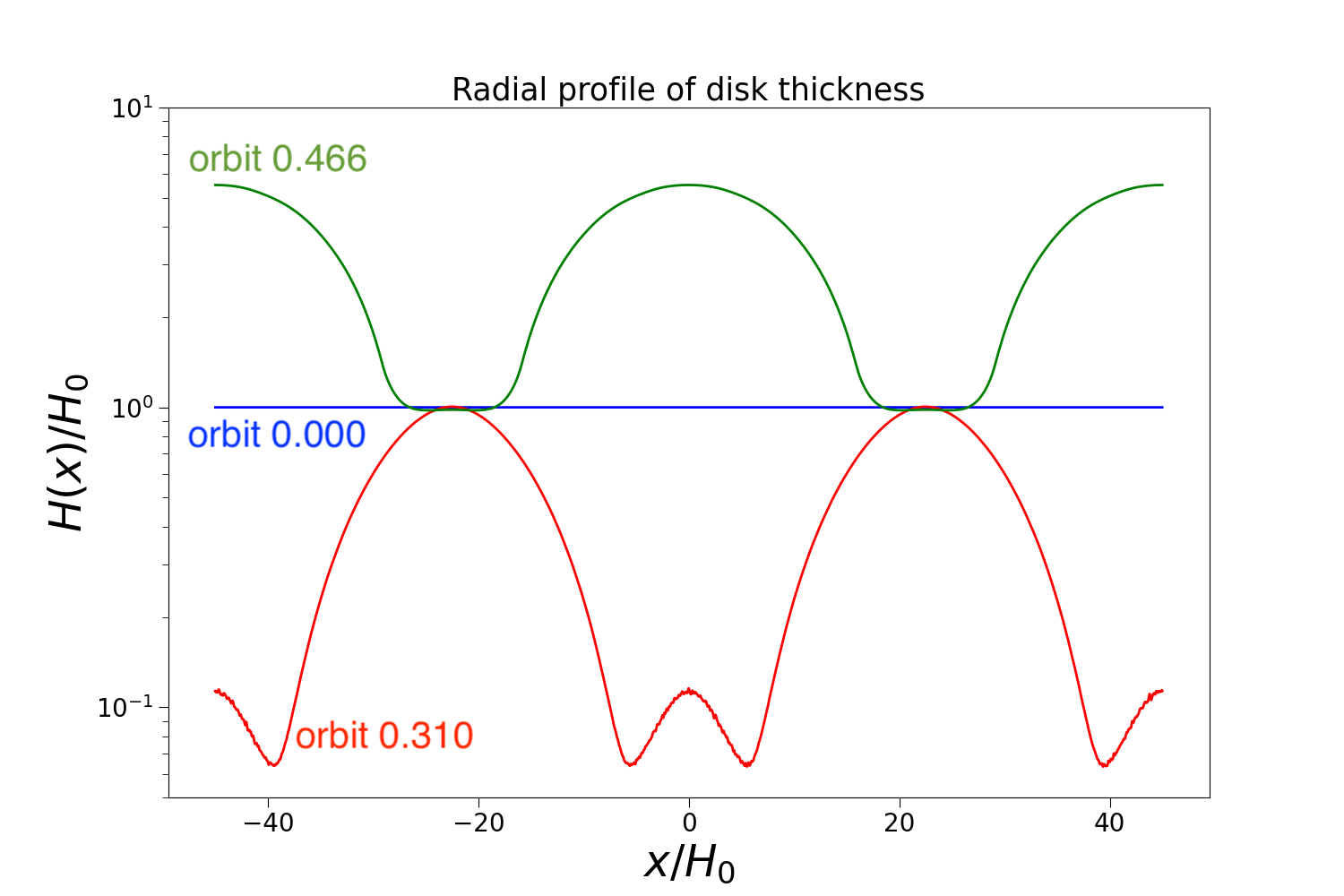}
    \caption{Radial variation in disc thickness at three different times during the first orbit. The disc bounces (i.e. becomes highly compressed) where the warp amplitude is large.}
    \label{FIGURE_FiducialSimulationVarianceInDynamicalH}
\end{figure}

At initialization the disc has uniform thickness and surface density. Due to the geometry of the warp the disc is not expected to be in vertical hydrostatic equilibrium. The combination of radial oscillations (due to the internal sloshing motions) and vertical oscillations (due to the tilt) result in a periodic compression and rarefaction of the disc at locations of non-zero warp amplitude twice per orbit (see  \cite{ogilvie2013local}). See Figure \ref{FIGURE_FiducialSimulationVarianceInDynamicalH}, which shows the instantaneous radial profile of disc thickness (Equation \ref{EQUATION_DiskThickness}) at initialization (orbit 0: blue curve), and at two other times during during the first orbit when the disc is thinnest (orbit 0.310) and thickest (orbit 0.466), respectively. Due to the large initial warp amplitude and the rapidly growing sloshing motions, the compression of the disc is very large during the first bounce at orbit 0.310 (red curve), and the disc thickness reaches as low as $H \sim 0.1H_0$ at the locations of maximum warp amplitude ($x/H_0\sim -45,0,45$). The disc subsequently thickens at these locations (green curve), reaching a maximum of $H \sim 5H_0$ at orbit 0.466. Thus the disc could be thought of as being in an extreme bouncing regime. Note that at the locations where the warp amplitude is zero (i.e. at $x/H_0 \sim -22.5,22.5$), which correspond to the peaks of the sinusoidal warp, the disc thickness remains fairly constant in time at $H \sim H_0$.

\subsubsection{Dissipation of tilt and flattening of warp}
By orbit 4, the disc has noticeably begun to flatten in the vicinity of the peaks of the sinusoidal warp ($x/H_0 \sim \pm22.5$): see top-right-hand panel of Figure \ref{FIGURE_FiducialSimulationFlowField}. The standing shocks that form below/above the peaks of the disc have significantly dissipated the tilt: at initialization the peaks oscillate between $z/H_0 \sim \pm 28$, whereas by orbit 4 the peaks oscillate between $z/H_0 \sim \pm 20$. The shocks have also disrupted the disc near the peaks, producing a sharp spike in surface density that can clearly be seen around $x/H_0 \sim -22.5$ in the top-right-hand panel of Figure \ref{FIGURE_Fiducial_OrbitAveragedRadialProfilesOrb0Orb1}, which shows the radial profile of orbit-averaged surface density over the first orbit. By orbit 13, the disc appears to have broken into two tilted rings, a left-ring spanning $-40 < x/H_0 < -5$ and a right-ring spanning $5 < x/H_0 < 40$: see bottom left-hand panel of Figure \ref{FIGURE_FiducialSimulationFlowField}.

\begin{figure}
    \centering
    \includegraphics[width=1.05\linewidth]{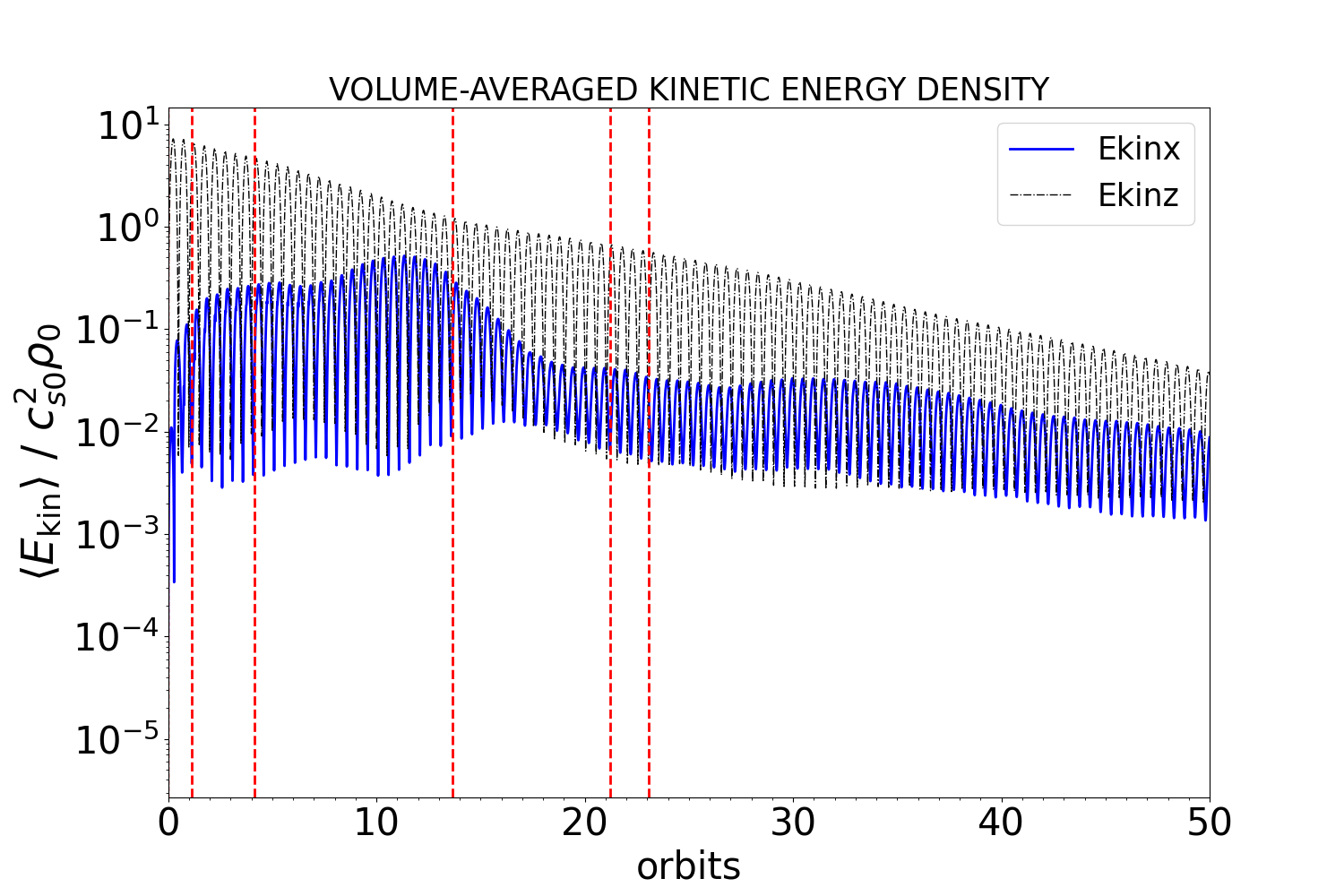}
    \caption{Time-evolution of volume-averaged vertical component (dot-dashed black line) and radial component (solid blue line) of kinetic energy density taken from our fiducial simulation of a free warp with initial warp amplitude $\psi_{\text{max}}  \sim 1.95$. Vertical dashed
    lines mark the times of the snapshots shown in Figure \ref{FIGURE_FiducialSimulationFlowField}.}
    \label{FIGURE_FiducialFreeWarpTimeSeriesOfEkinxEkinz}
\end{figure}

\subsubsection{Disc breaks into four rings}
\begin{figure*}
\centering
\includegraphics[scale=0.43]{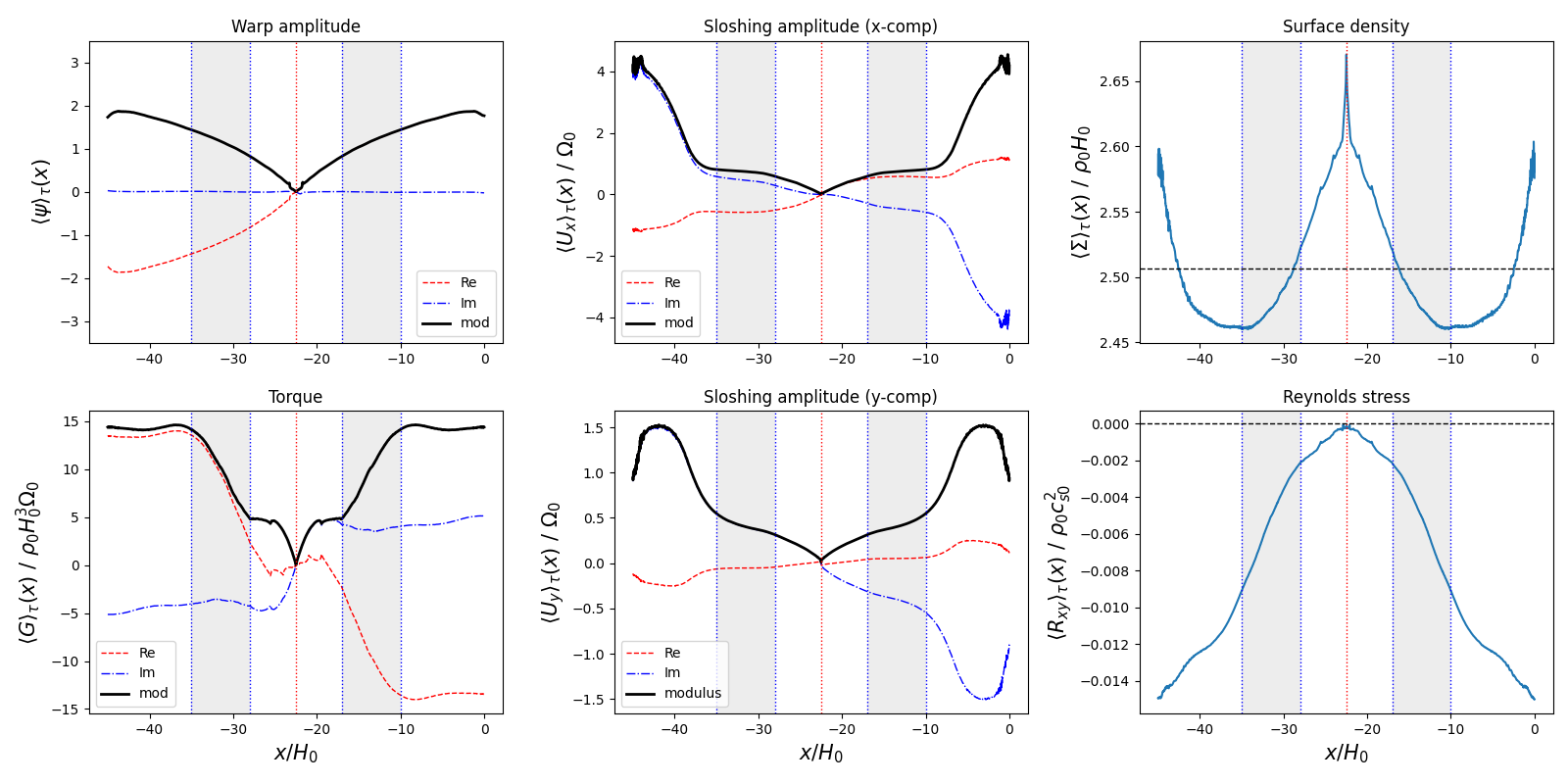}
\caption{Orbit-averaged radial profiles in the left-hand part of the disc from a simulation of a large-amplitude warp ($A=28H_0, \psi_{\text{max}} \sim 1.95$). See Section \ref{METHODS_Diagnostics} for definitions. The orbit-average was taken over the first orbit. \textit{Top-left}: warp amplitude $\langle \psi \rangle_{\tau}$. \textit{Top-middle}: shear amplitude ($x$-component) $\langle U_x \rangle_{\tau}$. \textit{Top-right}: surface density $\langle \Sigma \rangle_{\tau}$. \textit{Bottom-left}: torque $\langle G \rangle_{\tau}$. \textit{Bottom-middle}: shear amplitude ($y$-component) $\langle U_y \rangle_{\tau}$. \textit{Bottom-right}: $xy$-component of the Reynolds stress $\langle R_{xy} \rangle_{\tau}$. The dashed vertical line at $x/H_0 = -22.5$ shows the location of the left-hand peak in the sinusoidal disc (see top-left hand panel of Figure \ref{FIGURE_FiducialSimulationFlowField}). The shaded regions show the approximate region over which gaps will open later on. The dashed horizontal black line in the surface density plot shows the initial surface density.}
\label{FIGURE_Fiducial_OrbitAveragedRadialProfilesOrb0Orb1}
\end{figure*}

Finally, between orbit 13 and orbit 23 the disc breaks again, this time in the regions marked by the dashed vertical lines (see middle panel of Figure \ref{FIGURE_FiducialSimulationFlowField}). By orbit 21 the disc has broken into four rings: two \textit{untilted} rings (one in the middle of the the domain which is centred around $x/H_0 \sim 0$, and another at $x/H_0 \sim \pm 45$ which spans the shear-periodic radial boundaries), and two \textit{tilted} rings centred around $x/H_0 \sim \pm 22.5$. See bottom right-hand panel of Figure \ref{FIGURE_FiducialSimulationFlowField}. By the end of the simulation the four rings are about $15H_0$ wide and have a gap between them of around $7H_0$. The rings have considerable small-scale structure: the tilted rings (which were originally the peaks of sinusoidal warp before it broke) have a noticeable dip in the surface density at their centres (see top-right-hand panel of Figure \ref{FIGURE_Fiducial_OrbitAveragedRadialProfilesOrb0Orb1}), which can be traced back to successive interactions between these rings and the standing shocks that formed below or above the peaks of the sinusoidal warp.

\subsection{Other diagnostics}
\label{RESULTS_FiducialSimulation_Diagnostics}

\subsubsection{Time series of volume-averaged quantities}
\label{RESULTS_FiducialSimulation_TimeSeries}
In Figure \ref{FIGURE_FiducialFreeWarpTimeSeriesOfEkinxEkinz} we show the time-series of volume averaged radial and vertical kinetic energy density components. Both components are highly oscillatory on short time-scales, oscillating twice per orbit. Oscillations in the vertical kinetic energy (blue curve) are due to vertical oscillations of disc columns about $z = 0$, while oscillations in the radial kinetic energy are due to the horizontal sloshing motions.

On longer time-scales, we observe an irregular modulation in the radial kinetic energy (blue curve), whose envelope has a local maximum around orbit 3, and global maximum around orbit 11. This irregular beating is related to the period of the underlying growing bending wave (see section \ref{THEORY}). However, because the disc breaks in this large-warp amplitude simulation, the bending wave never fully develops, which explains the irregularity in the beating. This is in contrast to the more regular beat which we see at small warp amplitudes (see blue curve in Figure \ref{FIGURE_WarpAmplitudeComparison_EkinxEkinz}). Finally, we observe a secular decay in both energies over the course of the simulation, as both the disc tilt and the horizontal sloshing motions are damped by shocks.

\begin{figure*}
\centering
\includegraphics[scale=0.43]{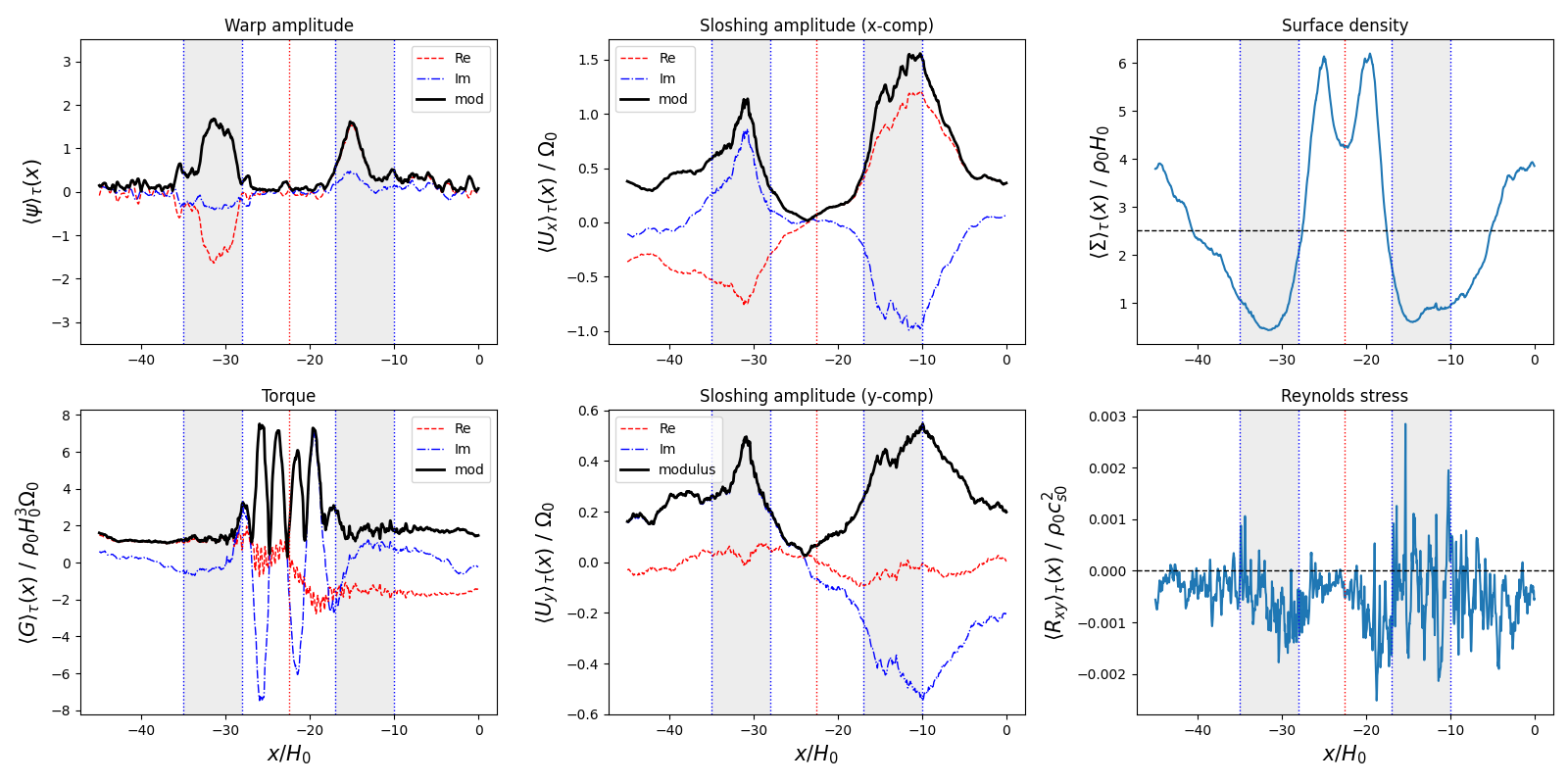}
\caption{Same diagnostics as in Figure \ref{FIGURE_Fiducial_OrbitAveragedRadialProfilesOrb0Orb1}, but the radial profiles have been orbit-averaged over orbit 25-26, by which time the disc has broken into four rings.}
\label{FIGURE_FiducialSimulation_OrbitAveragedProfilesOrb25Orb26}
\end{figure*}

\subsubsection{Orbit-averaged radial profiles}
\label{RESULTS_FiducialSimulation_OrbitAveragedProfiles}
Figure \ref{FIGURE_Fiducial_OrbitAveragedRadialProfilesOrb0Orb1} shows various radial profiles (in the left-hand part of the domain, only) averaged over the first orbit. The disc is smooth and sinusoidal during this orbit, except for the shocks that form near the peaks of the sinusoid (see top-middle panel of Figure \ref{FIGURE_FiducialSimulationFlowField}). In the top-left-hand panel of Figure \ref{FIGURE_Fiducial_OrbitAveragedRadialProfilesOrb0Orb1} we plot the radial profile of the smoothed warp amplitude $\langle \psi \rangle_{\tau}(x)$.\footnote{The actual profile of the warp amplitude is noisy due to small-scale features in the flow field, so we have smoothed it using a Savitzky-Golay filter with a window size of $101$ and a polynomial of degree $3$.} Its modulus is zero at the peak of the sine curve (red vertical dashed line) and increases monotonically with distance away from the peak, reaching maxima of $\sim 2$ at $x/H_0 = -45,0$. The warp gives rise to horizontal sloshing motions (middle panels of Figure \ref{FIGURE_Fiducial_OrbitAveragedRadialProfilesOrb0Orb1}), whose amplitudes are given by $\langle U_{x,y} \rangle_{\tau}$. The sloshing amplitude is zero where the warp amplitude is zero,  increases linearly with warp amplitude for $|\langle \psi \rangle_{\tau}| \lesssim 1$, is largely flat in the regions that will ultimately become gaps, and then increases again, peaking around $|\langle U_{x,y} \rangle_{\tau}| \sim (4,1.5) \Omega_0$ where the warp amplitude is greatest.

The shear mediates a torque $\langle G \rangle_{\tau}$ (strictly speaking, a radial flux of angular momentum), which is shown in the bottom-left-hand panel. Its modulus increases linearly with warp amplitude for small warp amplitudes before reaching a local maximum of $\sim 5 \rho_0 H_0^3 \Omega_0$ at $x/H_0 \sim -25,-20$. The torque then increases again, and peaks at $\sim 15 \rho_0 H_0^3 \Omega_0$ near what will become the outer edges of the gaps.

The top-right hand panel shows the radial profile of orbit-averaged surface density $\langle \Sigma \rangle_{\tau}$. At $t=0$ the surface density is uniform (dashed horizontal black line). By the end of the first orbit, however, the surface density has already developed two noticeable dips centred around $x \sim -35H_0$ and $x \sim -10H_0$. Between $x \sim -28 H_0$ and $x \sim -18H_0$, on the other hand, there is a noticeable \textit{enhancement} in the surface density. The change in surface density in these regions is small (around $\pm 2\%$, respectively), but the troughs in the surface density will persist and deepen as the warp evolves and the disc breaks. Finally, the orbit-averaged Reynolds stress $\langle R_{xy} \rangle_{\tau}$ (bottom-right-hand panel) is zero near the peak of the warp and negative otherwise. Note that the disc is laminar at this stage, and the only significant contributions to the Reynolds stress (which involves correlations between $u_x$ and $\delta u_y$) are due to the horizontal sloshing motions.

The evolution of these profiles over the next few orbits is complicated, but can be summarized as follows: the gaps in the surface density deepen and widen, and the sloshing amplitudes continue to grow and develop peaks within the gaps (shaded regions). The warp amplitude, on the other hand, decreases and flattens everywhere, except in the gaps where it develops a peak. See Figure \ref{FIGURE_FiducialSimulation_OrbitAveragedProfilesOrb25Orb26} which shows the profiles averaged between orbit 25 and 26, by which time the disc has broken into four flat rings. We can clearly see deep gaps in the surface density (top-right panel), which has decreased by around $80\%$ in the gaps compared to initialization. The warp amplitude (top-left panel) has decreased significantly outside the gap regions, indicating that the warp has been flattened in the rings (this can also be seen directly in the density field in Figure \ref{FIGURE_FiducialSimulationFlowField}). However, it \textit{peaks} in the gaps, reaching $\langle \psi \rangle_{\tau} \sim 2$ there. These peaks can also be seen in the sloshing motion amplitudes $\langle U_{x,y}\rangle_{\tau}$. The torque is mostly flat and of order unity, except inside the tilted ring centred around $x/H_0 \sim -22.5$, where it is highly oscillatory due to the complicated small-scale structure in the ring at those locations from repeated collisions between the disc and standing shocks.

\section{Parameter study}
\label{RESULTS_PARAMETERSTUDY}
In the previous section we analysed the evolution of a disc with a large initial warp. In Section \ref{RESULTS_ParameterStudy_InitialWarpAmplitudeStudy} below we discuss the evolution of the disc at smaller warp amplitudes. A key result is that at warp amplitudes $\psi_{\text{max}} \lesssim 1$ the disc no longer breaks, but instead becomes parametrically unstable. 

Another key parameter is viscosity. In Section \ref{RESULTS_ParameterStudy_ViscosityStudy} we first point out an important subtlety in the criterion for distinguishing between the wavelike and diffusion regimes in warped discs. We then include both small and large explicit viscosity in order to study its effects on the evolution of the warp in the wavelike regime. 

\begin{figure}
    \centering
    \includegraphics[width=1 \linewidth]{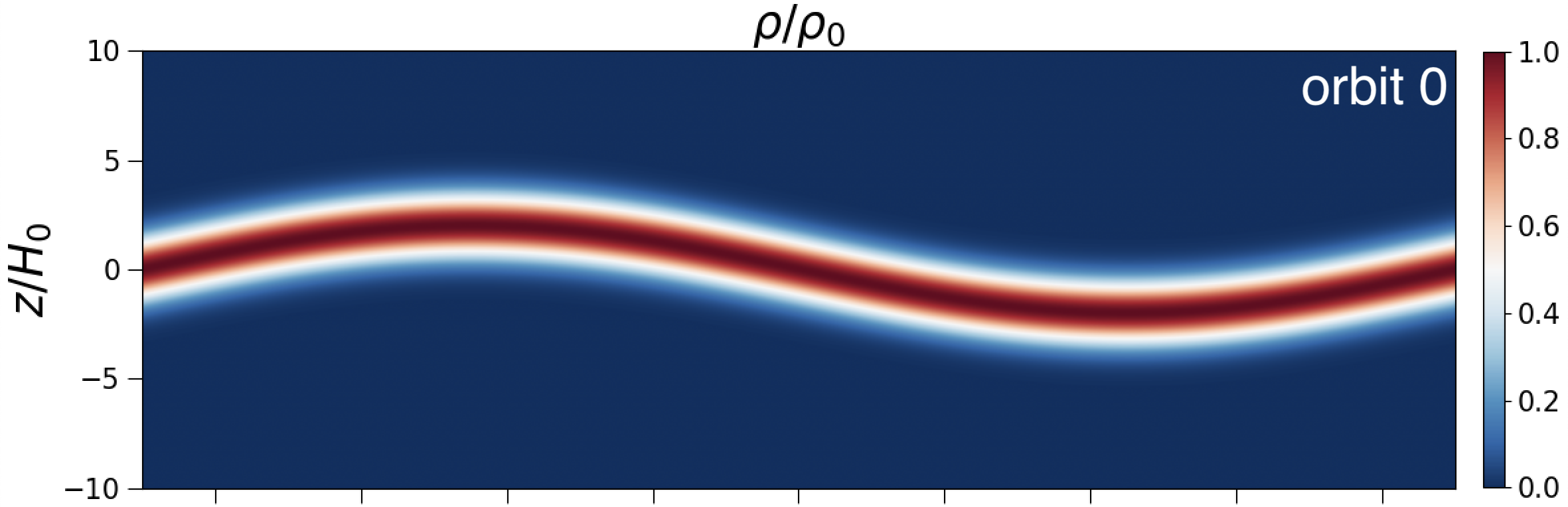}
    \includegraphics[width=1 \linewidth]{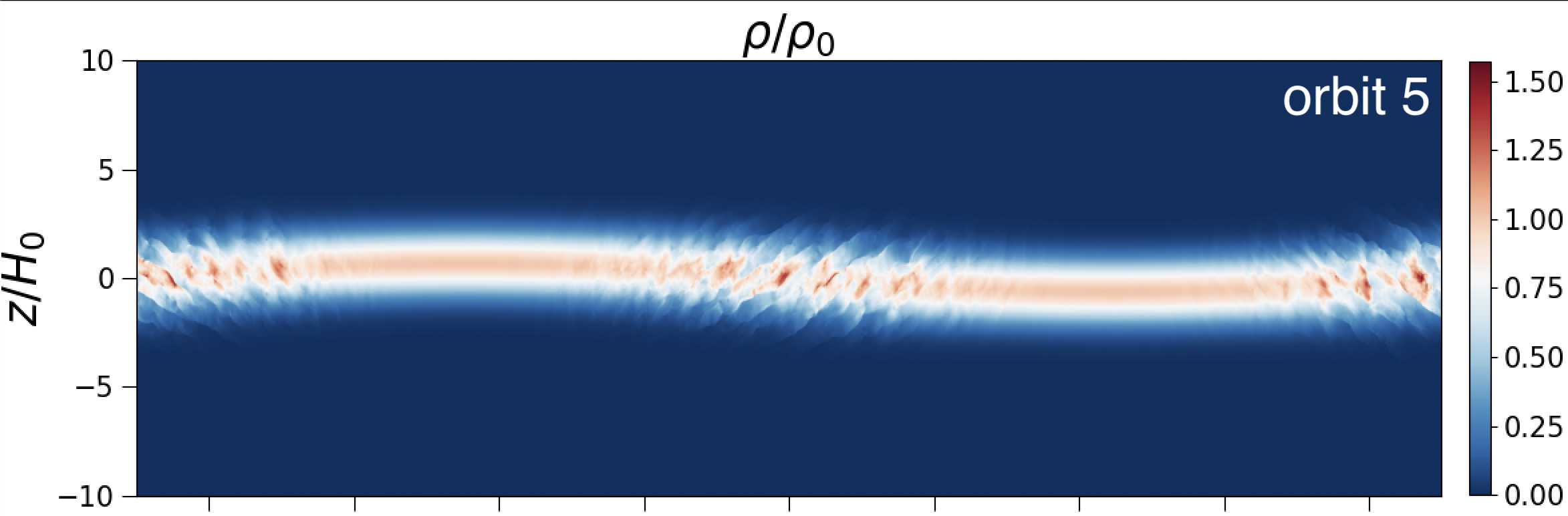}
    \includegraphics[width=1 \linewidth]{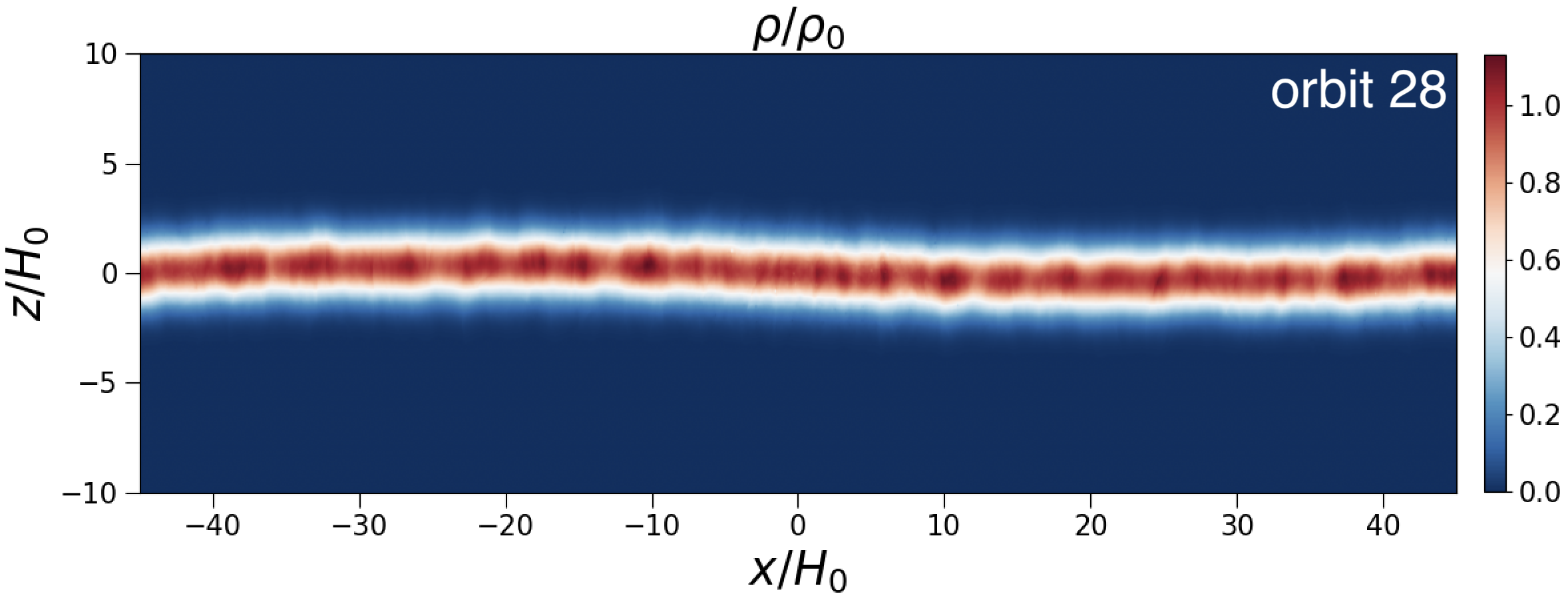}
    \caption{Evolution of density in the $xz$-plane of a small amplitude warp ($A = 2,\,\psi_{\text{max}}  \sim 0.14$). Top panel (orbit 0): initial condition. Middle panel (orbit 5): emergence of parametric instability at locations where the warp is steepest. Bottom panel: by orbit 28 the instability has spread throughout the entire disc.} \label{FIGURE_ParameterStudy_2DFlowFieldA18A2}
\end{figure}

\subsection{Warp amplitude study: parametric instability vs breaking}
\label{RESULTS_ParameterStudy_InitialWarpAmplitudeStudy}
In this section we present a warp amplitude study with initial (maximum) warp amplitudes of $A/H_0 = 2,9,18, 28$, corresponding to $\psi_{\text{max}} \sim 0.14, 0.63, 1.26$, and $1.95$, respectively.\footnote{In their semi-analytical study, \cite{dougan2018instability} investigated a wider range of warp amplitudes, from $\psi_{\text{max}}  \sim 0.1$ up to extreme amplitudes of $\psi_{\text{max}}  \sim 14$.}  All four simulations were run with the HLLC Riemann solver, parabolic reconstruction, and at a resolution of $64/H_0$. The simulations are listed in Appendix \ref{APPENDIX_TablesOfSimulations} as FreeWarp19, FreeWarp29, FreeWarp08b, and FreeWarp17, respectively.

\subsubsection{Small warp amplitudes: $\psi_{\text{max}}  \lesssim 1$}
\label{RESULTS_ParameterStudy_SmallWarpAmplitudeRuns}
First we consider a small warp with $A = 2H_0$ ($\psi_{\text{max}} \sim 0.14$). This run serves as a convenient benchmark against which to test the predictions of our analytical theory of small-amplitude bending waves (see Section \ref{THEORY}). 

In Figure \ref{FIGURE_ParameterStudy_2DFlowFieldA18A2} we show the evolution of the density in the $xz$-plane. The top panel shows the initial condition. A key result is that the disc does not break in this simulation. Instead parametric instability (PI) develops around orbit 3-4 and saturates around orbit 5 (middle panel). The instability develops where the warp amplitude is greatest ($x/H_0 \sim -45, 0$ and $45$) and appears in the density field (and also in $\rho v_x$, not shown) as modes with a wavelength of 1-2$H_0$ which tilt left and right twice per orbit in the frame of an observer moving along a shearing-box orbit.\footnote{Note the similarity of the modes of the PI in the middle panel of our Figure \ref{FIGURE_ParameterStudy_2DFlowFieldA18A2} with the middle panel of Figure 5 of \cite{gammie2000linear}.} An important point is that the instability arises \textit{locally}, particularly around $x = 0$, which is in the centre of the domain. This suggests that the instability is physical and not an artifact due to amplification of inertial waves that leave and re-enter the domain through the periodic radial boundaries.

After the instability saturates, the PI modes appear to spread throughout the entire disc (see bottom panel of Figure \ref{FIGURE_ParameterStudy_2DFlowFieldA18A2}). This could be because the modes are growing everywhere in the disc (the instability grows fastest at locations of maximum warp amplitude, $x/H_0 \sim -45,0,45$, but there is also non-zero growth rate at other locations, except where the warp amplitude is zero). Another reason could be because the PI modes themselves are propagating left and right in the disc; the PI consists of a pair of travelling inertial waves and occurs wherever there is a resonant three-mode coupling between these two inertial waves and the warp). The radial group velocities of the counter-propagating inertial waves involved in the lowest-order mode coupling that leads to parametric instability in a warped Keplerian disc are approximately $0.16$ and $0.18$ in units of $c_{\text{s}0}$. The waves could travel the distance $L_x/4$ from the locations of maximum warp amplitude to spread throughout the entire domain in 20 orbits. It is likely that the spreading of the parametric instability involves a combination of local growth (at an $x$-dependent growth rate) and radial propagation

\begin{figure}
    \centering
    \includegraphics[width=1.0\linewidth]{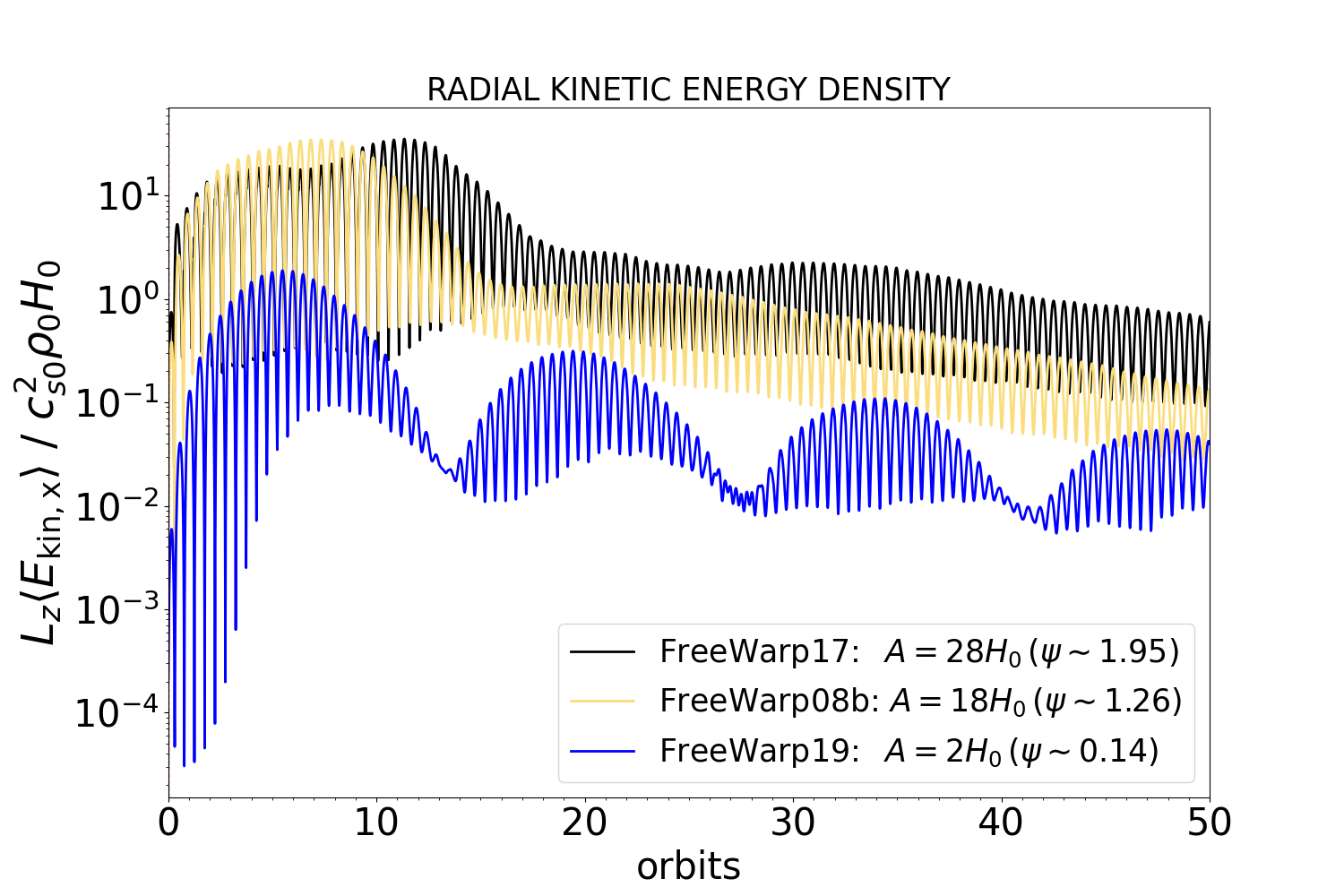}
    \includegraphics[width=1.0\linewidth]{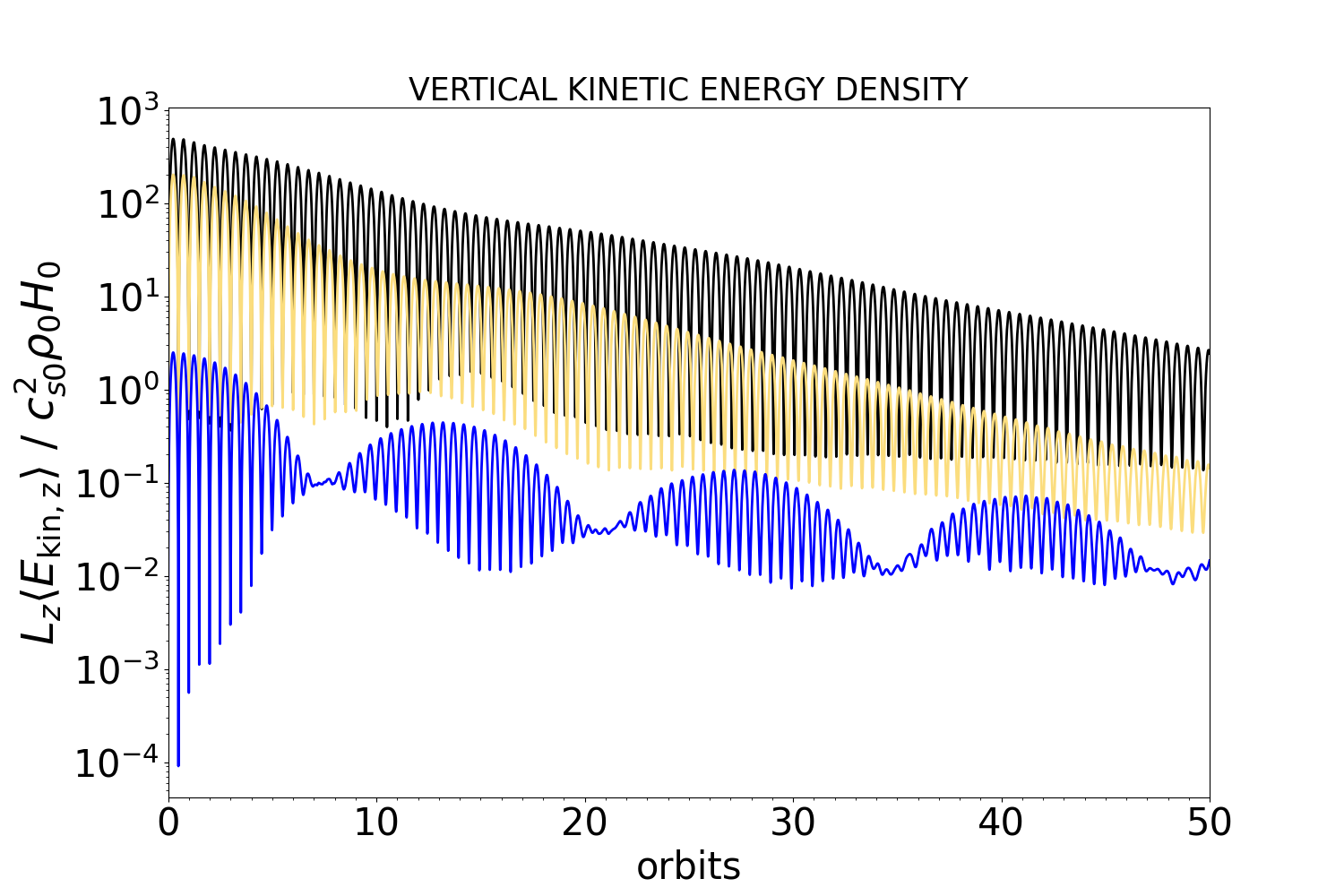}
    \caption{Warp amplitude study: comparison of volume-averaged radial kinetic energy density (top panel) and vertical kinetic energy density (bottom panel). Black curves: $A = 28 H_0$ ($\psi_{\text{max}}  \sim 1.95$). Gold curves: $A = 18 H_0$ ($\psi_{\text{max}}  \sim 1.26$). Blue curves: $A = 2 H_0$ ($\psi_{\text{max}} \sim 0.14$). Note that the $A=2H_0$ and $A=18H_0$ simulations were run in a box of vertical size $L_z = 48H_0$, while the $A=2H_0$ simulation was run in a box of vertical size $L_z=68H_0$, so we have multiplied the volume-averages by their respective vertical box sizes to facilitate easier comparison between the simulations, and with the theory.}
    \label{FIGURE_WarpAmplitudeComparison_EkinxEkinz}
\end{figure}

In  Figure \ref{FIGURE_WarpAmplitudeComparison_EkinxEkinz} we show the corresponding volume-averaged kinetic energy time-series (multiplied by the vertical box size $L_z$) from three of the four simulations in our warp amplitude study: the blue curve corresponds to the smallest warp amplitude run ($A=2H_0$). Both the radial (top panel) and vertical (bottom panel) components of the energy oscillate twice per orbit and are exactly 90 degrees out of phase. Their envelopes exhibit longer oscillations with a clear beat period on time-scales of around 14-15 orbits. Finally, over the course of the simulation (50 orbits) there is a steady decay in both components of the kinetic energy.

The beat in the kinetic energy components is related to the propagation of a bending wave through the disc. The measured beat period of 14 orbits in the kinetic energy agrees well with the predicted period of $\frac{1}{2}\times1/(\frac{1}{2}c_{s0} k) \approx 14.3$ orbits) of a bending wave in our low-amplitude bending wave theory (see Section \ref{THEORY}). At large warp amplitudes (gold and black curves in Figure \ref{FIGURE_WarpAmplitudeComparison_EkinxEkinz}) the beat is more irregular and only really significant within the first 10 orbits, which suggests that there might be a maximum warp amplitude (likely around $\psi_{\text{max}}  \sim 1$) that a bending wave can support.

\begin{figure*}
    \centering
    \includegraphics[width=1\linewidth]{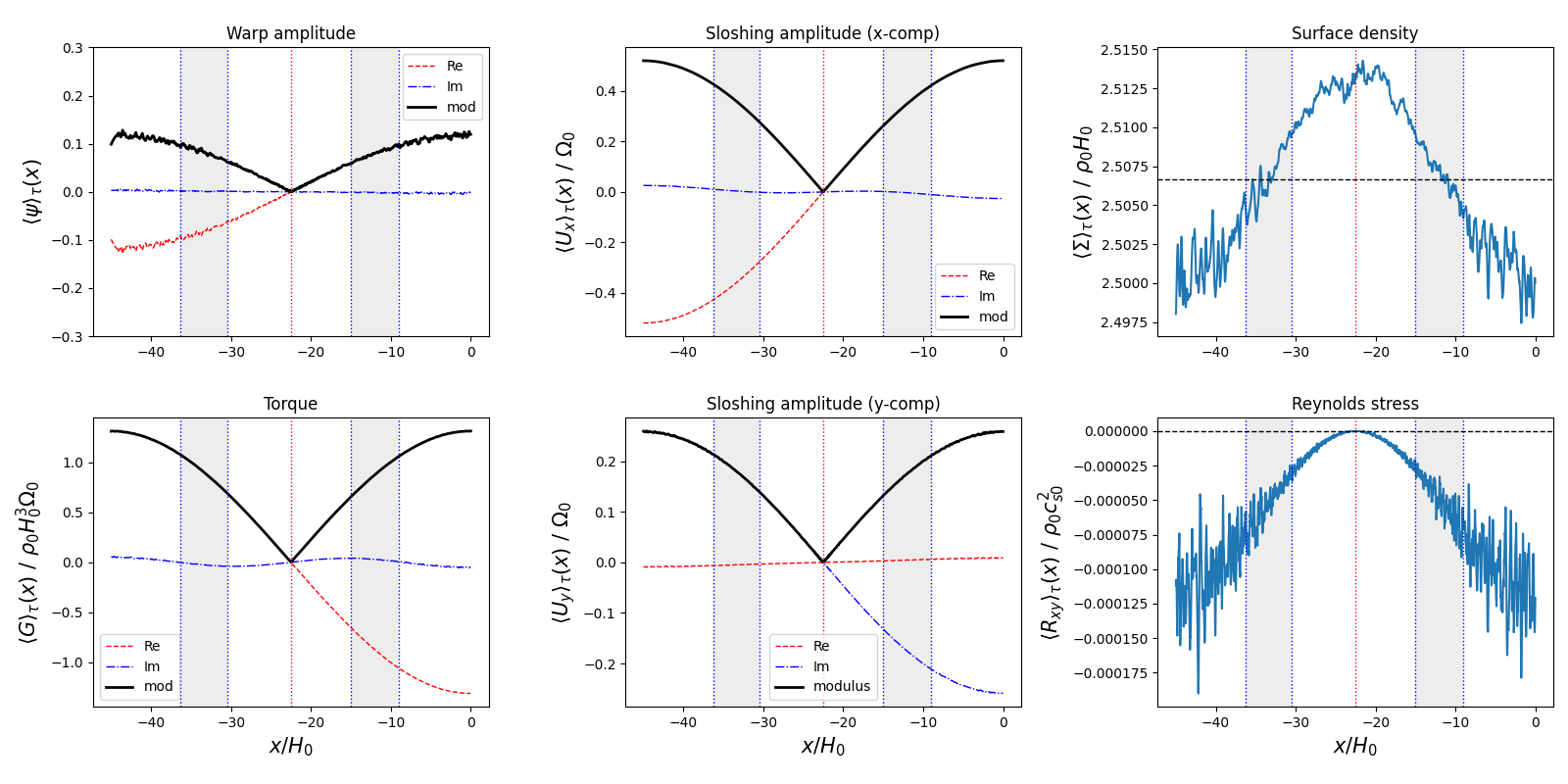}
    \caption{Orbit-averaged radial profiles in the left-hand part of the disc from a simulation of a small warp amplitude ($A=2H_0, \psi_{\text{max}}  \sim 0.14$).  The profiles have been averaged over orbit 2-3 (before the parametric instability saturates). See Section \ref{METHODS_Diagnostics} and Figure \ref{FIGURE_Fiducial_OrbitAveragedRadialProfilesOrb0Orb1} for definitions. Note that we have included the shaded areas, which mark the regions over which gaps open in the large-warp amplitude ($A=28H_0, \psi_{\text{max}}  \sim 1.95$) run, but these are shown for reference, only: the disc does not break in the $A=2H_0$ run.}
    \label{FIGURE_ParameterStudy_OrbitAveragedProfilesA2}
\end{figure*}

We do not see any evidence of shocks in the small-warp amplitude simulation. Furthermore, the kinetic energy evolution for this (inviscid) simulation is almost identical to that in a simulation with a small explicit viscosity of $\alpha = 0.003$ (compare black and gold curves in Figure \ref{FIGURE_ParameterStudy_A2EkinTimeSeriesViscosityStudy}),which suggests that the the numerical dissipation is small in the inviscid run. This leaves the parametric instability as the dominant dissipation mechanism at small warp amplitudes, since the instability takes energy from the bending mode and puts it in small scale modes.

In the low-warp amplitude run, the warp does not steepen after initialization even at the locations of maximum initial warp amplitude ($x/H_0 \sim -45,0,45$), in contrast to what we find at large warp amplitudes (see Figure \ref{FIGURE_FiducialSimulationCenterOfMassProfile}). Over the first six orbits, the oscillation (i.e. the tilt) is damped and the warp flattens around the locations where the warp amplitude was initially zero ($x/H_0 = -22.5, 22.5$). After orbit 6 the vertical oscillations begin to grow again, reaching about $45\%$ of their initial value around orbit 14, before they are dampened once more. This `re-invigoration' of the tilt (which coincides with peaks in the envelope in the vertical kinetic energy) is related to the propagation of the aforementioned bending wave through the disc.\footnote{Note that in \cite{held2024instability} we found that the re-invigoration of the bounce in a strongly bouncing disc was due to a three-mode coupling between the vertical bouncing and oppositely directed travelling waves that formed the warp or standing wave. In our simulation here, the bouncing is weak ($H(t)$ oscillates between around $0.9H_0$ and $1.1H_0$), but ultimately the mechanism is similar as both situations involve a (partly) reversible transfer of energy between a vertical oscillation and some other oscillator.}

In Figure \ref{FIGURE_ParameterStudy_OrbitAveragedProfilesA2} we show the orbit-averaged radial profiles of various diagnostics. The orbit-average was taken over orbit 2-3, before the parametric instability saturated. The radial structure, and particularly the evolution, of these profiles is considerably simpler than in the large-warp amplitude case (the reader may wish to compare to Figure \ref{FIGURE_Fiducial_OrbitAveragedRadialProfilesOrb0Orb1}). The torque $\langle G\rangle_{\tau}$ (bottom-left panel) increases monotonically with warp amplitude. Over the first six orbits (not shown), the shear components $\langle U_{x,y} \rangle_{\tau}$ (middle columns) and the torque both grow and reach maxima of $\langle U_{x,y} \rangle_{\tau} \sim (0.8,0.4) \Omega_0$ and $\langle G\rangle_{\tau} \sim 2\rho_0 \Omega_0 H_0^3$, respectively. The warp amplitude $\langle \psi \rangle_{\tau}$ (top-left-hand panel) maintains its original profile, but decreases significantly over time, meaning the warp is being flattened. 

The orbit-averaged surface density $\langle \Sigma \rangle_{\tau}$ (top-right-hand panel) increases by around $\sim 0.5\%$ around the location of the peak of the sine curve ($x/H_0 \sim -22.5$) over the first three orbits, which is in excellent agreement with our theoretical prediction (see Section \ref{RESULTS_DISCUSSION}). Finally, the $xy$-component of the Reynolds stress (bottom-right-hand panel) is zero at the peak of the sine curve and becomes increasingly more negative outside away from the peak. However, over the course of the simulation (50 orbits), we find that  the radial profiles of surface density and Reynolds stress are generally noisy without any notable structure. In particular, we do not observe the development of gaps in the surface density like we did at larger warp amplitudes. We conclude that the disc does not break at this small warp amplitude.

Finally, we have also carried out a simulation at a moderate warp amplitude of $A = 9H_0$ ($\psi_{\text{max}} \sim 0.63$). This run is more similar to the $A=2H_0$ ($\psi_{\text{max}} \sim 0.14$) run than to the runs with warp amplitude $\psi_{\text{max}} > 1$, so we will omit a detailed discussion for the sake of brevity. The key results are that the disc becomes parametrically unstable and does not break, though it does produce some shocks at early times. By the end of the simulation the disc in the $A = 9H_0$ run looks similar to the disc shown in the bottom panel of Figure \ref{FIGURE_ParameterStudy_2DFlowFieldA18A2} for the  $A = 2H_0$ run. The similarity of this run to the $\psi_{\text{max}}  \sim 0.14$ run suggests that the transition from parametric instability to disc breaking likely occurs around $\psi_{\text{max}} \sim 1$, at least for warps of wavelength $\lambda \sim 90H_0$.

\subsubsection{Large warp amplitudes: $\psi_{\text{max}}  \gtrsim 1$}
\label{RESULTS_ParameterStudy_LargeWarpAmplitudeRuns}
Next we compare our fiducial simulation with $A=28H_0$ ($\psi_\text{max} \sim 1.95$) to a run with a somewhat smaller warp amplitude of $A=18H_0$ ($\psi_\text{max}  \sim 1.26$). Note that the latter run has almost the same maximum warp amplitude as the large warp amplitude simulation in \cite{lodato2010diffusive}. 

In Figure \ref{FIGURE_WarpAmplitudeComparison_EkinxEkinz} we show the volume-averaged time-series of the kinetic energy from these runs (black and gold curves for $A=28H_0$ and $A=18H_0$, respectively).\footnote{We have multiplied each time-series by the simulation's corresponding vertical box size ($L_z=48H_0$ for the  $A=18H_0$ run and $L_z=68H_0$ for the $A=28H_0$ run) in order to eliminate the box-size-dependence in the vertical average.} The time-series for the large-warp amplitude runs have similar shapes: the radial kinetic energy increases initially as the sloshing motions grow, and decreases after around 8 orbits as the sloshing motions (and the warp giving rise to the sloshing motions) are damped by shocks. We do not observe parametric instability in these runs. At these large warp amplitudes the horizontal sloshing motions are so strong that the parametrically unstable modes are likely sheared out. Interestingly, the peak in the radial kinetic energy (as well as the dissipation of vertical kinetic energy) is larger between around orbit 2 and orbit 9 in the $A=18H_0$ simulation compared to the $A=28H_0$ simulation. This could be because in the $A=28H_0$ case the sloshing motions are more effectively damped by the stronger shocks or by the stronger bouncing.

The flow field and orbit-averaged radial profiles in the $A=18H_0$ simulation (not shown) are qualitatively similar to those in our fiducial simulation at $A = 28H_0$ (see Figures \ref{FIGURE_Fiducial_OrbitAveragedRadialProfilesOrb0Orb1} and \ref{FIGURE_FiducialSimulation_OrbitAveragedProfilesOrb25Orb26}). We see transient steepening of the warp at early times where the warp amplitude is maximum (at $x/H_0 \sim -45,0,45$), and the horizontal sloshing motions periodically converge into shocks at the peaks of the sine curve ($x/H_0 \sim -22.5,22.5$). As in our fiducial simulation, the disc flattens around these locations and eventually breaks into four rings. However, in the $A=18H_0$ run the gaps are not as deep: we measure $\langle \Sigma \rangle_{\tau} \sim \rho_0 H_0$ at orbit 25, compared to $\langle \Sigma \rangle_{\tau} \sim 0.5 \rho_0 H_0$ at the same time in the $A = 28H_0$ run. The warp has also been noticeably damped by this stage, with $\langle \psi \rangle_{\tau}$ peaking at around $0.5$ in the gaps in the $A=18H_0$ run compared to $\sim 2$ in the $A=28H_0$ run.

\subsection{Viscosity study}
\label{RESULTS_ParameterStudy_ViscosityStudy}
So far, our simulations have not included explicit viscosity. Given the high resolutions that we can reach in the shearing box (our highest resolution run was at $128$ cells per scale-height $H_0$) the numerical dissipation is likely small in our inviscid runs and they should be in wavelike regime ($\alpha < H_0/r$). In order to check this, we first compare our inviscid runs to runs with a small explicit viscosity  ($\alpha = 0.003$). We also look at the effect of an even larger explicit viscosity of $\alpha = 0.03$. We compare small and large viscosities at a large warp amplitude ($A = 28H_0, \psi_{\text{max}}  \sim 2$) and also at a small warp amplitude ($A = 2H_0, \psi_{\text{max}}  \sim 0.14$).

First, we point out an important subtlety (relevant to both local and global simulations) regarding the criterion for distinguishing between the \textit{wavelike} and \textit{diffusive} regimes of warp propagation. These are usually distinguished by comparing the viscosity $\alpha$ to the disc aspect ratio $H_0/r$, i.e. $\alpha < H_0/r$ (wavelike) and $\alpha > H_0/r$ (diffusive). In the shearing box it is tempting to use the radial wavelength of the warp $\lambda$ in place of $r$. In our simulations this is equal to the radial box size $L_x = 90H_0$, giving $H_0/\lambda \sim 0.01$, which suggest that our inviscid and small viscosity runs are the in wavelike regime ($\alpha \lesssim 0.003 \ll H_0/\lambda$), while our large viscosity run is in the diffusive regime ($\alpha = 0.03 \gtrsim H_0/\lambda$). However, an implicit assumption of comparing $\alpha$ with $H/r$ is that radial wavenumber of the global warp $k = 2\pi/\lambda$ is roughly $1/r$. If that is not the case, the relevant comparison is in fact of $\alpha$ with $kH$. In that light, $kH_0 \sim 0.07$ for our set-up, and so all of our warps are actually in the wavelike regime ($\alpha < kH_0$), though our large-viscosity simulation ($\alpha = 0.03$) is only marginally so.

\begin{figure}
    \centering
    \includegraphics[width=1\linewidth]{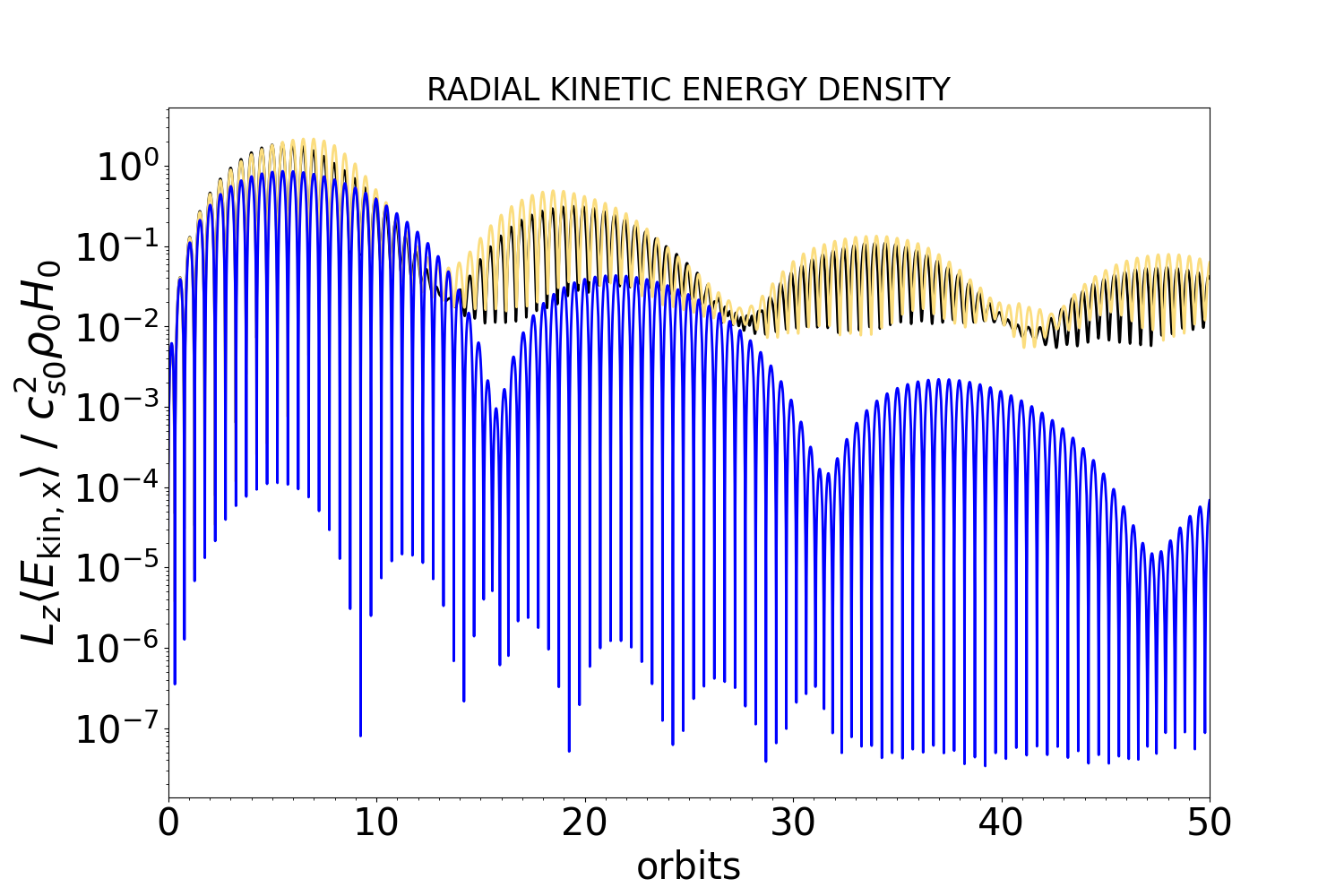}
    \includegraphics[width=1\linewidth]{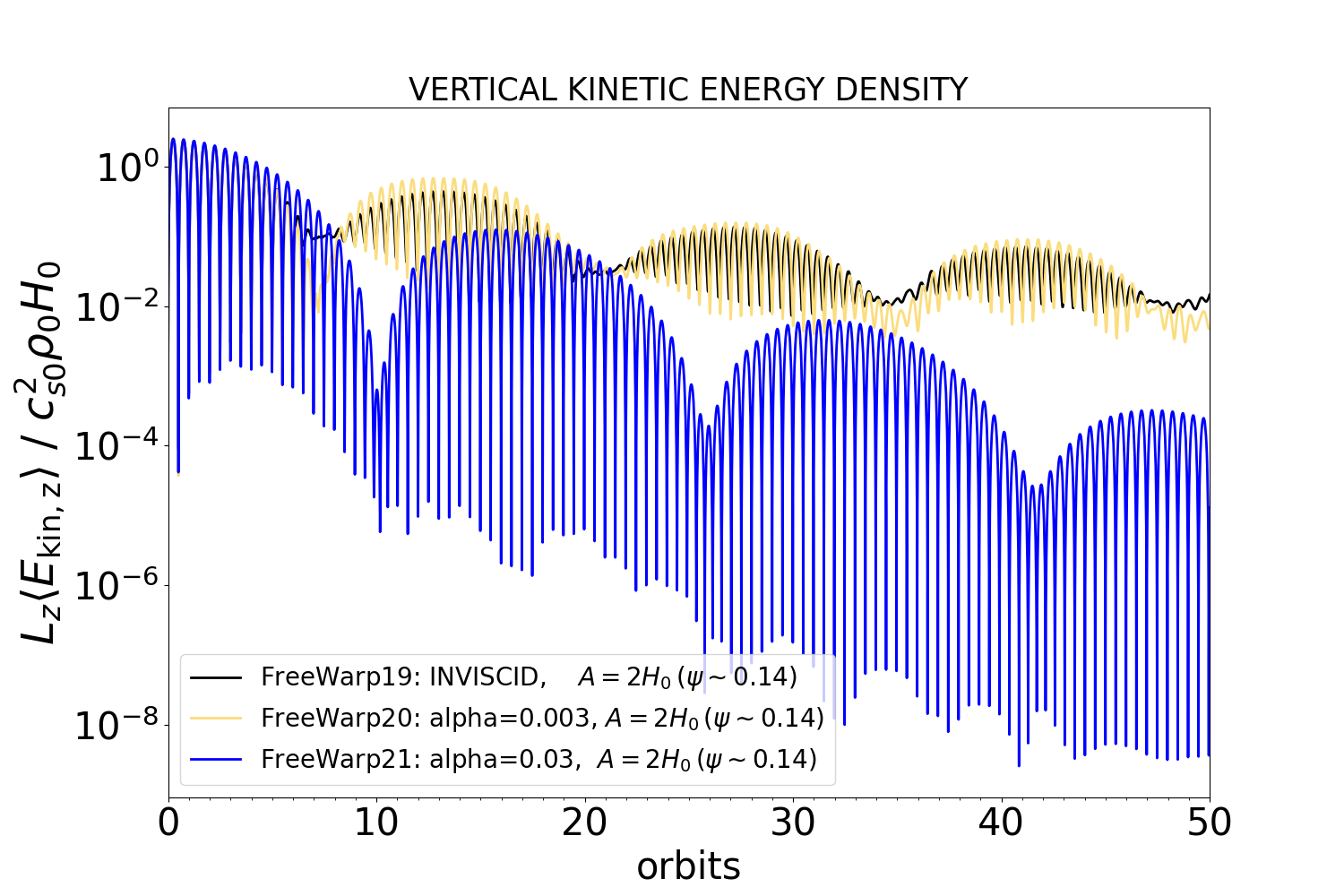}
    \caption{Viscosity study at a small warp amplitude ($A = 2, \psi_{\text{max}}  \sim 0.14$). Comparison of radial (top panel) and vertical (bottom panel) kinetic energy density time-series from three different simulations: inviscid (black curve), $\alpha = 0.003$ (gold curve), $\alpha = 0.03$ (blue curve). We have multiplied
    the volume-averages by the vertical box size ($L_z = 48H_0$) to facilitate comparison with the theory (see Figure \ref{FIGURE_Theory_EkinxEkinzTimeSeries}).}
    \label{FIGURE_ParameterStudy_A2EkinTimeSeriesViscosityStudy}
\end{figure}

\subsubsection{Effect of viscosity at small warp amplitudes}
We start by comparing the effect of explicit viscosity on a small-amplitude warp ($A = 2H_0, \psi_{\text{max}}  \sim 0.14$). All simulations were run at a resolution of $64/H_0$ for 50 orbits and employed the HLLC Riemann solver and parabolic reconstruction. 

In Figure \ref{FIGURE_ParameterStudy_A2EkinTimeSeriesViscosityStudy} we compare the time-series of volume-averaged kinetic energy density components in the inviscid, $\alpha = 0.003$, and $\alpha = 0.03$ runs. The kinetic energy time-series broadly share the same three characteristics and are very similar to those predicted by our linear bending wave theory (see Figure \ref{FIGURE_Theory_EkinxEkinzTimeSeries}): fast oscillations on orbital time-scales, a longer modulation (beat) on time-scales of around 10-15 orbits, and a steady decay over the course of the simulation (50 orbits). Considering the evolution as a whole, we see that the black and gold curves (corresponding to the inviscid and $\alpha = 0.003$ simulations) lie almost on top of each other, which demonstrates that the warp evolves in a similar manner in the inviscid and small-viscosity regimes. This implies that, at least at resolutions of $64/H_0$ and above, the numerical dissipation is negligible in our inviscid runs.\footnote{Note that the viscous timescale (on scales of order $H_0$) is given by $t_{\text{visc}} \sim 1/(\alpha \Omega_0)$. This works out to be around 53 orbits for the $\alpha = 0.003$ run, which is slightly longer than the duration of the simulation.  However, small-scale features in flow (due to parametric instability) mean there are length-scales that are shorter than $H_0$, so viscosity is likely to act there.} Furthermore, there is no decay in kinetic energy in the inviscid and low viscosity runs in our theoretical prediction (Figure \ref{FIGURE_Theory_EkinxEkinzTimeSeries}). The theory did not include parametric instability, so this shows that the dissipation in the corresponding simulations is almost entirely dominated by the PI. 

All three simulations exhibit beating on longer time-scales which corresponds to the frequency of the bending wave propagating in the warped disc (see Figure \ref{FIGURE_ParameterStudy_A2EkinTimeSeriesViscosityStudy}). The beat period is very similar in the inviscid and low viscosity runs. In the large-viscosity run (blue curve), the beat period is slightly longer ($\sim 16$ orbits, compared to $\sim 14$ orbits in the inviscid and low-viscosity runs. This can be explained by adding viscous damping to the long-wavelength bending-wave theory. The beat frequency (twice bending wave frequency $\Omega_b$, see Equation \ref{EQUATION_PhaseOfBendingWave}) is supposed to be $\sqrt{c_s^2k^2-\gamma^2}$,
where $\gamma=\alpha\Omega$ is the viscous damping rate of the sloshing motions. That gives a beat period of $14.3$ orbits for $\alpha=0$ and $15.9$ orbits for $\alpha=0.03$. Thus our small-amplitude large viscosity simulation is also in excellent agreement with the theoretical prediction.

Turning next to the flow fields (i.e., $\rho$ and $\rho v_x$ in the $xz$-plane; not shown), we find that parametric instability develops in the inviscid and low-viscosity runs, but it is killed in the large-viscosity run. For the largest viscosity we investigated, the disc remains smooth for the entirety of the simulation. Thus the only source of dissipation in this run is the explicit viscosity itself, and we note the close agreement between the kinetic energy time-series predicted by theory and those measured in the simulation (compare blue curves in Figures \ref{FIGURE_Theory_EkinxEkinzTimeSeries} and \ref{FIGURE_ParameterStudy_A2EkinTimeSeriesViscosityStudy}). By the end of the large viscosity simulation the warp has been almost entirely smoothed out.

Finally we compare the orbit-averaged radial profiles in the three runs (note shown). The profiles for the $\alpha = 0.03$ are qualitatively similar to those in the inviscid run (see discussion at the end of Section \ref{RESULTS_ParameterStudy_SmallWarpAmplitudeRuns} and Figure \ref{FIGURE_ParameterStudy_OrbitAveragedProfilesA2}). In the large viscosity run, the tilt is rapidly damped after initialization, dropping from $\langle \psi \rangle_{\tau} \sim  0.14$ to $\sim 0.05$ at $x/H_0=-45,0,45$ after ten orbits. The shear and torque grow after initialization, but at a slower rate than in the wavelike runs and reach maxima of $\langle U_x \rangle_{\tau} \sim 0.6\Omega_0$ and $ \langle G \rangle_{\tau} \sim 1.5 \rho_0 H_0^3\Omega_0 $ after seven orbits, compared to $\sim \Omega_0$ and $\sim 2 \rho_0 H_0^3\Omega_0 $ in the wavelike runs. The percentage increase in surface density at $x \sim -22.5H_0$ over the first few orbits is about an order of magnitude smaller in the large viscosity run ($\Delta \Sigma/\Sigma_0 \sim 0.0007$ compared to $\sim 0.005$).

\subsubsection{Effect of viscosity at large warp amplitudes}
Next we compare the effect viscosity on the the evolution of the warp in the wavelike regime in a disc with a large initial warp amplitude. We compare three simulations which were all initialized with a maximum warp amplitude of $A = 28H_0$ ($\psi_{\text{max}}  \sim 1.95$). Each simulation was run with a resolution of $32/H_0$ using the HLL Riemann solver and parabolic interpolation.

\begin{figure}
    \centering
    \includegraphics[width=1\linewidth]{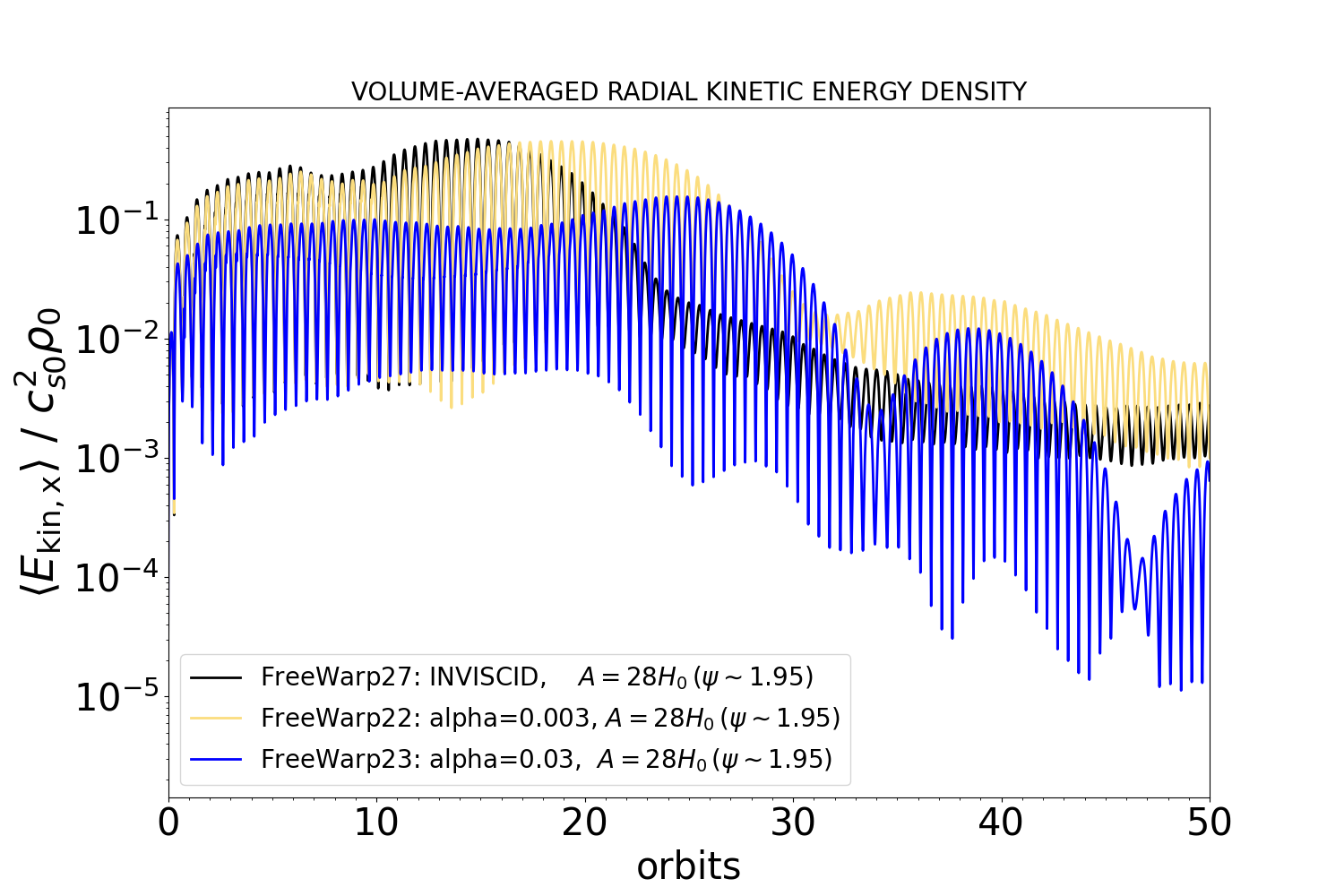}
    \includegraphics[width=1\linewidth]{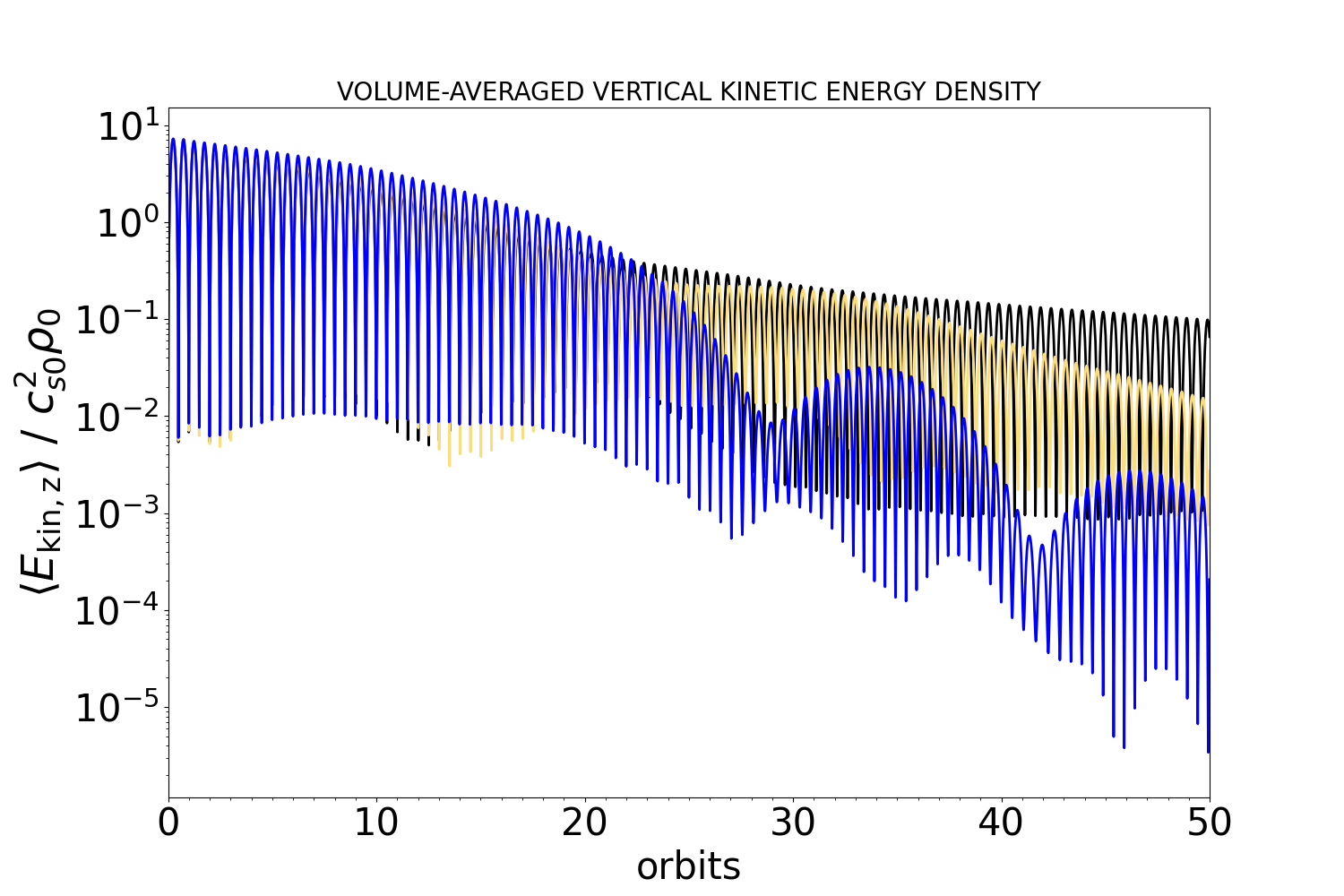}
    \caption{Viscosity study at a large warp amplitude ($A = 28H_0, \psi_{\text{max}}  \sim 1.95$). Comparison of radial (top panel) and vertical (bottom panel) kinetic energy density time-series from three different simulations: inviscid (black curve), $\alpha = 0.003$ (gold curve), $\alpha = 0.03$ (blue curve).} \label{FIGURE_ParameterStudy_A28EkinTimeSeriesViscosityStudy}
\end{figure}

In Figure \ref{FIGURE_ParameterStudy_A28EkinTimeSeriesViscosityStudy} we show the kinetic energy time-series from the three runs. Similar to the small-warp amplitude runs described in the previous section, the kinetic energy evolution in the small-viscosity run (gold curve) is similar to that in the inviscid run (black curve), although the match is not as close as it was at small warp amplitudes, especially in the radial kinetic energy after the first ten orbits or so. In particular, in the $\alpha = 0.003$ run, the envelope in the radial kinetic energy peaks around five orbits later compared to the inviscid run, and the beating of the envelope is more pronounced.

In the density field (not shown), the evolution of the disc in the $\alpha = 0.003$ run looks very similar to that of the inviscid run whose density field evolution is shown in Figure \ref{FIGURE_FiducialSimulationFlowField}. We observe many of the same features, such as shocks, steepening of the warp, etc. The same goes for the orbit-averaged radial profiles (not shown), although the gaps are somewhat wider (by around $\sim 1$-$2H_0$) and deeper ($\text{min}\langle \Sigma \rangle_{\tau} \sim 0.5 \rho_0 H_0$) in the inviscid run compared to the small-viscosity run where we measure $\text{min}\langle \Sigma \rangle_{\tau} \sim 0.9 \rho_0 H_0$.

With the exception of the first few orbits, the large viscosity run at $\alpha = 0.03$ (blue curve in Figure \ref{FIGURE_ParameterStudy_A28EkinTimeSeriesViscosityStudy}) differs significantly from its low-viscosity counterparts. The vertical kinetic energy (bottom panel) is damped at a greater rate after orbit 20, and the beating is also much more prominent than in the wavelike runs. The density field in the large viscosity run (not shown) resembles that in the wavelike runs only over the first few orbits, which is not surprising given that this is before a viscous timescale ($t_{\text{visc}} \sim 1/(\alpha \Omega_0) \sim 5$ orbits) has elapsed. The horizontal sloshing motions converge into shocks above (or below) the peaks of the sinusoidal warp, although the flow field generally looks smoother in the large viscosity run (note that the viscous timescale is smaller than 5 orbits on length-scales smaller than $H_0$, and so the viscosity will impact thin structures such as shocks). Over the first ten orbits or so, the disc flattens near the peaks of the sinusoidal warp, and steepens around $x/H_0 \sim -45,0,45$. 

In the inviscid and small-viscosity simulations, the disc eventually breaks into four rings: two tilted rings centered around $x/H_0 \sim -22.5, 22.5$ and two untilted rings at $x/H_0 \sim \pm 45H_0$ and $0H_0$, respectively (see bottom right-hand panel of Figure \ref{FIGURE_FiducialSimulationFlowField}). In the large viscosity run, however, we observe only two untilted rings by the end of the simulation: a central ring located at $x \sim 0 H_0$ and a second ring at $x = -45H_0, 45H_0$ which is connected by the left- and right-hand radial boundaries and is, in fact, the same ring. Between the rings there is a broad and shallow gap with a minimum orbit-averaged surface density of $\text{min}\langle \Sigma \rangle_{\tau} \sim 1.5 \rho_0 H_0$, compared to the initial surface density of $\sqrt{2\pi}\rho_0H_0$. This gap is shallower (and broader) than the gaps in the corresponding inviscid and low-viscosity simulations in which we measured gaps of depth $\text{min}\langle \Sigma \rangle_{\tau} \sim 0.5 \rho_0 H_0$ and $\sim 0.9 \rho_0 H_0$, respectively. In fact, following the evolution of the surface density profile (not shown), we see that viscous diffusion in the large-viscosity run actually destroys the surface density structure that forms the two tilted rings (i.e. it wipes out the central peak in surface density, as seen, for example, in the top-right hand panel of Figure \ref{FIGURE_FiducialSimulation_OrbitAveragedProfilesOrb25Orb26} from the inviscid run). Crucially, by the end of the simulation, the orbit-averaged warp amplitude is very small ($\psi \sim 0.01$) and almost uniform in radius. We do not observe peaks in the warp amplitude where there are gaps in the surface density, in contrast to wavelike runs (see top-left and top-right hand panels of Figure \ref{FIGURE_FiducialSimulation_OrbitAveragedProfilesOrb25Orb26}). We conclude that viscous diffusion prevents the disc from fully breaking in the large-viscosity case, at least at the warp amplitude ($\psi \sim 1.95$) and wavelength ($\lambda \sim 90H_0$) which we have investigated here.

\section{Discussion}
\label{RESULTS_DISCUSSION}
In Section \ref{RESULTS_FiducialSimulation_DiskBreakingMechanism} we summarize our key results, compare them to our theoretical predictions (for small warp amplitudes), and elaborate on the physical mechanisms at play during the evolution of a free (unforced) warp in the wavelike regime at both small ($\psi_{\text{max}}\sim 0.14$) and large ($\psi_{\text{max}} \sim 1.95$) warp amplitudes, and under the action of small ($\alpha \lesssim 0.003$) and large  ($\alpha = 0.03$) viscosities. Our results suggest that the critical warp amplitude required for disc breaking to occur is $\psi_{\text{max}} \sim 1$, so we have separated our summary into the $\psi_{\text{max}} \lesssim 1$ and $\psi_{\text{max}} \gtrsim 1$ regimes, accordingly. We caution, however, that is unclear whether this critical value is universal since all our simulations were limited to a warp of wavelength $\lambda = 90H_0$. Finally, we discuss how our results relate to existing theoretical mechanisms of disc breaking in the literature (Section \ref{RESULTS_Discussion_ComparisonWithTheoreticalBreakingModels}).

\subsection{Summary of results}
\label{RESULTS_FiducialSimulation_DiskBreakingMechanism}

\subsubsection{Evolution at small warp amplitudes ($\psi_{\text{max}} \lesssim 1$)}
At small warp amplitudes the disc does not break. Instead a long-wavelength ($\lambda \sim 90H_0$) bending wave propagates through the disc, and modulates the shorter-period oscillations which occur once per orbit (or twice per orbit in the kinetic energy density). In Section \ref{THEORY} we developed a linear theory of small-amplitude bending waves which predicts that a bending wave with wavenumber $k = 2\pi/\lambda$ should oscillate with a frequency of $\Omega_b = (1/2)c_{\text{s}0} k \sim 0.0349 \Omega_0$ or $1/\Omega_b \sim 28.6$ orbits. The corresponding beat frequency for the kinetic energy is twice that, so $\Omega_b \sim 0.0698\Omega_0$ or 14.3 orbits. This is in excellent agreement with the beat of about 14 orbits which we measure in the kinetic energy time-series in our small-warp amplitude simulation ($A=2H_0, \psi_{\text{max}} \sim 0.14$): see blue curves in Figure \ref{FIGURE_WarpAmplitudeComparison_EkinxEkinz}.

Over the first few orbits the warp evolves as an undamped  bending wave. The radial pressure imbalance in the warp gives rise to oscillatory horizontal sloshing motions which change sign twice per orbit. The sloshing amplitude grows over the first four orbits, and reaches a maximum orbit-averaged value of $\langle U_x \rangle_{\tau} \sim 0.8 \Omega_0$. Unlike at large warp amplitudes (see below), the sloshing motions do not converge into shocks. However, the sloshing motions involve a correlation between $u_x$ and $\delta u_y$ and thus give rise to a Reynolds stress. We measure a minimum orbit- (between orbit 3 and orbit 4) and vertically-integrated stress of $L_z\langle R_{xy} \rangle_{\tau} \sim -0.0072$ at the left-most location of maximum warp amplitude ($x = 0H_0$). We also measure a fractional increase in orbit-averaged surface density $\Delta \langle \Sigma \rangle_{\tau}/\Sigma_0 \sim (2.519-\sqrt{2\pi})/\sqrt{2\pi} \sim 0.0049$ at the left-hand peak of the sinusoidal warp ($x \sim -22H_0$. Using our linear, inviscid, bending wave theory (see Section \ref{THEORY}), we can predict the expected fractional change in surface density (Equation \ref{EQUATION_FractionalChangeSurfaceDensity}) and also the sign and magnitude of the Reynolds stress (Equation \ref{EQUATION_ReynoldsStressxy}). The predicted fractional increase in surface density at the peaks of the sinusoidal warp after four orbits is $ \Delta\ln\Sigma\approx-\f{\psi_{\text{max}}^2}{4}\cos(2kx)[1-\cos(c_{\text{s}0}kt)] \sim 0.00473$\footnote{Here we have used: $\psi_{\text{max}} \sim 0.14$, $k = 2\pi/L_x$, $L_x = 90H_0$, $x = -22.5H_0$, $c_{\text{s}0} =1$, $t=3.5$orbits ($\sim 22\Omega_0^{-1}$).}, which is in excellent agreement with the value measured in the simulation. The predicted $xy$-component of the vertically-integrated Reynolds stress is $\approx \f{\psi_{\text{max}} ^2\Sigma\Omega_0^2H_0}{16k}\cos^2kx\sin(c_{s0}kt) \sim -0.017$. However, note that the surface density change is not simply explained by $R_{xy}$ (see Appendix \ref{APPENDIX_InviscidBendingWaveTheory}).

Another important outcome of the low-amplitude warp simulations is that 4-5 orbits after initialization the warp becomes parametrically unstable. The instability first develops locally, in the vicinity of the locations of maximum warp amplitude ($x/H_0 \sim -45, 0, 45$). By around orbit 10 the instability has spread throughout the entire warp. As there are no shocks, the parametric instability acts as the dominant source of dissipation because it takes energy out of the warp and transfers it to small scales, where it is lost through numerical dissipation (in inviscid case) or else through the explicit viscosity (in the viscous case).

In the large viscosity run ($\alpha = 0.03$) the parametric instability is suppressed, and the disc evolves as a pure bending wave which is damped only by the explicit viscosity. The kinetic energy evolution measured in the simulation (blue curve in Figure \ref{FIGURE_ParameterStudy_A2EkinTimeSeriesViscosityStudy}) is in very good agreement with that predicted by our linear bending wave theory with dissipation (blue curve in Figure \ref{FIGURE_Theory_EkinxEkinzTimeSeries}), and the measured beat period of 16 orbits is in excellent agreement with that predicted by theory of $1/\sqrt{c_s^2k^2-\gamma^2} \sim 15.9$ orbits, where $\gamma = \alpha\Omega$.

\subsubsection{Evolution at large warp amplitudes ($\psi_{\text{max}} \gtrsim 1$)}
At large warp amplitudes the evolution is significantly more complicated than at small warp amplitudes. The dominant processes are the development of very strong horizontal sloshing motions which reach supersonic speeds in the radial direction several scale-heights above the local disc midplane, as well as shocks, flattening of the warp at some locations in the disc and steepening of the warp at other locations, and ultimately the breaking of the disc into four flat rings, of which two are tilted.

Let us first summarize the outcome in our inviscid and small viscosity runs ($\alpha = 0.003$). We see clear evidence of warp steepening at locations of maximum warp amplitude ($x/H_0 \sim -45,0,45$) shortly after initialization, but the effect is transient and probably not directly related to the breaking mechanism. Instead, the breaking is facilitated by the the horizontal sloshing motions, which are driven by the large radial pressure imbalance in the warp. The sloshing motions transport both mass and the misaligned component of angular momentum (i.e. the tilt). If there were no dissipation, the bending wave would just continue oscillating and the transport would be reversible. However, the crucial difference at large warp amplitudes is that these sloshing motions are so strong (reaching $\langle U_x \rangle_{\tau} \sim 2 \Omega_0$ within two orbits of initialization) that they converge into standing shocks twice per orbit several scale-heights below (or above) the peak (trough) of the sinusoidal warp. These shocks thus act as a source of dissipation, as  foreshadowed by \cite{nelson1999hydrodynamic} who first identified such shocks in early simulations of bending waves. Together with the sloshing motions, the shocks thus facilitate an irreversible accumulation of mass, and also of the flux of the tilt, at the peaks of the sinusoidal warp.

The result is a gradual opening of several gaps in the disc which are centred around $x/H_0 \sim -30, -15, 15$ and $30$ as shown in the bottom right-hand panel of Figure \ref{FIGURE_FiducialSimulationFlowField}, and a flattening of the warp around the peaks of the sinusoid ($x/H_0 \sim -22.5, 22.5$). By the end of the simulation the disc has broken into four rings of width $\sim 15H_0$ separated by gaps of width $\sim 7H_0$. A quantitative signature of breaking is that the \textit{troughs} (gaps) in the orbit-averaged surface density (see top left-hand panel of Figure \ref{FIGURE_FiducialSimulation_OrbitAveragedProfilesOrb25Orb26}) coincide with \textit{peaks} in the orbit-averaged warp amplitude (top left-hand panel in the same figure), which is otherwise small and radially uniform.

In the large viscosity run ($\alpha = 0.03$) we also observe the formation of shocks and a moderate decrease in the surface density at the same locations as we do in the low-vo regime. However, after a viscous timescale of around $1/(\alpha \Omega) \sim 5$ orbits, the large viscosity run begins to differ appreciably from its small-viscosity counterparts. In the low viscosity runs, the disc broke into four rings, of which two were untilted and two were tilted. In the large viscosity run, however, the tilted rings are subsumed into the gaps with most of the mass piling up in the two untilted rings. In other words, the multiple ring-gap proto-structure that starts to emerge at early times in the large viscosity run gets eliminated by viscous diffusion. Although we note a moderate decrease in surface density between the remaining untilted rings, we hesitate to call this outcome a `break' because we do not see a corresponding peak in the orbit-averaged warp amplitude $\langle \psi \rangle_{\tau}$ within the gap. 

We note that our large viscosity result is different to that found in the SPH simulations of \cite{lodato2010diffusive}: they find that the disc breaks cleanly at $\alpha = 0.03$ (see their Figure 11). Their warp has the same radial wavelength ($\lambda \sim 90H_0$) as the one we considered in our simulations, and the timescale over which the break occurs is similar to the timescale over which the break occurs in our small viscosity runs ($\sim 20$ orbits). We speculate that if a gap opens up early on in the SPH simulations this might result in a significant decrease in viscous diffusion in the gap region, because there are no particles in that region. This could inhibit the gap from filling up again due to the effects of viscous diffusion.

\subsection{Comparison with disc breaking models}
\label{RESULTS_Discussion_ComparisonWithTheoreticalBreakingModels}
Several mechanisms for disc breaking have been suggested in the literature. They can broadly be classified as follows:
(i) precession-driven arguments, 
(ii) disc breaking due to steepening of the warp (equivalently, an anti-diffusion of the \textit{warp}),
(iii) disc breaking facilitated by an angular momentum flux reversal (equivalently, an anti-diffusion of the \textit{surface density}). We review each of these mechanisms in turn, and compare them to our own simulations.

\subsubsection{Precession arguments}
The first argument, most commonly found in literature, invokes differential precession of the disc to explain breaking via twisting of the disc. Note that we do not include precession in our (Keplerian) simulations, and thus the breaking mechanism in our runs must be due to a different mechanism. However, we summarize the arguments here for the convenience of the reader.

At its simplest, this mechanism involves a comparison of the timescale over which the warp can propagate by means of a bending wave (i.e. the \textit{radial communication timescale}) to the \textit{precession timescale} \citep{papaloizou1995dynamicsoftilted,lubow2018linear,rabago2024warps}. Following \cite{papaloizou1995dynamicsoftilted}, who considered the case of a circumstellar disc with a slightly tilted external companion, if the bending wave timescale across the disc is short compared to the typical precession timescale, then the disc will behave like a rigid body. If the bending wave timescale is comparable to the precession timescale, however, then a significant warp will be induced in the sense that the tilt will vary with radius, although it can still be small in amplitude. For larger tilts, the disc is expected to break where the precession timescale is smaller than the radial communication timescale \citep{lubow2018linear,rabago2024warps}. The break in this case is caused by a \textit{twisting} of the disc as a result of the differential precession.\footnote{Note that locally the disc cannot distinguish between a twist and warp.} Note that, although the timescale argument itself is independent of the warp amplitude, the consequence of poor radial communication probably does depend on the amplitude of the tilt.

A similar approach compares the precessional \textit{torque} on the disc to the disc's internal viscous torque \citep{nixon2012broken}. Suppose we tilt the disc through some angle. A precessional torque will then act on the disc. For the disc to behave coherently (i.e. somewhat like a rigid body) it has to provide an internal torque which increases with tilt angle. If at some critical tilt (or warp amplitude) the internal torque can no longer cancel the precession torque, the disc will break. This argument assumes that the internal torque by which the disc will resist the warping or breaking is the same viscous torque which transports vertical (i.e. aligned) angular momentum in the disc. In reality, however, the torque that resists warping or breaking is the horizontal (i.e. misaligned) component of the torque which is really due to pressure (or horizontal sloshing motions, depending on one's perspective) rather than viscosity.

\subsubsection{Instability of the warp amplitude (anti-diffusion of warp amplitude)}
Another disc breaking mechanism, foreshadowed by \cite{ogilvie2000alpha}, and analyzed in detail by \cite{dougan2018instability}, involves an instability of the warp evolution that leads to runaway steepening of the warp. The mechanism can be understood physically as follows: for small warp amplitudes the torque scales linearly with the warp amplitude. Suppose at some critical warp amplitude the torque begins to \textit{decrease} as the warp amplitude is increased. Then a small increase in the warp amplitude results in a further decrease in the torque which is thus no longer able to flatten the warp. This results in further steepening of the warp, and a runaway process occurs. Eventually the warp steepens so much that the warp amplitude becomes discontinuous and the disc breaks. Note that this behavior was also found in the 1D models of \cite{nixon2012broken}.

While we do see clear evidence of warp steepening in our large-warp amplitude simulations (both for small and large viscosities\footnote{Dogan's theory is valid in the diffusive regime, whereas our simulations are in the wavelike regime. Our large viscosity, large-warp amplitude simulation with $\alpha = 0.03$ and $\psi_{\text{max}} \sim 1.95$ marginally falls within the unstable region in the bottom left-hand of their Figure 5.}), the effect is transient, typically taking place over the first four or five orbits of the simulation. In addition, in our simulations the disc does not break where the warp amplitude is \textit{greatest}, but somewhere in between locations of $\psi_{\text{max}}$ and $\psi_{\text{min}} = 0$. As discussed previously, the breaking mechanism in our simulations is due to an irreversible transport of mass and tilt by the horizontal sloshing motions to the location of minimum warp amplitude where the sloshing motions converge into a shock. These non-linear dynamics are not present in the linear theory of \cite{dougan2018instability}.

\subsubsection{Angular momentum flux reversal (anti-diffusion of surface density)}
Finally, breaking could also be facilitated by an instability involving a reversal in the sign of the the angular momentum flux in the disc, which would in turn result in an anti-diffusion of the surface density. In other words, rather than the usual viscous spreading of a ring into a disc, mass is instead concentrated into a particular annulus while gaps open in the surrounding regions until the disc has broken at those locations. Note that this flux reversal was first noted by \cite{ogilvie1999non} and \cite{ogilvie2000alpha}, but was not related there to breaking. 

Changes in surface density are usually assumed to require angular momentum transport. It is possible to transport the vertical component of angular momentum in the radial direction through a correlation between $u_x$ and $\delta u_y$: this can occur in the horizontal sloshing motions if there is a phase relation between $u_x$ and $\delta u_y$. If $u_x$ and $\delta u_y$ are not exactly 90 degrees out of phase there will be a Reynolds stress component which will transport vertical angular momentum in the radial direction. In our linear bending wave theory applicable to small warp amplitudes (see Appendix \ref{APPENDIX_InviscidBendingWaveTheory}), we do indeed find that a bending wave is associated with a negative Reynolds stress and a corresponding change in surface density. However, in the absence of dissipation (due to shocks, parametric instability, or explicit viscosity, for example) we emphasize that this is a reversible process. Furthermore, zonal flows are involved and need to be included in the angular-momentum budget. At large warp amplitudes ($\psi_\text{max} \gtrsim 1$), on the other hand, we have found that the formation of shocks is inevitable because the sloshing motions grow to very large amplitudes. These shocks then act as a source of dissipation, and the aforementioned surface density evolution is likely part of the breaking process.

\section{Conclusions}
\label{CONCLUSIONS}
We have investigated disc breaking, in which an initially continuous warped accretion disc breaks into multiple rings. Such broken discs have been imaged directly (and inferred indirectly) in discs around young stars. Disc breaking has also been found in many global numerical simulations of warped discs with various configurations, ranging from tilted circumbinary protoplanetary discs to tilted circumstellar discs around spinning black holes. 

Relative to an observer moving along an untilted circular orbit, a tilted disc will appear to oscillate about the orbital plane once per orbit (see \cite{ogilvie2022hydrodynamics} for details). We carried out quasi-2D (in the $xz$-plane), isothermal, local (shearing box) simulations using a grid-based Godunov code (\textsc{PLUTO}). We followed the evolution of a free warp of radial wavelength $\lambda = 90H_0$ over fifty orbits, a set-up which is somewhat reminiscent of early, global smoothed-particle-hydrodynamic (SPH) simulations of disc breaking carried out by \cite{lodato2010diffusive}. The local approximation enabled us to model the breaking region, as well as other small-scale processes (such as shocks and parametric instability), at much higher resolutions than are currently accessible in global simulations, and also allowed for a wide exploration of the relevant parameter space. Altogether we investigated warp amplitudes from $\psi_{\text{max}} \sim 0.1$ to $\psi_{\text{max}} \sim 2$ at three viscosities (inviscid, $\alpha = 0.003$, and $\alpha = 0.03$). All our simulations are in the wavelike regime, though the largest viscosity run is only marginally so. Our resolutions ranged from $16$ to $128$ cells per scale-height $H_0$.

Our key finding is that, at large warp amplitudes ($\psi_{\text{max}} \gtrsim 1)$, and in the wavelike regime, the disc breaks into four rings, of which two are untilted with respect to the local orbital plane, and two are tilted. After initialization we observe the growth of a strong (supersonic) horizontal sloshing motions, and steepening of the warp at locations of maximum warp amplitude. A key and novel ingredient of the breaking mechanism is the formation of shocks due to these converging horizontal sloshing motions. The shocks act as a source of dissipation, and therefore facilitate an irreversible transfer of mass and the misaligned component of angular momentum (i.e. the tilt) \textit{from} locations where warp amplitude is large \textit{to} locations where it is small. The resultant rings are flat and are around $15$ scale-heights $H_0$ wide and separated by gaps of around $7H_0$. The disc also begins to break via the same mechanism when a large viscosity is included ($\alpha = 0.03$), however, we find that the emerging gaps in the disc are filled over several viscous diffusion timescales.

At small warp amplitudes ($\psi_{\text{max}}\lesssim 1)$, on the other hand, the disc does not break. Instead a bending wave can propagate through the disc, and we developed a linear theory of small-amplitude bending waves 
to complement the numerical simulations. A key finding is that (at sufficiently small viscosities) we observe the emergence of the hydrodynamic parametric instability (PI) which acts as the main source of dissipation. We have compared the results of our low-warp amplitude simulations to our linear bending wave theory, and found that the simulations are in excellent agreement with the theory, which correctly predicts the expected bending wave timescale, the fractional change in surface density, and the sign and magnitude of the Reynolds stress. In the large viscosity run the parametric instability is suppressed and the warp evolves as a pure bending wave, albeit one that is damped by the explicit viscosity.

\section*{Acknowledgments}
We thank Giuseppe Lodato for useful discussions. This research was funded by STFC through grant ST/X001113/1. Simulations were run on the CSD3 cluster at the University of Cambridge through an STFC DiRAC computing grant, and on the Swirles cluster at DAMTP, University of Cambridge.

\section*{Data availability}
The data underlying this article will be shared on a reasonable request to the corresponding author.




\bibliographystyle{mnras}
\bibliography{BrokenDisksBib} 




\appendix

\section{Numerical convergence}
\label{APPENDIX_ParameterStudy_NumericalParameters}

\begin{figure}
    \centering
    \includegraphics[width=1\linewidth]{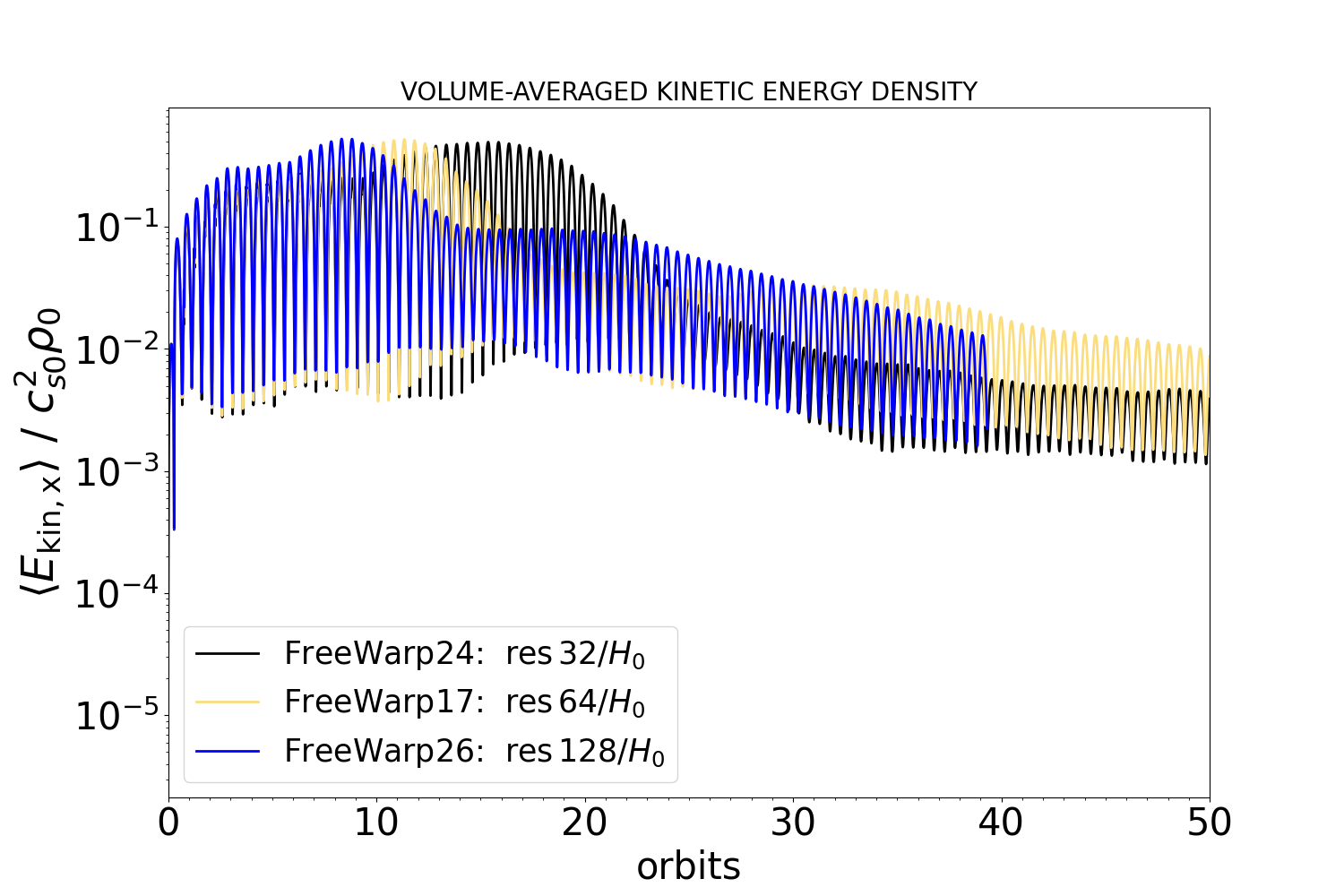}
     \includegraphics[width=1\linewidth]{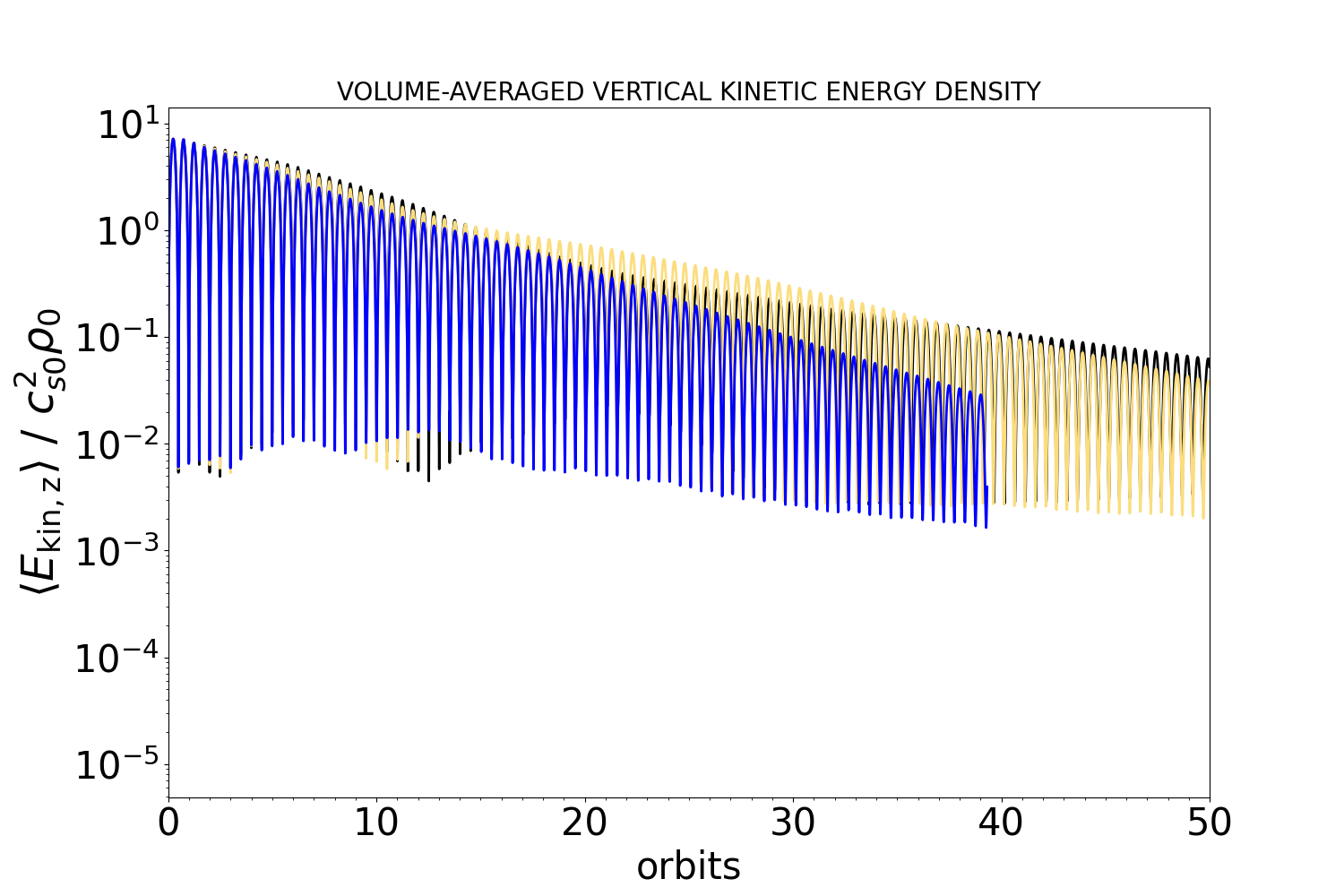}
    \caption{Resolution study for large warp amplitude simulation ($A = 28H_0$, $\psi_{\text{max}}\sim 1.95$). Top: radial kinetic energy density. Bottom: vertical kinetic energy density. Black curve: $32/H_0$. Gold curve: $64/H_0$. Blue curve: $128/H_0$.}
    \label{FIGURE_Appendix_ResolutionStudyTimeSeries}
\end{figure}

\subsection{Resolution study}
\label{APPENDIX__ParameterStudy_ResolutionStudy}
In Figure \ref{FIGURE_Appendix_ResolutionStudyTimeSeries} we compare the results of our fiducial simulation with a large warp amplitude ($A = 28H_0, \psi_{\text{max}} \sim 1.95$) at three different resolutions: $32/H_0$, $64/H_0$, and $128/H_0$. All three simulations were run with the HLLC Riemann solver and parabolic reconstruction.\footnote{At this large warp amplitude the disc enters an extreme bouncing regime at the locations where the warp amplitude is maximum ($x/H_0 \sim -45, 0, 45$). During a bounce the disc can be compressed to as little as $\sim 0.1 H_0$. This means that at the instant the disc is highly compressed it is only resolved by $3$ cells at our lowest resolution of $32/H_0$. At the higher resolutions the disc is resolved by $\sim 6$ and $\sim 13$ cells, respectively.} As the resolution is increased the horizontal sloshing motions saturate earlier (the peak in the envelope of the radial kinetic energy moves to the left), but the simulations are otherwise qualitatively similar.

We also carried out a resolution study for a small warp amplitude  ($A = 2H_0, \psi_{\text{max}} \sim 0.14$) at resolutions of $16/H_0$, $32/H_0$, and $64/H_0$. The kinetic energy time-series (not shown) are very similar, and the parametric instability develops at all three resolutions.

\subsection{Riemann solver and reconstruction scheme study}
\label{APPENDIX__ParameterStudy_RiemannSolverInterpolationStudy}
To check the effects of numerical scheme on the results, we repeated the large warp amplitude simulation  ($A = 28H_0, \psi_{\text{max}} \sim 1.95$) with different combinations of Riemann solver (HLLC vs HLL) and reconstruction scheme (linear vs parabolic). The four combinations yield qualitatively similar results, and the disc breaks in all four cases.

\subsection{Boundary condition study}
\label{APPENDIX__ParameterStudy_BoundaryConditionStudy}
In our fiducial simulation ($A = 28H_0,\psi_{\text{max}} \sim 1.95$) we employed reflective vertical boundaries for simplicity. More realistic outflow boundary conditions allow the fluid to escape through the vertical boundaries. We find very little difference in the time-series of radial and of vertical kinetic energy (not shown). The mass loss in the outflow zBC simulation is negligible (around $5\times10^{-5}\%$ of the initial mass in the box).

\section{Inviscid bending wave theory}
\label{APPENDIX_InviscidBendingWaveTheory}
Here we derive two predictions of inviscid linear bending wave theory: the Reynolds stress and the expected change in surface density, both to second order in the wave amplitude $A$ For simplicity, in this section we work in units such that  $\Omega_0=c_{\text{s}0}=1$. The enthalpy is given by $h=\ln(\rho/\rho_0)$, where $\rho_0$ is the initial density in the midplane.

\subsection{Expansion to second order in wave amplitude}
The governing equations are given by Equations (\ref{theory1})--(\ref{theory4}). Let us start by expanding each fluid quantity to second order in the wave amplitude $A$:
\begin{align}
  &u_x=Au_1(x,z,t)+A^2u_2(x,z,t)+\cdots,\\
  &\delta u_y=Av_1(x,z,t)+A^2v_2(x,z,t)+\cdots,\\
  &u_z=Aw_1(x,z,t)+A^2w_2(x,z,t)+\cdots,\\
  &h=h_0(z)+Ah_1(x,z,t)+A^2h_2(x,z,t)+\cdots.
\end{align}
We substitute the expansions above into the governing equations, and equate coefficients at different orders in the wave amplitude $A$.

\subsection{Solution at zeroth and first order}
At zeroth order, $O(A^0)$, we obtain $0=-z-\p_zh_0$, which has solution $h_0=-\f{1}{2}z^2$ and corresponds to the initial Gaussian density distribution.

At first order, $O(A^1)$, the solutions satisfy the linearized equations (\ref{linearized1})--(\ref{linearized4}). As explained in Section \ref{THEORY_BendingWaveAnsatz}, we propose ansatzes of the form
\begin{align}
  &u_1=\hat u_1(t)z\cos(kx),\label{firstorderansatz1}\\
  &v_1=\hat v_1(t)z\cos(kx),\label{firstorderansatz2}\\
  &w_1=\hat w_1(t)\sin(kx),\label{firstorderansatz3}\\
  &h_1=\hat h_1(t)z\sin(kx)\label{firstorderansatz4}.
\end{align}
Substituting these ansatzes into the linearized equations, we obtain a set of evolution equations for the amplitudes $\hat u_1(t)$, etc. As explained in Section \ref{THEORY_ModeAmplitudes}, the solutions to these equations which satisfy the initial conditions $\hat u_1(0)=\hat v_1(0)=\hat w_1(0)=0$, $\hat h_1(0)=-1$ are
\begin{align}
  &\hat u_1=\f{\omega_+\sin(\omega_+t)-\omega_-\sin(\omega_-t)}{\omega_++\omega_-},\\
  &\hat v_1=\f{\cos(\omega_+t)-\cos(\omega_-t)}{2(\omega_++\omega_-)},\\
  &\hat w_1=\f{\sin(\omega_+t)+\sin(\omega_-t)}{\omega_++\omega_-},\\
  &\hat h_1=-\left[\f{\omega_+\cos(\omega_+t)+\omega_-\cos(\omega_-t)}{\omega_++\omega_-}\right],
\end{align}
where $\omega_\pm$ are given by Equation \ref{EQUATION_Theory_DispesionRelation2}.

\subsection{Solution to second order}
At second order, $O(A^2)$, the solutions must satisfy
\begin{align}
  &\p_tu_2+(u_1\p_x+w_1\p_z)u_1-2v_2=-\p_xh_2,\label{secondorder1}\\
  &\p_tv_2+(u_1\p_x+w_1\p_z)v_1+\tfrac{1}{2}u_2=0,\label{secondorder2}\\
  &\p_tw_2+(u_1\p_x+w_1\p_z)w_1=-\p_zh_2,\label{secondorder3}\\
  &\p_th_2+(u_1\p_x+w_1\p_z)h_1+u_2\p_zh_0=-(\p_xu_2+\p_zw_2)\label{secondorder4}.
\end{align}
Here the second-order quantities are forced by products of first-order quantities. Since the first-order solution involves the first antisymmetric mode ($n=1$) of the disc, at second order the symmetric modes $n=0$ and $n=2$ are forced. In an isothermal disc these involve the Hermite polynomials $\text{He}_0(z)=1$ and $\text{He}_2(z)=z^2-1$, which are orthogonal with respect to a mass-weighted vertical integration. The general form of the second-order solution is
\begin{align}
  &u_2=u_{20}(x,t)+u_{22}(x,t)(z^2-1),\\
  &v_2=v_{20}(x,t)+v_{22}(x,t)(z^2-1),\\
  &w_2=w_{22}(x,t)z,\\
  &h_2=h_{20}(x,t)+h_{22}(x,t)(z^2-1).
\end{align}
Evolution equations for each of the coefficients $u_{20}(x,t)$, etc., can be obtained by substituting these forms into equations (\ref{secondorder1})--(\ref{secondorder4}) and decomposing them into Hermite polynomials. We are most interested in the $n=0$ components (denoted $u_{20}$, $v_{20}$ and $h_{20}$) because they determine the changes in surface density and horizontal mass fluxes at second order in the wave amplitude. Note in particular that $z^2$ (e.g.\ the product of two sloshing velocity components) is decomposed as $1+(z^2-1)$, which involves both the $n=0$ and $n=2$ modes.

The forcing terms that are products of first-order quantities [which are $\propto\cos(kx)$ or $\sin(kx)$] involve the wavenumbers $k+k=2k$ and $k-k=0$. Bearing in mind also the reflectional symmetry in $x$, it follows that the solution at second order has the structure
\begin{align}
  &u_{20}(x,t)=\hat u_{20}(t)\sin(2kx),\\
  &v_{20}=\hat v_{20}(t)\sin(2kx),\\
  &h_{20}=\hat h_{20}(x,t)\cos(2kx)+\bar h_{20}(t).\label{h20}
\end{align}
Evolution equations for each of the amplitudes $\hat u_{20}(t)$, etc., can be obtained by a Fourier decomposition of the equations for $u_{20}$, etc. They are
\begin{align}
  &\p_t\hat u_{20}-\tfrac{1}{2}k\hat u_1^2+\tfrac{1}{2}\hat w_1\hat u_1-2\hat v_{20}=2k\hat h_{20},\\
  &\p_t\hat v_{20}-\tfrac{1}{2}k\hat u_1\hat v_1+\tfrac{1}{2}\hat w_1\hat v_1+\tfrac{1}{2}\hat u_{20}=0,\\
  &\p_t\hat h_{20}+\tfrac{1}{2}k\hat u_1\hat h_1-\tfrac{1}{2}\hat w_1\hat h_1=-2k\hat u_{20},\\
  &\p_t\bar h_{20}+\tfrac{1}{2}k\hat u_1\hat h_1+\tfrac{1}{2}\hat w_1\hat h_1=0.
\end{align}
Since $\hat u_1$ and $\hat w_1$ are combinations of $\sin(\omega_+t)$ and $\sin(\omega_-t)$, while $\hat v_1$ and $\hat h_1$ are combinations of $\cos(\omega_+t)$ and $\cos(\omega_-t)$, the solution of these equations is such that $\hat u_{20}$ is a combination of $\sin(2\omega_+t)$, $\sin(2\omega_-t)$, $\sin((\omega_++\omega_-)t)$ and $\sin((\omega_+-\omega_-)t)$, while $\hat v_{20}$, $\hat h_{20}$ and $\bar h_{20}$ are combinations of the equivalent cosines, together with constant terms. All of the coefficients can be determined analytically from these equations and the initial conditions\footnote{The initial values for our set-up are $u_{20}=v_{20}=0$ and $h_{20}=-\f{1}{2}\sin^2(kx)$, i.e.\ $\hat u_{20}=\hat v_{20}=0$ and $\hat h_{20}=-\bar h_{20}=\f{1}{4}$; the Eulerian enthalpy perturbation at second order in our purely vertical displacement is $\f{1}{2}\xi_z^2\p_z^2h_0$.}; we omit the details.

\subsection{Surface density change to second order}
We are now in a position to derive an expression for the surface density change to second order in the warp amplitude, i.e. $\Sigma_2$. Let us start with the definition of the density in terms of the enthalpy:

\begin{align}
  \rho&=\rho_0\,\ee^h=\rho_0\,\ee^{-z^2/2}\left[1+Ah_1+A^2\left(\f{1}{2}h_1^2+h_2\right)+\cdots\right].
\end{align}
The surface density can then be found from $\Sigma=\int\rho\,\dd z$. Since $h_1\propto z$, there is no first-order change in surface density. To second order, using the orthogonality of the Hermite polynomials, we find $\Sigma=\Sigma_0+A^2\Sigma_2+\cdots$, where $\Sigma_0=\sqrt{2\pi}\,\rho_0 H_0$ is the initial uniform surface density and
\begin{align}
    \f{\Sigma_2}{\Sigma_0} = \f{1}{2}\hat h_1^2\sin^2(kx)+\hat h_{20}\cos(2kx)+\bar h_{20}.
\end{align}
Note that the sine term comes from substituting Equation \ref{firstorderansatz4} for $h_1$, and the cosine term from from substituting Equation \ref{h20} for $h_2$. The term $\hat h_{22}(z^2-1)$ vanishes upon mass-weighted vertical integration. The mean surface density should be invariant: $\f{1}{4}\hat h_1^2+\bar h_{20}=\cst$. Indeed, the time-derivative of this quantity, $\f{1}{2}\hat h_1\,\p_t\hat h_1+\p_t\hat h_{20}$, vanishes by virtue of the evolution equations for $\hat h_1$ and $\bar h_{20}$, and the initial value of this quantity is zero. Therefore the fractional surface density variation is
\begin{equation}
  \f{A^2\Sigma_2}{\Sigma_0}=A^2\left(\hat h_{20}-\f{1}{4}\hat h_1^2\right)\cos(2kx).
\end{equation}
The quantity in brackets can be expressed, using the solution of the first- and second-order problems, as a combination of $\cos(2\omega_+t)$, $\cos(2\omega_-t)$, $\cos((\omega_++\omega_-)t)$ and $\cos((\omega_+-\omega_-)t)$, plus a constant. We omit details of this calculation. Of most interest are the last two terms, which have low or zero frequency and are the dominant ones after an orbit-averaging of the surface density. When these coefficients are evaluated in the long-wavelength limit $k\ll1$, the leading approximation gives
\begin{equation}
    \frac{\langle A^2\Sigma_2 \rangle_{\tau}}{\Sigma_0} \approx-\f{A^2}{4}k^2 [1-\cos(kt)]\cos(2kx),
\end{equation}
which is the same as Equation \ref{EQUATION_FractionalChangeSurfaceDensity}, once we substitute for $A = \psi_\text{max}/k$, put back $c_{\text{s}0}$ and write the argument of the cosine term in terms of the phase of the bending wave $t_b = (1/2)c_{\text{s}0}k t$.

\subsection{Reynolds stress to second order}
Next let us derive the Reynolds stress to second order in the warp amplitude, i.e. $R_{xy}=A^2R_{xy2}+\cdots$. The vertically-averaged Reynolds stress to second-order in the amplitude is given by
\begin{align}
     \int\rho u_x\delta u_y\,\dd z=\Sigma_0A^2\hat u_1\hat v_1\cos^2(kx)+\cdots.
\end{align}
Substituting the expressions for $\hat u_1$ and $\hat v_1$ in the long-wavelength limit and orbit-averaging (which allows us to drop all but the low-frequency components), we find that
\begin{equation}
  \langle R_{xy} \rangle_{\tau} \approx A^2\left(-\f{1}{16}k\right)\cos^2(kx)\sin(kt).
\end{equation}

It turns out that the change in surface density cannot be accounted for simply by the angular-momentum transport associated with the Reynolds stress $R_{xy}$. The orbit-averaged and vertically integrated angular-momentum equation at second order in the wave amplitude has an important contribution from an azimuthal mass flux (a zonal flow) that evolves along with the bending wave. This zonal flow is in nearly geostrophic balance with a radial force associated with the Reynolds stress component $R_{xx}$, which is larger than $R_{xy}$. Remarkably, if the zonal flow is neglected in the angular-momentum equation, the expected change in surface density due to $R_{xy}$ is (in the long-wavelength limit) equal and opposite to that calculated above and measured in the low-amplitude warp simulations.

\section{Tables of Simulations}
\label{APPENDIX_TablesOfSimulations}

\begin{table*}
\centering
\caption{Large warp amplitude simulations: $A = 28H_0$, $\psi_{\text{max}} \sim 1.95$. The fiducial simulation is FreeWarp17, shown in red. The columns indicate: box size $[L_x,L_y,L_z]$  (in units of the initial mid-plane scale-height $H_0$), resolution (in cells per $H_0$), viscosity, whether the simulation is in the diffusive or wavelike regime, maximum initial warp amplitude $\psi_{\text{max}}$ and $A=\psi_{\text{max}}/k$ (where $k=2\pi/L_x$ is the radial wavenumber of the warp), vertical boundary conditions (zBC), Riemann solver, interpolation scheme, runtime, and final outcome of the simulation.}
\label{TABLE_LargeWarpAmplitudeSimulations}
 	\begin{tabular}{lcccccccccr}
		\hline
		Run	& Box Size & Res & $\alpha$ & Regime & $\psi_{\text{max}}\, (A/H_0)$ & zBC &  Riemann & Reconstr. & Runtime & Outcome \\ 
        \hline
        FreeWarp24 & [90,0.08,68] & $32/H_0$ & inviscid & wavelike & 1.95 (28) & reflective & HLLC & parabolic & 50 & breaking \\
        FreeWarp17 & [90,0.08,68] & $64/H_0$ & inviscid & wavelike & 1.95 (28) & reflective & HLLC & parabolic & 50 & breaking  \\
        FreeWarp26 & [90,0.08,68] & $128/H_0$ & inviscid & wavelike & 1.95 40 & reflective & HLLC & parabolic & 40 & breaking  \\
        \hline
        FreeWarp28 & [90,0.08,68] & $32/H_0$ & inviscid & wavelike & 1.95 (28) & reflective & HLLC & linear & 50 & breaking \\
        FreeWarp27 & [90,0.08,68] & $32/H_0$ & inviscid & wavelike & 1.95 (28) & reflective & HLL & parabolic & 50 & breaking \\
        FreeWarp11 & [90,0.08,68] & $32/H_0$ & inviscid & wavelike & 1.95 (28) & reflective & HLL & linear & 50 & breaking \\
        \hline
        FreeWarp25 & [90,0.08,68] & $64/H_0$ & inviscid & wavelike & 1.95 (28) & outflow & HLLC & parabolic & 50 & breaking  \\   
        \hline
	\end{tabular}
\end{table*}

\begin{table*}
\centering
\caption{Warp amplitude study.}
\label{TABLE_WarpAmplitudeStudy}
 	\begin{tabular}{lcccccccccr}
		\hline
		Run	& Box Size & Res & $\alpha$ & Regime & $\psi_{\text{max}}\, (A/H_0)$ & zBC &  Riemann & Reconstr. & Runtime & Outcome \\ 
        \hline
        FreeWarp02 & [90,0.08,48] & $16/H_0$ & inviscid & wavelike & 1.26 (18) & reflective & HLL & linear & 25 & breaking \\
        FreeWarp18 & [90,0.08,48] & $32/H_0$ & inviscid & wavelike & 1.26 (18) & reflective & HLL & parabolic & 40 & breaking \\
        FreeWarp08b & [90,0.08,48] & $64/H_0$ & inviscid & wavelike & 1.26 (18) & reflective & HLLC & parabolic & 50 & breaking \\
        \hline
        FreeWarp09 & [90,0.08,48] & $16/H_0$ & inviscid & wavelike & 0.63 (9) & reflective & HLL & linear & 40 & instability \\
        FreeWarp29 & [90,0.08,48] & $64/H_0$ & inviscid & wavelike & 0.63 (9) & reflective & HLLC & parabolic & 50 & instability \\
        \hline
        FreeWarp10 & [90,0.08,48] & $16/H_0$ & inviscid & wavelike & 0.14 (2) & reflective & HLL & linear & 40 & instability \\
        FreeWarp19 & [90,0.08,48] & $64/H_0$ & inviscid & wavelike & 0.14 (2) & reflective & HLLC & parabolic & 50 & instability \\
	\end{tabular}
\end{table*}

\begin{table*}
\centering
\caption{Viscosity study. Note that the diffusive regime is characterized by $\alpha > k H_0 $. Here $k = 2\pi/\lambda$, with $\lambda = L_x = 90H_0$, is the radial wavenumber of the sinusoid used to initialize the warp. Thus simulations with $\alpha \lesssim  0.07$ fall within the wavelike regime.}
\label{TABLE_ViscosityStudy}
 	\begin{tabular}{lcccccccccr}
		\hline
		Run	& Box Size & Res & $\alpha$ & Regime & $\psi_{\text{max}}\, (A/H_0)$ & zBC &  Riemann & Reconstr. & Runtime & Outcome \\  
        \hline
        FreeWarp22 & [90,0.08,68] & $32/H_0$ & $0.003$ & wavelike & 1.95 (28) & reflective & HLL & parabolic & 50 & breaking \\
        FreeWarp23 & [90,0.08,68] & $32/H_0$ & $0.03$ & wavelike & 1.95 (28) & reflective & HLL & parabolic & 50 & breaking inhibited \\
        \hline
        FreeWarp20 & [90,0.08,48] & $64/H_0$ & $0.003$ & wavelike & 0.14 (2) & reflective & HLLC & parabolic & 50 & instability \\
        FreeWarp21 & [90,0.08,48] & $64/H_0$ & $0.03$ & wavelike & 0.14 (2) & reflective & HLLC & parabolic & 50 & damped bending wave  \\
        \hline
	\end{tabular}
\end{table*}

\bsp	
\label{lastpage}
\end{document}